\documentclass[useAMS,usenatbib]{mn2e}
\usepackage{graphicx}

\newcommand{\ml}{$ M/L$}

\newcommand{\simlt}{\lower.5ex\hbox{$\; \buildrel < \over \sim \;$}}
\newcommand{\simgt}{\lower.5ex\hbox{$\; \buildrel > \over \sim \;$}}
\newcommand{\cahk}{$\rm CaHK$}
\newcommand{\mgf}{$\rm Mg4780$}
\newcommand{\nad}{$\rm NaD$}
\newcommand{\tioi}{$\rm TiO1$}
\newcommand{\tioiio}{$\rm TiO2_{SDSS}$}
\newcommand{\tioii}{$\rm TiO2$}

\newcommand{\naii}{$\rm Na8190_{SDSS}$}
\newcommand{\cat}{$\rm CaT$}
\newcommand{\cai}{$\rm Ca1$}
\newcommand{\caii}{$\rm Ca2$}
\newcommand{\caiii}{$\rm Ca3$}
\newcommand{\mgfep}{$\rm [MgFe]'$}
\newcommand{\mgfe}{$\rm [MgFe]$}
\newcommand{\fet}{$\rm Fe3$}
\newcommand{\hbo}{$\rm H\beta_o$}
\newcommand{\hb}{$\rm H\beta$}
\newcommand{\mgb}{$\rm Mgb5177$}
\newcommand{\hgf}{$\rm H\gamma_F$}
\newcommand{\caf}{$\rm Ca4227$}
\newcommand{\cafr}{$\rm Ca4227r$}
\newcommand{\kms}{\,km\,s$^{-1}$}
\newcommand{\afe}{$[\alpha/{\rm Fe}]$}
\newcommand{\afep}{$[Z_{Mg}/Z_{Fe}]$}
\newcommand{\mlstar}{$M_\star/L$}
\title[Constraining the IMF of Early-Type Galaxies] {SPIDER VIII --
  Constraints on the Stellar Initial Mass Function of Early-type
  Galaxies from a Variety of Spectral Features.}

\author[La Barbera et al.]{F. La Barbera$^{1}$\thanks{E-mail:
    labarber@na.astro.it (FLB); i.ferreras@ucl.ac.uk (IF)},
I. Ferreras$^{2\star}$,
A. Vazdekis$^{3,4}$,
I.G. de la Rosa$^{3,4}$,
R.R. de Carvalho$^{5}$, 
\newauthor
M. Trevisan$^{5}$,
J. Falc\'on-Barroso$^{3, 4}$,
E. Ricciardelli$^{6}$
\\ 
$^{1}$INAF -- Osservatorio Astronomico di Capodimonte, Napoli, Italy\\ 
$^{2}$MSSL, University College London, Holmbury St Mary,
  Dorking, Surrey RH5 6NT, UK\\
$^{3}$Instituto de Astrof\'\i sica de Canarias (IAC), E-38200
  La Laguna, Tenerife, Spain\\
$^{4}$Departamento de Astrof\'\i sica, Universidad de La
  Laguna, E-38205, Tenerife, Spain\\
$^{5}$Instituto Nacional de Pesquisas Espaciais/MCT,
  S. J. dos Campos, Brazil\\
$^{6}$Departament d'Astronomia i Astrofisica, Universitat de
  Valencia, C/Dr Moliner 50, E-46100, Burjassot, Valencia, Spain\\
}
\begin{document}

\date{Accepted 2013 May 27. Received 2013 May 27; in original form 2013 February 27}

\pagerange{\pageref{firstpage}--\pageref{lastpage}} \pubyear{2013}

\maketitle

\label{firstpage}

\begin{abstract}
We perform a spectroscopic study to constrain the stellar Initial Mass
Function (IMF) by using a large sample of $24,781$ early-type galaxies
from the SDSS-based SPIDER survey.   { Clear evidence is found of a
  trend  between  IMF and  central  velocity dispersion  ($\sigma_0$),
  evolving  from  a  standard  Kroupa/Chabrier  IMF  at  $\sigma_0\sim
  100$\,\kms\  towards   a  more  bottom-heavy   IMF  with  increasing
  $\sigma_0$,   becoming  steeper  than   the  Salpeter   function  at
  $\sigma_0\simgt  220$\,\kms.   We  analyze  a  variety  of  spectral
  indices,  combining   gravity-sensitive  features,  with   age-  and
  metallicity-sensitive indices,  and we  also consider the  effect of
  non  solar abundance  variations.  The  indices, corrected  to solar
  scale   by  means   of  semi-empirical   correlations,   are  fitted
  simultaneously  with   the  (nearly  solar-scaled)   extended  MILES
  (MIUSCAT)  stellar  population   models.   Similar  conclusions  are
  reached when analyzing the spectra with a hybrid approach, combining
  constraints from  direct spectral fitting in the  optical with those
  from IMF-sensitive indices.   Our analysis suggests that $\sigma_0$,
  rather than  \afe, drives  the variation of  the IMF.   Although our
  analysis cannot  discriminate between a single  power law (unimodal)
  IMF  and a  low-mass ($\simlt  0.5$\,M$_{\odot}$)  tapered (bimodal)
  IMF, robust constraints can be inferred for the fraction in low-mass
  stars at birth.   This fraction (by mass) is  found to increase from
  $\sim 20  \%$ at $\sigma_0  \sim 100$\,\kms, up  to $\sim 80  \%$ at
  $\sigma_0 \sim  300$\,\kms.  However, additional  constraints can be
  provided with stellar  mass-to-light ($M/L$) ratios: unimodal models
  predict $M/L$ significantly larger  than dynamical $M/L$, across the
  whole $\sigma_0$  range, whereas a  bimodal IMF is  compatible.  Our
  results  are  robust  against  individual abundance  variations.  No
  significant  variation is  found in  Na and  Ca in  addition  to the
  expected change  from the correlation between  \afe\ and $\sigma_0$.
}
\end{abstract}

\begin{keywords}
galaxies: stellar content -- galaxies: fundamental parameters -- galaxies: formation
\end{keywords}

%%%%%%%%%%%%%%%%%%%%%%%%%%%%%%%%%%%%%%%%%%%%%%%%%%%%%%%%%%%%%%%%%%%%%%%%
\section{Introduction}
\label{sec:intro}

The formation  of stars in a  gas cloud is a  problem of extraordinary
physical  complexity, remaining  one  of the  major  open problems  in
modern astrophysics.  A way to  constrain this fundamental issue is to
examine   the  stellar   Initial  Mass   Function  (IMF),   i.e.   the
distribution of  the masses  of stars at  birth. The IMF  has critical
implications in the framework of galaxy formation and evolution, as it
sets  the  overall  mass-scale   of  galactic  systems,  controls  the
intensity of  the stellar feedback  processes and drives  the chemical
enrichment abundance patterns.

To date,  the IMF  can only be  directly constrained from  star number
counts in the disk of  the Galaxy. Current observations are consistent
with a universal Kroupa/Chabrier-type IMF, i.e.  a power-law behaviour
at high  mass ($>1$\,M$_\odot$) with  a turn-off  at lower
masses \citep{Kroupa01,Chabrier03},  
%although with a  large scatter at
{ although with a significant uncertainty in the region at masses
  around a few tenths of M$_\odot$ \citep[see][and references
    therein]{Bastian:10}.
%  { Although not well constrained  so far, 
However, the low-mass portion of the IMF is extremely important because }
the mass
fraction in  stars below $1$\,M$_\odot$  varies from $\sim 1/3$  for a
Kroupa IMF up  to $\sim 2/3$ for a \citet{salp:55}  IMF, i.e.  a large
mass fraction of  a stellar population is indeed  expected in the form
of low-mass stars.

An  indirect method to  study the  IMF involves  the use  of dynamical
modelling techniques to constrain  the total and stellar mass-to-light
ratios of  a stellar  system.  These techniques  have been  applied to
star    clusters,   favouring    a    Kroupa-like   IMF    \citep[see,
  e.g.,][]{Bastian:06}.   In  early-type  galaxies  (hereafter  ETGs),
detailed dynamical  modelling of the kinematic properties  in a nearby
sample revealed a systematic increase in the stellar $M/L$ with galaxy
mass  \citep{Cappellari:12a},  a result  confirmed  over a  260-strong
sample    of     ETGs    from    the     ATLAS$^{\rm    3D}$    survey
\citep{Cappellari:12c}, where  the trend is expected to  evolve from a
Kroupa IMF at low  velocity dispersion ($\sigma\sim 80$\,\kms) towards
a Salpeter  function at $\sigma\sim  260$\,\kms. Independent dynamical
studies    have   arrived    at   similar    {conclusions   \citep[see,
  e.g.,][]{Thomas:11,  WCT:12,  Cappellari:12a, Dutton:12,Tortora:13}.}
Strong gravitational lensing over  galaxy scales can also be exploited
to  derive stellar  $M/L$  especially when  the  Einstein radius  only
extends  over  the  central   regions,  where  dark  matter  does  not
complicate  the  issue  of  transforming  total  into  stellar  $M/L$.
Although  strong  lensing studies  of  low-mass  spheroids rejected  a
Salpeter IMF  \citep{FSW:05,FSB:08,ECross:10}, recent work  covering a
wider  range of galaxy  mass hinted  at systematic  variations towards
higher  stellar $M/L$  with increasing  mass~\citep{Auger:10, Treu:10,
  Barnabe:11}.  Hence,  both lensing  and dynamical studies  require a
scenario consistent  with either a  bottom- or a top-heavier  IMF than
Kroupa/Chabrier in  massive ETGs.  In  fact, a stellar  population can
have large $M/L$ because of  a large fraction of either low-mass stars
or  remnants  of massive  stars,  and  neither  dynamical nor  lensing
modelling    is     capable    of    distinguishing     between    the
two~\citep{Cappellari:12a}.

A unique opportunity to investigate more directly the IMF is offered
by gravity-sensitive features in the integrated light of unresolved
stellar populations, as originally proposed
by~\citet{Spinrad:62}. Some of the most discriminant features include
the Na\,I doublet feature at
$\lambda\lambda8183,8195$\,\AA\ \citep[hereafter
  NaI8200]{FaberFrench:80, SchiavonFeH:97}, the Wing-Ford FeH band
\citep{WingFord:69, SchiavonNaD:97} at $9,900$\,\AA,
%which are
prominent in  the atmospheres of low-mass  dwarves, as well  as the Ca
triplet lines  at $\lambda \sim 8600$\,\AA (hereafter  CaT; see, e.g.,
\citealt{Diaz:89,  Cenarro:03}),  which are  strong  in giants,  while
barely detectable in dwarves.   Several other features, mostly related
to Na, Ca, and TiO lines, have been also identified in the literature,
as  potentially  useful  tools  to  constrain  the  dwarf/giant  ratio
(\citealt{CvD12a},       \citealt[hereafter      STK12]{Spiniello:12},
\citealt{Smith:12}). Another  potentially useful tool  is the analysis
of (optical-NIR) broad-band colours~\citep{Vazdekis:1996, Vazdekis:12,
  Ricciardelli:12}.  In this respect,  ETGs offer a unique opportunity
to study the IMF, as  they host homogeneous, old, stellar populations,
with little  amount of  dust, allowing for  a clean analysis  of their
stellar  content.  Nevertheless,  this kind  of analysis  is  far from
being  trivial.   All   dwarf/giant  sensitive  features  represent  a
contribution at the  level of a few percent to  the total galaxy flux,
and these features are also  sensitive to age, metallicity, as well as
abundance  ratio  of   (several)  chemical  elements.   Moreover,  NIR
features,  like NaI8200  and  CaT,  fall in  a  spectral region  often
severely contaminated by sky  emission and telluric absorption.  These
problems  have  hampered  the  use of  gravity-sensitive  features  to
investigate the IMF for a long time.

The first observational attempts to constrain the giant/dwarf ratio in
the  IMF  were  made  by~\citet{Cohen:78}  and~\citet{FaberFrench:80},
towards the centres of M31 and M32, using the NaI8200 feature.  Later,
\citet{Carter:86} extended the study to  a sample of massive ETGs, and
found that NaI8200 was  enhanced, especially in massive galaxies, with
strong  radial gradients.   These studies  concluded in  favour  of an
excess of dwarves (relative to  giants) in the galaxy central regions.
Opposite      conclusions      were     drawn      by~\citet{Hardy:88}
and~\citet{Delisle:92},  based  on NaI8200A,  CaT,  and the  Wing-Ford
band, who explained the radial variation and central strength of these
features as  a metallicity (rather  than IMF) effect. All  these early
works were in fact plagued  by many difficulties, in particular, small
sample sizes, low signal-to-noise ratio and resolution of the spectra,
or  uncertainties in  the  available stellar  population models.   The
recent  availability of  dedicated  instrumentation and  sophisticated
reduction techniques  has opened up  new opportunities to  exploit the
constraining  power  of  (NIR)  IMF-sensitive features,  by  means  of
high-quality spectroscopy. \citet{Cenarro:03} proposed a trend towards
an excess of  low-mass stars in massive galaxies, from  a study of the
CaT  region.  More recently,  \citet{vDC:10,vDC:11} used  the NaI8200A
and Wing-Ford band  to conclude that the IMF  is bottom-heavier than a
Kroupa/Chabrier distribution in a sample  of eight massive ETGs in the
Virgo  and  Coma clusters.   This  result  has  been confirmed,  using
(theoretical) stellar population models with varying element abundance
ratios  and a  full spectral  fitting analysis  by~\citet{CvD12a}, and
further  extended  to  a  set  of  34  ETGs  from  the  SAURON  survey
\citep{CvD12b}.    At  the   same  time,   STK12  and~\citet[hereafter
  FLD13]{Ferreras:13} have  gone beyond a simple  test of universality
for  the IMF, analyzing  large samples  of ETGs  drawn from  the Sloan
Digital Sky Survey  (SDSS), focusing on Na and  TiO spectral features.
Remarkably, FLD13 relied  on completely independent stellar population
models      with     respect     to~\citet{CvD12a},      using     the
MIUSCAT~\citep{Vazdekis:12,   Ricciardelli:12}   spectrally   extended
version    of     the    stellar    population     synthesis    models
MILES~\citep{Vazdekis10}.   Although   all  these  studies   favour  a
non-universal  IMF, with more  massive ETGs  harbouring bottom-heavier
distributions than Kroupa/Chabrier, the debate lingers on.  In fact, a
different picture has been  proposed whereby the variation of spectral
features with  galaxy mass  in ETGs  is driven by  a change  of single
element  (mostly Na  and Ca)  abundance patterns,  rather than  an IMF
variation  \citep[see,   e.g.,][and  references  therein]{worthey:11}.
According to this interpretation, for instance, the NaI8200 (CaT) line
strength  would increase  (decrease) with  galaxy mass  because  of an
over-(under-)abundance  of Na  (Ca)  relative to  Mg  in more  massive
systems. As shown by~\citet{CvD12a}, a way out to break the degeneracy
is by  combining spectral  features sensitive to  IMF with  those more
sensitive to the abundance of single chemical species.  To this effect
one has  to rely  on theoretical models,  rather than  fully empirical
stellar  population libraries, to  model the  impact of  variations in
both   IMF    and   abundance   ratio,   over   a    wide   range   of
wavelength. Therefore,  it is  important to consider  simultaneously a
variety of spectral features from different chemical species.

In this paper, we follow  this approach.  We extend our previous
work (FLD13),  analyzing stacked spectra  of $\sim 24,781$  nearby ($z
\sim 0.07$)  ETGs with exceptionally  high S/N-ratio, spanning  a wide
range of  central velocity dispersion, from $100$  to $300$\,\kms.  As
in  FLD13, we  mainly rely  on extended  MILES  (MIUSCAT) state-of-the-art
stellar population models for the analysis.  The main novelties of the
present contribution  are the  following: (i) we  adopt a wide  set of
spectral  indices (with respect  to, e.g.,  STK12 and  FLD13), fitting
several TiO, Ca, and Na features simultaneously, to constrain the IMF;
(ii)  we  combine   age,  metallicity,  IMF-  and  abundance-sensitive
features in the  analysis; (iii) we use a  semi-empirical technique to
correct spectral  indices to solar-scale,  hence allowing for  a clean
comparison to  the base (nearly  solar-scaled) models and  avoiding to
rely completely  on theoretical  models; (iv) we  adopt a wide  set of
fitting techniques, including a  pure spectral index fitting approach,
and a  hybrid approach, where  direct spectral fitting is  combined to
the constraints from IMF-sensitive indices (as in FLD13); (v) we allow
for  a wide  range  of star-formation  histories,  including one-  and
two-simple   stellar   populations  (hereafter   SSP),   as  well   as
exponentially declining star-formation  models.  All results presented
in  this work  point  consistently to  a  steepening of  the IMF  with
velocity dispersion in ETGs.

The layout of the paper is as follows: Sec.~\ref{sec:data} describes
the sample of ETGs, drawn from the SDSS-based SPIDER survey.
Sec.~\ref{sec:stacks} details our stacking procedures. The stellar
population models used to analyze the stacks are described in
Sec.~\ref{sec:models}.  Sec.~\ref{sec:indices} presents the spectral
indices used in this work, showing their sensitivity to age,
metallicity, and IMF.  In Sec.~\ref{sec:afe}, we explore the
correlation of the targeted spectral indices with abundance ratio.
The different fitting methods to interpret the spectra are described
in Sec.~\ref{sec:fitting}, while results from different fits are
discussed in Sec.~\ref{sec:results}.  Sec.~\ref{sec:col_ml} shows the
ability of the best-fit models, with a varying IMF, to match
independent constraints, i.e.  mass-to-light ratios and optical-NIR
colours.  A summary is given in Sec.~\ref{sec:summary}.

%%%%%%%%%%%%%%%%%%%%%%%%%%%%%%%%%%%%%%%%%%%%%%%%%%%%%%%%%%%%%%%%%%%%%%%
\begin{figure*}
\begin{center}
\leavevmode
\includegraphics[width=15cm]{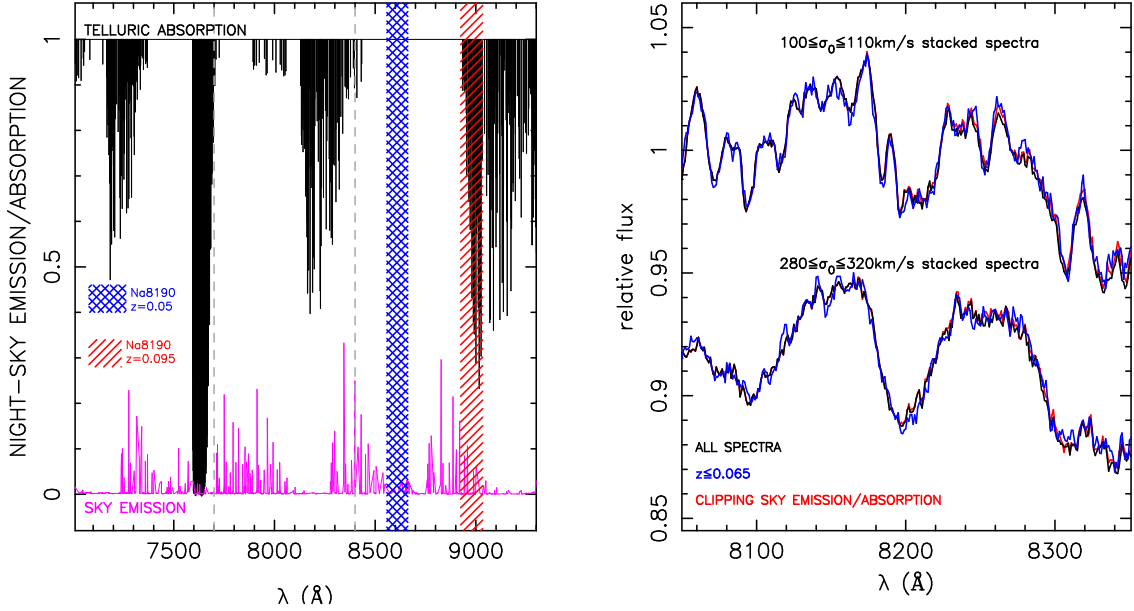}
\end{center}
\caption{Left: telluric (black) and emission (red) lines of the
  night-sky are plotted in the wavelength range from $\sim 7000$ to
  $\sim 9300$\,\AA.  The blue and red shaded regions mark the
  observed spectral window of \naii\ for the lowest ($z\sim 0.05$) and
  highest ($z\sim0.095$) redshift limits of our sample. Notice that at
  $z\sim 0.05$, the \naii\ is unaffected by sky contamination. Right:
  stacked spectra with $\sigma \sim 100$\,\kms\ (upper curves) and
  $\sigma \sim 300$ (lower curves; as labelled) are plotted over a
  $300$\,\AA\ width region, centered on the \naii\ feature.  All
  spectra have been normalized by the median flux in the given
  spectral region.  The high-$\sigma_0$ stacks have been arbitrarily
  shifted downwards (by $-0.085$) for displaying purposes. Different
  colours correspond to the cases where all (black), only $z \le 0.065$
  (blue) spectra are stacked, and the case where sky lines are clipped
  out of the stacking procedure (red).  }
%\contcaption{.}
\label{fig:sky_stacks}
\end{figure*}
%%%%%%%%%%%%%%%%%%%%%%%%%%%%%%%%%%%%%%%%%%%%%%%%%%%%%%%%%%%%%%%%%%%%%%%

%%%%%%%%%%%%%%%%%%%%%%%%%%%%%%%%%%%%%%%%%%%%%%%%%%%%%%%%%%%%%%%%%%%%%%%%
\section{Data}
\label{sec:data}

%%%%%%%%%%%%%%%%%%%%%%%%%%%%%%%%%%%%%%%%%%%%%%%%%%%%%%%%%%%%%%%%%%%%%%%%
\subsection{Sample}
\label{sec:sample}

The SPIDER\footnote{Spheroids Panchromatic Investigation in Different
  Environmental Regions~\citep{SpiderI}} sample consists of 39,993
nearby (0.05$<$z$<$0.095) ETGs, selected from Data Release 6 of the
Sloan Digital Sky Survey (SDSS-DR6;~\citealt{SDSS:DR6}), as described
in~\citet[hereafter Paper I]{SpiderI}.  Galaxies are selected to have
r-band Petrosian magnitude brighter than $-20$, corresponding to an
absolute B-band magnitude of $-19$, where the separation between the
two families of {\it bright} and {\it ordinary} ellipticals
occurs~\citep{capaccioli1992, graham&guzman2003, Trevisan:12}.  The
sample covers two orders of magnitude in dynamical mass, from $2\times
10^{10}$ to $10^{12}$M$_\odot$~\citep{delaRosa:12}.  Following
~\citet{BER03}, we have defined ETGs as bulge-dominated systems (i.e.
SDSS attribute $fracDev_r \!  > \!  0.8$, where $fracDev_r$ measures
the fraction of galaxy light better fitted by a de~Vaucouleurs, rather
than an exponential, law), featuring passive spectra within the SDSS
fibres (SDSS attribute $eClass \!  < \!  0$, where $eClass$ indicates
the spectral type of a galaxy based on a principal component
analysis).  All galaxies have spectroscopy as well as central velocity
dispersions, $\sigma_0$, available from SDSS, and we select those with
no spectroscopic warning on (i.e.  SDSS $zWarning$ attribute set to
zero). The spectra, ranging from $3800$ to $9200$\,\AA, are retrieved
from SDSS-DR7~\citep{SDSS:DR7}, de-redshifted to a common rest-frame
and corrected for foreground Galactic extinction (see Paper I).  The
resulting restframe spectral range varies from $3620$--$8760$\,\AA, at
lowest redshift ($z=0.05$), to $3470$--$8400$\,\AA\ at the upper
redshift limit of the sample ($z=0.095$). This spectral coverage gives
us the opportunity to study several IMF-sensitive spectral features in
ETGs (see Sec.~\ref{sec:indices}), like the Na doublet at $\lambda
\sim 8190$\,\AA\ \citep[hereafter Na8190]{SchiavonNaD:97}, and the
Calcium triplet (CaT) at $\lambda \sim
8600$\,\AA\ (see~\citealt{Cenarro2001} and references therein). Also,
the superb quality of SDSS spectra, with a flux calibration accuracy
at the $1\%$ level~\footnote{~\citet{SDSS:DR6} tested the
  spectrophotometric accuracy of SDSS data by estimating the relative
  difference between observed and model spectra of white dwarfs. The
  difference is smaller than $\sim 1\%$ from $\sim 3800$ to $8000$\,
  \AA, except for regions possibly affected by interstellar
  absorption.  }, is suitable to detect the expected variations -- at
the level of a few percent -- from a varying IMF.  We come back
to this point at the end of Sec.~\ref{sec:stacks}.

For the present study, we select the subsample of $38,447$ ETGs with
$\sigma_0 \ge 100$\,\kms, as the SPIDER sample becomes significantly
incomplete below this limit (see~\citealt{SpiderII}).  Following our
previous work~\citep[hereafter FLD13]{Ferreras:13}, we further limit
the analysis to objects ($N_{ETGs}=33,095$) with low internal
extinction, i.e.  a colour excess $E(B-V) < 0.1$~mag~\footnote{ We
  note that relaxing the threshold in colour excess does not affect our
  results.  For instance, for $E(B-V) < 0.2$, the number of objects in
  each $\sigma_0$ bin changes by a few percent. Since we
  median-combine the spectra in each bin, the impact of this variation
  on line strengths is completely negligible (e.g.  the
  \naii\ equivalent width changes by less than $0.01$~\AA\ in all
  bins).  }.  The $E(B-V)$ is measured by fitting the SDSS spectrum of
each ETG with the spectral fitting code {\tt STARLIGHT}~\citep{CID05},
assuming a~\citet{Cardelli} extinction law (see~\citealt{Swindle},
hereafter Paper V, for details).  Each spectrum is fitted in the
spectral range of $3900$ to $7350$\,\AA, with three different sets of
simple stellar population (SSP) models, namely, \citet[BC03]{BrC03},
updated Charlot \& Bruzual (often refereed to as CB07), and MILES
models~\citep[MI10]{Vazdekis10}. Each set of models spans a wide range
of ages and metallicities (see Paper V for details).  The IMFs used
are Scalo (BC03), Chabrier (CB07), and Kroupa (MI10).  We estimate
$E(B-V)$ by averaging output values from the three {\tt STARLIGHT}
runs~\footnote{Runs with different models provide fairly consistent
  extinction estimates.  The mean difference of $E(B-V)$ amounts to
  about $-0.03 $ and $-0.035$ for BC03-MI10 and CB07-MI10,
  respectively.}.  Notice that this procedure is aimed at removing objects
with significant internal extinction from the analysis, while not
accounting for possible IMF variations within the sample (i.e.  the
hypothesis we want to test).

Finally, we restrict the sample to spectra with the highest
signal-to-noise (S/N) ratio (see Sec~\ref{sec:stacks}), resulting in a
dataset comprising $24,781$ galaxies (i.e. 62\% of the original SPIDER
catalogue). In addition, for the analysis of the \cat\ index the
sample is further reduced to $3,877$~ETGs as we need to exclude
galaxies with redshift $z>0.06$, where the index falls close to the
red limit of the SDSS spectra.

%Since  we   further  restrict  the  stacking  to   spectra  with  best
%signal-to-noise (S/N) ratio and  we exclude galaxies at redshift above
%$z = 0.06$  in the \cat\ spectral region  ($\lambda > 8470$~$\AA$; see
%below),  the  final  sample  of  ETGs consists  of  $24,781$  (out  of
%$33,095$) objects,  this number dropping  down to $3,877$ ETGs  in the
%\cat\ spectral range.

%%%%%%%%%%%%%%%%%%%%%%%%%%%%%%%%%%%%%%%%%%%%%%%%%%%%%%%%%%%%%%%%%%%%%%%%
\subsection{Stacked spectra}
\label{sec:stacks}
Dwarf (relative to giant) stars contribute only up to a few percent to
the integrated light of  galaxies. Hence, constraining the stellar IMF
from  gravity-sensitive  features   requires  spectra  with  exquisite
signal-to-noise    ratio,     typically    a    few     hundreds    or
more~\citep[hereafter  CvD12a]{CvD12a}.   Unfortunately,  single  SDSS
spectra usually do not have such high S/N.  For the SPIDER sample, the
average  S/N over  the Na8190  passband ($8180-8200$\,\AA)  amounts to
$\sim  15$ (hereafter  quoted per  \AA ),  ranging from  $\sim  12$ at
$\sigma_0 \sim 100$\,\kms\ to  $\sim 25$ at $\sigma_0 \sim 300$\,\kms.
In  order to  test  for variations  of  the IMF,  we  rely on  stacked
spectra.

%%%%%%%%%%%%%%%%%%%%%%%%%%%%%%%%%%%%%%%%%%%%%%%%%%%%%%%%%%%%%%%%%%%%%%%
\begin{figure}
\begin{center}
\leavevmode
\includegraphics[width=8cm]{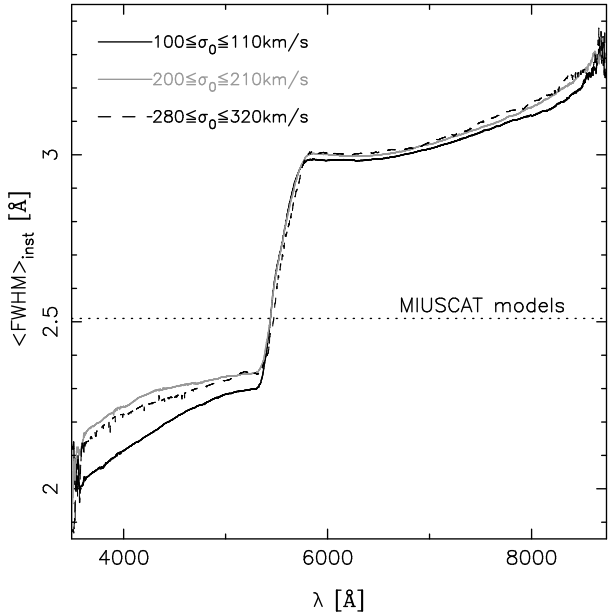}
\end{center}
\caption{Median resolution of stacked spectra of ETGs, for three bins
  of $\sigma_0$, as labelled. The horizontal dotted line marks the
  FWHM spectral resolution of the extended MILES (MIUSCAT)
  models~\citep{Jesus:11}.  }
%\contcaption{.}
\label{fig:res}
\end{figure}
%%%%%%%%%%%%%%%%%%%%%%%%%%%%%%%%%%%%%%%%%%%%%%%%%%%%%%%%%%%%%%%%%%%%%%%

As described in~FLD13, we have assembled 18 stacked spectra in narrow
bins of $\sigma_0$, over the available range of $100$ to $320$\kms
(see Sec.~\ref{sec:sample}).  The bins have a width of 10\kms, except for
the last two bins, where, because of the smaller number of galaxies,
we adopt the range [260,280], and [280,320]\kms, respectively.  We bin
the sample according to $\sigma_0$, because the underlying stellar
populations of ETGs are known to correlate strongly with velocity
dispersion \citep{Bernardi:05}. For each $\sigma_0$ bin, we median
stack the available spectra, considering only pixels with no SDSS flag
raised\footnote{i.e. no bad pixels, flat field issues, etc. For
  details see {\tt http://www.sdss.org/dr6/dm/flatFiles/spSpec.html}}. In
order to avoid possible biases related to differences in $S/N$ within
the $\sigma_0$ bins, we exclude spectra in the lowest quartile of the
$S/N$ distribution in each bin, resulting in a total number of
$24,781$ ETGs used to create the stacks.  Moreover, in the spectral
region of CaT ($\lambda > 8470$\,\AA), we exclude spectra of galaxies
at redshift $z > 0.06$, for which the CaT feature is redshifted beyond
the SDSS upper spectral limit of $\sim 9200$\,\AA.  This selection
leads to a smaller, but still significant, sample of $3,877$ ETGs used
for the stacking at $\lambda > 8470$\,\AA.  Notice that excluding
spectra at $z > 0.06$ over the entire available spectral range does
not change significantly the other relevant IMF-sensitive features
explored in the present work.  For instance, averaging over all stacks,
the equivalent width of \naii\ would change by less than 1~$\sigma$
(between the full stacks and those with $z \le 0.06$).  This proves
that the EWs of CaT can be meaningfully compared to those of spectral
features at bluer wavelengths.  Relevant properties of each stack are
summarized in Tab.~\ref{tab:stacks}, where we report the $\sigma_0$
range of all bins, the number of ETGs per bin, and the median S/N
ratio of stacked spectra, computed within the central passband of five
representative spectral indices used in this
work~(Sec.~\ref{sec:indices}).  The stacked spectra feature a
remarkably high $S/N$, larger than a few hundreds throughout the whole
spectral range.  At $\sigma_0 \sim 150$\,\kms, the $S/N$ of the stacks
peaks up to a maximum value of $\sim 1800$ ($\sim 800$) for the
IMF-sensitive \tioii\ (\naii) spectral feature, at $\lambda \sim
6200$\,\AA\ ($8200$\,\AA). In addition, in Sec.~\ref{sec:afe}, we
probe the effect of variations in \afe\ on our analysis by further
splitting the sample according to \afe, at fixed $\sigma_0$, we refer
the reader to that section for details.

%%%%%%%%%%%%%%%%%%%%%%%%%%%%%%%%%%%%%%%%%%%%%%%%%%%%%%%%%%%%%%%%%%%%%%%
\begin{figure}
\begin{center}
\leavevmode
\includegraphics[width=8.5cm]{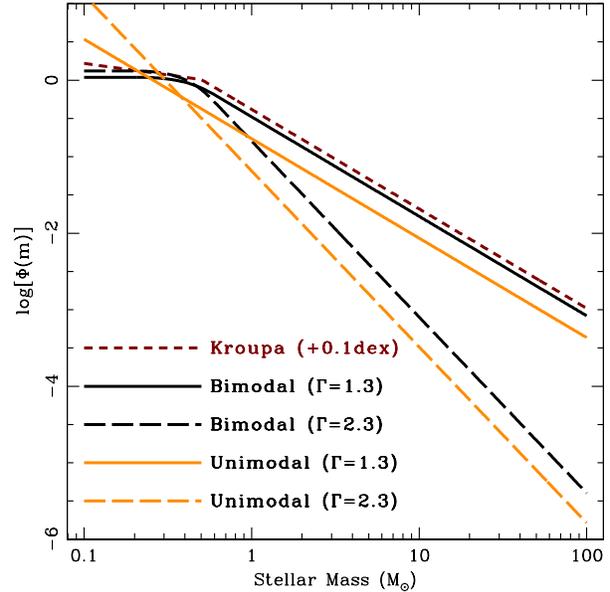}
\end{center}
\caption{ { Examples of the unimodal (orange) and bimodal (black) Initial Mass
  Functions used in this paper. Note the $\Gamma=1.3$ unimodal case
  (orange solid line) matches the \citet{salp:55} IMF (defined as a
  power law with index 1.35), whereas the bimodal case with
  $\Gamma=1.3$ (black solid line) maps the Kroupa Universal (2001) IMF
  (red dashed line), shifted by 0.1\,dex in the figure for
  illustration purposes.} }
\label{fig:imf}
\end{figure}
%%%%%%%%%%%%%%%%%%%%%%%%%%%%%%%%%%%%%%%%%%%%%%%%%%%%%%%%%%%%%%%%%%%%%%%

A  major  source  of  concern  when  studying  NIR  spectral  features
(e.g. \naii\ and  \cat) is the possible sky  contamination of observed
spectra,  including telluric  absorption and  emission lines  from the
night sky.  We performed extensive tests, all of them showing that sky
contamination does not affect at  all our EW estimates. As an example,
Fig.~\ref{fig:sky_stacks} (left panel) plots telluric lines ({ from
  Hanuschik~2006,  unpublished})   and  sky  emission   lines  in  the
wavelength region  of $7000$ to  $9200$\,\AA. The red and  blue shaded
areas mark the observed wavelength  range of the Na8190 feature at the
lower  ($z\sim 0.05$;  blue)  and upper  ($z\sim0.095$; red)  redshift
limit of the SPIDER sample.  At $z\sim0.05$, the Na8190 is observed in
a  region almost  unaffected by  sky contamination,  while at  $z \sim
0.095$, the feature overlaps with  a strong (H$_2$O) telluric band (at
$\lambda \sim 9000$\,\AA). The right panel of the same Figure compares
the  stacked spectra with  $\sigma \sim  100$\,\kms\ and  $\sigma \sim
300$\,\kms\ (black curves), to the case  where (i) only ETGs at $z \le
0.065$  are combined, with  Na8190 being  virtually unaffected  by sky
contamination,  and (ii)  a more  ``aggressive'' stacking  approach is
adopted, where only flux values more than 1\AA\ away from any telluric
line and sky  emission are combined.  The resulting  spectra (not used
for  the present  analysis but  for sky  contamination tests)  show an
excellent  agreement, with  differences at  the subpercent  level.  To
obtain  a more  quantitative  estimate of  how  sky contamination  may
affect our  results, we split  the sample into low-  and high-redshift
bins, with  $0.05<z<0.06$ and $0.085<z<0.095$,  respectively.  We have
produced stacked spectra corresponding  to these two $z$ intervals for
the lowest  and highest velocity  dispersion bins of our  stacks, $100
\le  \sigma_0 \le 110$\,\kms\  and $280  \le \sigma_0  \le 320$\,\kms,
respectively.      At    highest     $\sigma_0$,     the    equivalent
width~\footnote{These equivalent widths are  computed on the stacks at
  their nominal  resolution, i.e.  no  correction is applied  to bring
  them   to  the  same   $\sigma_0$.}   of   the  \naii\   index  (see
Sec.~\ref{sec:indices})   is   $0.51   \pm  0.03$\,\AA\   ($0.52   \pm
0.04$\,\AA) at  lower (higher)  redshift, while at  lowest $\sigma_0$,
the corresponding equivalent widths are  $0.75 \pm 0.04$ and $0.78 \pm
0.07$  at  lower  and  higher  redshift.  Consistent  with  FLD13,  we
conclude  that   sky  contamination  does   not  affect  at   all  the
\naii\  feature. The  same result  holds true  for the  other relevant
features.  Regarding the \cat\ index, one may notice that part of this
feature always falls within the H$_2$O telluric band, at $\lambda \sim
9000$\,\AA.  To  test the  impact of this  on the stacked  spectra, we
split  the  \cat\  sample  of   ETGs  into  two  redshift  bins,  with
$0.05<z<0.055$    and   $0.055<z<0.060$,   respectively.     For   the
$\sigma_0=100$--$110$ and  $200$--$210$\,\kms\ stacks, we  find that a
difference of \cat\ equivalent width  between the two $z$ intervals of
$0.3  \pm  0.25$  and  $0.24  \pm  0.21$,  respectively,  i.e.   fully
consistent with zero within the errors.

In order  to compare stellar  population models to stacked  spectra of
ETGs,  we  characterize how  they  are  affected  by the  instrumental
resolution  of the  SDSS spectrograph.   For each  spectrum,  the SDSS
pipeline  provides the  instrumental resolution,  $FWHM_{inst}$,  as a
function  of observed  wavelength.  We  de-redshifted and  stacked the
$FWHM_{inst}$  curves   in  the  same  way  as   the  galaxy  spectra.
Fig.~\ref{fig:res} compares the resulting stacked resolution, $\langle
FWHM\rangle_{inst}$, for  three representative $\sigma_0$  bins ($\sim
100$,   $200$,   and   $300$\,\kms).    The  variation   of   $\langle
FWHM\rangle_{inst}$  with $\lambda$  reflects the  fact that  the SDSS
resolution  is wavelength dependent  (see e.g.~\citealt{Bernardi:03}),
increasing from $\sim 2.2$\,\AA\ for  the blue arm of the spectrograph
($\lambda < 6000$\,\AA)  to $\sim 3.1$\,\AA\ for its  red arm.  Notice
that  the  variation  of  $\langle FWHM\rangle_{inst}$  among  stacked
spectra  is completely negligible  ($< 0.2$\,\AA\  in the  blue).  For
this reason, throughout the present  study, we use the average stacked
resolution of all  spectra in the sample. Notice  also that the actual
resolution of all stacked spectra analyzed in the present study (given
by  the  SDSS instrumental  resolution  plus  the  $\sigma_0$ of  each
stack~\footnote{ Hereafter, with this  expression, we mean that the
  actual  resolution  of  a  given  spectrum is  obtained by  adding  in
  quadrature   the  $\sigma_0$   to   the  resolution   of  the   SDSS
  spectrograph.   })  is  well  above that  of  our reference  stellar
population  models  (i.e.   the  extended MILES  models,  see  below),
allowing for an accurate comparison of models and data to be performed.

%%%%%%%%%%%%%%%%%%%%%%%%%%%%%%%%%%%%%%%%%%%%%%%%%%%%%%%%%%%%%%%%%%%%%%%
\begin{figure*}
\begin{center}
\leavevmode
\includegraphics[width=14cm]{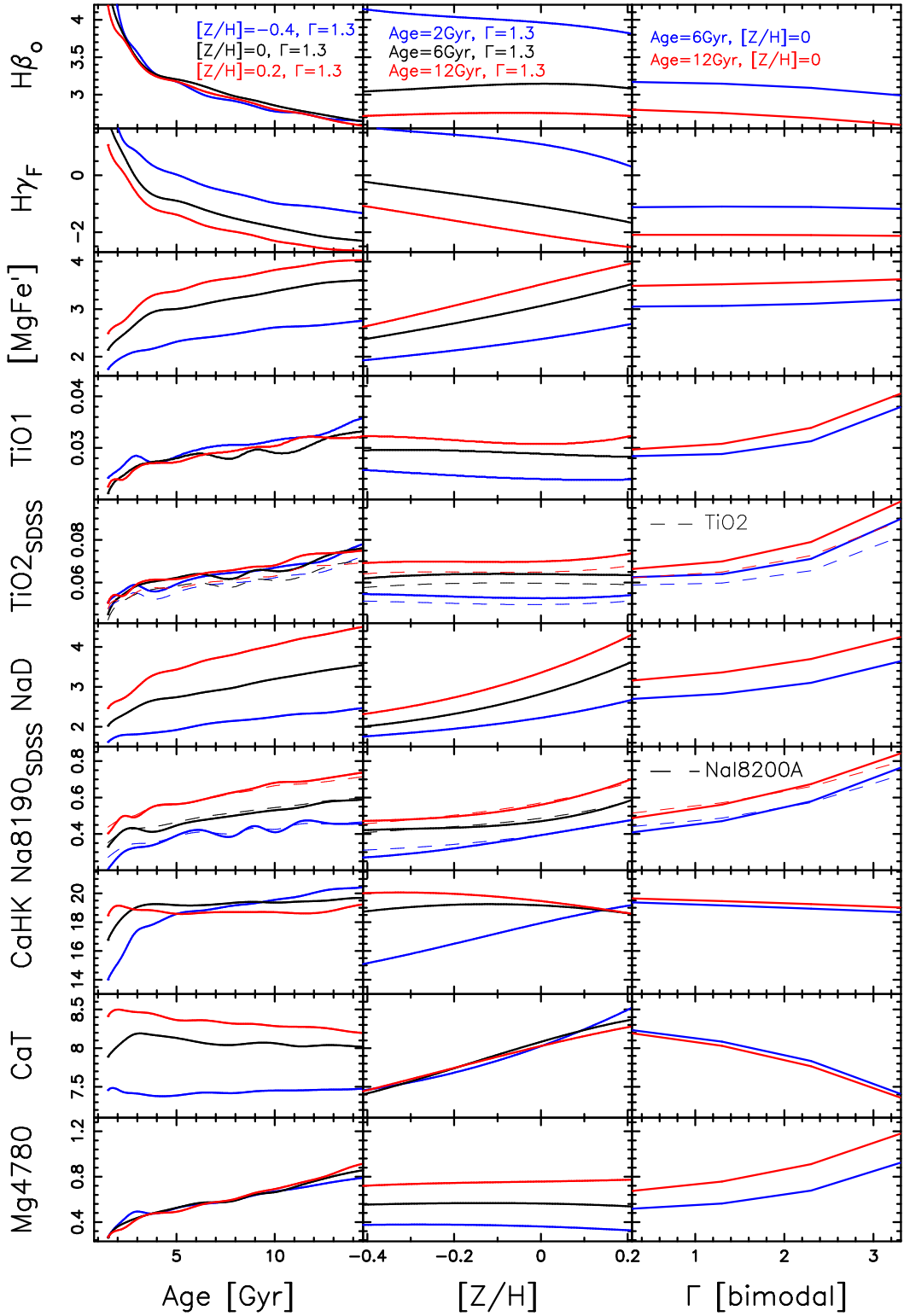}
\end{center}
\caption{Sensitivity  of  selected  spectral  indices to  age  (left),
  metallicity  (middle), and  (bimodal) IMF  slope  (right). Different
  colours correspond  to different MILES extended  (MIUSCAT) SSPs, for
  age, metallicity, and IMF slope,  as labelled in the top panels.  In
  particular, we plot three  SSPs with $\Gamma=1.3$, and metallicities
  $[Z/H]=-0.4$, $0$ (solar), and $+0.2$ (left panels); three SSPs with
  ages  $2$,  $6$,  and  $12$~Gyr, and  bimodal  $\Gamma=1.3$  (middle
  panels); and  two SSPs  with solar metallicity  and ages of  $6$ and
  $12$~Gyr (right panels).  In  the panels showing \tioiio\ and \naii,
  we  overplot the  trends for  the  Lick-based \tioii\  and the  $\rm
  NaI8200A$  spectral  indices,   respectively.   Notice  the  similar
  behaviour of the new (``SDSS'') and the Lick-based indices (see text
  for details).}
%\contcaption{.}
\label{fig:sens_indices}
\end{figure*}
%%%%%%%%%%%%%%%%%%%%%%%%%%%%%%%%%%%%%%%%%%%%%%%%%%%%%%%%%%%%%%%%%%%%%%%

Finally, we notice that our sample of ETGs is likely contaminated by a
small fraction of  early spiral systems, i.e., spiral  galaxies with a
prominent bulge component. As discussed in Paper I, this contamination
fraction  is about  15$\%$,  falling to  less  than 5$\%$  for a  high
quality subsample defined on the basis of visual image classification.
We found that restricting the  stacking procedure to this high quality
selected  sample  does not  lead  to  any  significant change  in  the
relevant spectral indices.  For  instance, the \naii\ index changes by
less than $0.02$\,\AA\ for all stacks.  This might be expected because
we combine the  spectra using the median, instead  of the average, and
confirms that  morphological contamination does not affect  at all our
conclusions.

\begin{table}
\centering
\small
 \caption{Properties of stacked spectra of ETGs in bins of central
   velocity dispersion ($\sigma_0$).}
  \begin{tabular}{c|c|c|c|c|c|c}
   \hline
 $\sigma_0$ range & $N_{\rm ETGs}$ &  \multicolumn{5}{|c}{median signal-to-noise ratio (\AA$^{-1}$)} \\
$\rm [km/s]$ & & ${\rm CaHK}$ & ${\rm H\beta_o}$ & ${\rm TiO2}$ &${\rm Na8190}$ & ${\rm CaT}$ \\
(1) & (2) & (3) & (4) &(5) &(6) & (7)\\
   \hline
$100$--$110$&  $1062$&  $ 134$ & $ 446$ & $ 955$ & $ 401$ &  $ 160$\\
$110$--$120$&  $1864$&  $ 190$ & $ 604$ & $1318$ & $ 561$ &  $ 223$\\
$120$--$130$&  $2452$&  $ 226$ & $ 732$ & $1564$ & $ 690$ &  $ 273$\\
$130$--$140$&  $2604$&  $ 245$ & $ 766$ & $1692$ & $ 772$ &  $ 298$\\
$140$--$150$&  $2662$&  $ 255$ & $ 799$ & $1790$ & $ 807$ &  $ 309$\\
$150$--$160$&  $2516$&  $ 263$ & $ 798$ & $1797$ & $ 830$ &  $ 324$\\
$160$--$170$&  $2494$&  $ 277$ & $ 824$ & $1837$ & $ 852$ &  $ 333$\\
$170$--$180$&  $2003$&  $ 255$ & $ 767$ & $1716$ & $ 802$ &  $ 305$\\
$180$--$190$&  $1711$&  $ 247$ & $ 706$ & $1636$ & $ 761$ &  $ 282$\\
$190$--$200$&  $1376$&  $ 233$ & $ 655$ & $1519$ & $ 702$ &  $ 263$\\
$200$--$210$&  $1087$&  $ 215$ & $ 601$ & $1422$ & $ 633$ &  $ 248$\\
$210$--$220$&   $824$&  $ 196$ & $ 550$ & $1279$ & $ 605$ &  $ 245$\\ 
$220$--$230$&   $645$&  $ 182$ & $ 498$ & $1172$ & $ 539$ &  $ 209$\\ 
$230$--$240$&   $487$&  $ 159$ & $ 457$ & $1042$ & $ 486$ &  $ 171$\\ 
$240$--$250$&   $340$&  $ 135$ & $ 374$ & $ 868$ & $ 444$ &  $ 166$\\ 
$250$--$260$&   $239$&  $ 122$ & $ 328$ & $ 787$ & $ 378$ &  $ 138$\\ 
$260$--$280$&   $255$&  $ 127$ & $ 371$ & $ 850$ & $ 417$ &  $ 163$\\ 
$280$--$320$&   $160$&  $ 106$ & $ 268$ & $ 693$ & $ 347$ &  $ 126$\\ 
   \hline
  \end{tabular}
\label{tab:stacks}
\end{table}

%%%%%%%%%%%%%%%%%%%%%%%%%%%%%%%%%%%%%%%%%%%%%%%%%%%%%%%%%%%%%%%%%%%%%%%
\begin{figure*}
\begin{center}
\leavevmode
\includegraphics[width=15cm]{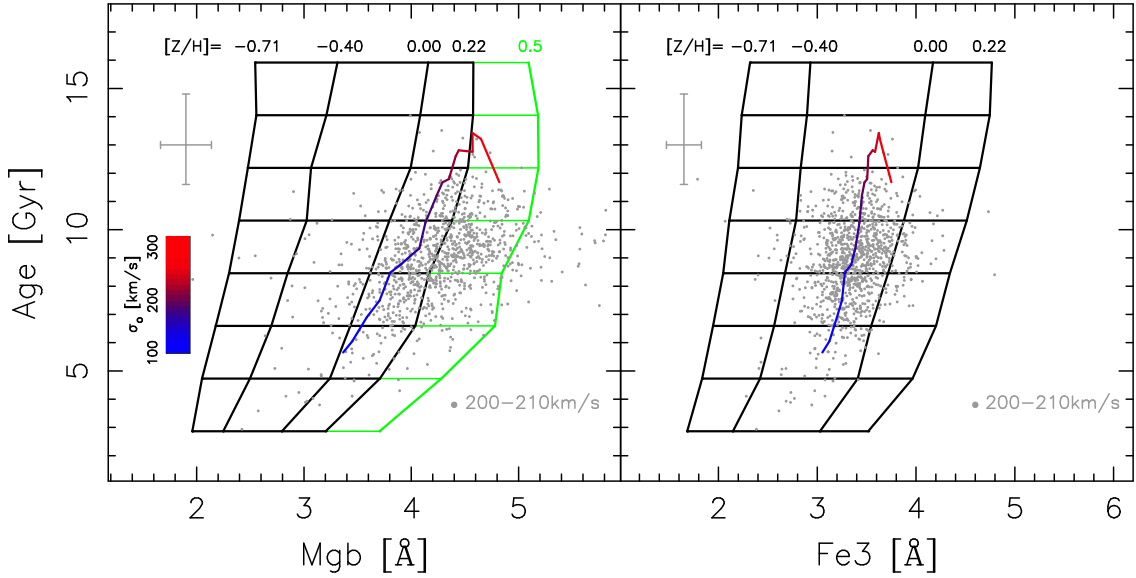}
\end{center}
\caption{Illustration of the \afe\ proxy obtained from base (nearly
  solar-scaled) extended MILES (MIUSCAT) models.  The $Age$ parameter
  is plotted against Mg$b$ (left) and Fe$3$ (right) spectral indices.
  Grey dots are ETGs within the $\sigma_0\sim 200-210$\,\kms\ bin,
  while the blue-through-red curve connects data points for the 18
  $\sigma_0$ stacks, with colour encoded as shown by the bar on the
  left panel.  The grey error bars show typical uncertainties on age,
  as well as on the Mg$b$ and Fe$3$ indices.  The grids show the
  effect of varying age and metallicity of extended MILES (MIUSCAT)
  SSPs, with the green portion, in the left plot, resulting from a
  (linear) extrapolation of the model Mg$b$ to high ($[Z/H]=+0.5$)
  metallicity.  The proxy is defined as the difference of
  metallicities estimated from the Mg$b$ and Fe$3$ grids, respectively.
  Notice that for the grey points (i.e. single ETGs) the age is estimated from
  spectral fitting (as obtained from the {\tt STARLIGHT} code), while
  for the stacks -- because of the high $S/N$ ratio -- we used \hbo-based
  age estimates (see text for details).  }
%\contcaption{.}
\label{fig:proxy_estimate}
\end{figure*}
%%%%%%%%%%%%%%%%%%%%%%%%%%%%%%%%%%%%%%%%%%%%%%%%%%%%%%%%%%%%%%%%%%%%%%%

%%%%%%%%%%%%%%%%%%%%%%%%%%%%%%%%%%%%%%%%%%%%%%%%%%%%%%%%%%%%%%%%%%%%%%%%
\section{Stellar population models}
\label{sec:models}

We  analyze the  spectra of  ETGs  with the  extended MILES  (MIUSCAT)
stellar population models~\footnote{The  models are publicly available
  at   {\tt  http://miles.iac.es/pages/ssp-models/miuscat-models.php}.
}~\citep[hereafter          MIUSCAT-I,         and         MIUSCAT-II,
  respectively]{Vazdekis:12,        Ricciardelli:12},        combining
CaT~\citep{Cenarro2001}   and    MILES~\citep{ps06}   with   Indo-U.S.
empirical  stellar libraries over  the spectral  range $\lambda\lambda
3465-9469$\,\AA,       at      a      nominal       resolution      of
$2.51$\,\AA\  FWHM~\citep{Jesus:11}. MIUSCAT  models are  identical to
MILES  and  CaT  models  in  the  spectral  ranges  covered  by  these
libraries,   $\lambda\lambda   3525-7500$\,\AA\  and   $\lambda\lambda
8350-9020$\,\AA, respectively,  while the Indo-U.S. is  used to ``fill
the gap''  between MILES and CaT  and extend blueward  and redward the
wavelength coverage of the models (see MIUSCAT-I for details).  Notice
also that MIUSCAT models  rely on solar-scaled isochrones with stellar
spectra  following   the  abundance   pattern  of  our   Galaxy,  i.e.
approximately  solar-scaled at  solar metallicity.   The  MIUSCAT SSPs
cover  a wide range  of ages,  from $0.06$  to $17.78$~Gyr,  and seven
metallicity  bins,  i.e.   $[Z/H]=\{-2.32, -1.71$,  $-1.31$,  $-0.71$,
$-0.4$, $0$, $+0.22\}$.   The SSPs are provided for  several IMFs.  We
use here  two power-law IMFs,  described in~\citet{Vazdekis:1996}, i.e
unimodal and bimodal, both characterized  by their slope $\Gamma$ as a
single free parameter { (see Fig.~\ref{fig:imf})}. The bimodal IMFs
are similar to the unimodal ones for stars with mass above
$0.6$\,M$_{\odot}$, but significantly shallower at lower masses,
becoming flat below $\sim 0.2$\,M$_\odot$.  For $\Gamma \sim 1.3$, the
bimodal IMF gives a good representation of the Kroupa IMF, while the
unimodal case coincides, for the same slope, with a \citet{salp:55}
IMF.  The lower and upper mass-cutoff of the IMFs are set to $0.1$ and
$100$\,M$_\odot$, respectively.  We refer the reader to appendix~A
of~\citet{Vazdekis:2003} for a detailed description of unimodal and
bimodal IMFs. For the present study, we interpolate the extended MILES
(MIUSCAT) SSPs, for each IMF, over the age (metallicity) range from
$1$ ($-0.4$) to $17.78$~Gyr ($+0.22$), where the models are expected
to be safe (see MIUSCAT-I).  Younger ages, as well as lower
metallicities, than those considered here are not relevant to study
our sample of luminous ETGs.  The interpolation is done for $200$ and
$150$ steps in age and metallicity, respectively. For bimodal IMFs, we
consider the cases $\Gamma=\{0.3$, $0.8$, $1.0$, $1.3$, $1.5$, $1.8$,
$2.0$, $2.3$, $2.8$, $3.3\}$.  For unimodal IMFs, we restrict the
analysis to cases with $\Gamma \le 2.3$.  At larger slopes, current
uncertainties in the modelling of very low Main Sequence stars may
impact significantly the synthetic SEDs (see~MIUSCAT-I).

%%%%%%%%%%%%%%%%%%%%%%%%%%%%%%%%%%%%%%%%%%%%%%%%%%%%%%%%%%%%%%%%%%%%%%%%
\section{Spectral indices}
\label{sec:indices}

{ Our analysis  is mainly based on constraining the  IMF by the use
  of targeted line  strengths. We mainly focus on  a detailed analysis
  of the  spectral indices, although we also  include information from
  spectral fitting to  make the results more robust  (as in FLD13, see
  \S\S\ref{subsec:hybrid})}
%We constrain the  IMF of ETGs following two  different approaches, (i)
%the analysis of line strengths (as,  e.g, in STK12), and (ii) a hybrid
%approach, combining full spectral  fitting at optical wavelengths with
%constraints from  IMF-sensitive spectral  indices (as in  FLD13).  
The use of line strengths brings some well-known drawbacks into the
study of unresolved stellar populations.  First, indices are defined
relative to pseudo-continua, often affected themselves by absorption
features.  Second, line strengths may reflect the abundance of
individual elements in a stellar population, in addition to age,
metallicity, and IMF, complicating the interpretation of their
variation. On the other hand, spectral indices have the advantage of
summarizing the spectral information in a few numbers. In principle,
one can define specific indices that efficiently constrain metallicity
and age, or -- most importantly for this work -- single element
abundance patterns and the IMF. We select in this paper a variety of
spectral indices, bringing the observed line strengths to solar scale
by means of their correlations with the abundance ratio (\afe,
Sec.~\ref{sec:afe_corr}).  This has the advantage of allowing for a
direct comparison with the expectations from nearly solar-scaled,
fully empirical -- rather than theoretical -- stellar population
models (MIUSCAT).  In contrast, CvD12a and~\citet{CvD12b} adopted a
direct spectral fitting approach, using $\alpha$-enhanced stellar
population models, based on synthetic stellar spectra to describe the
effect of abundance patterns.  We notice that most pixels of a galaxy
spectrum are sensitive to age, metallicity, and abundance patterns,
while only a small number of absorption features are sensitive to the
IMF.  Therefore, a direct comparison to stellar population models may
be driven by differences between models and data, rather than true IMF
variations.  This further motivates our choice to rely on line
strengths, and complement the analysis with the hybrid approach.

%%%%%%%%%%%%%%%%%%%%%%%%%%%%%%%%%%%%%%%%%%%%%%%%%%%%%%%%%%%%%%%%%%%%%%
\begin{figure}
\begin{center}
\leavevmode
\includegraphics[width=8cm]{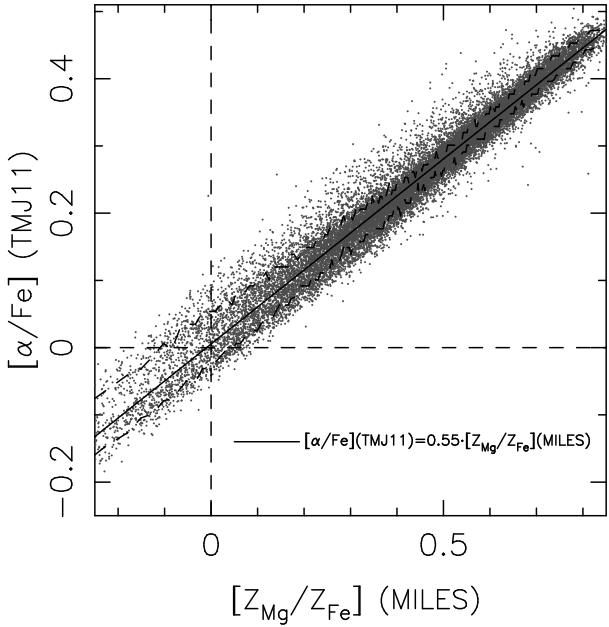}
\end{center}
\caption{Comparison of the ``true'' \afe\ parameter with the
  (approximately) solar-scaled proxy, \afep, obtained from extended
  MILES (MIUSCAT) models (see the text).  The \afe\ is estimated with
  TMJ11 stellar population models.  Grey points correspond to the
  sample of $24,781$ SPIDER ETGs selected for the present study.
  Dashed horizontal and vertical lines mark the value of zero on both
  axes.  The solid curve is the best-fitting line, obtained with a
  least-squares fitting procedure, assuming \afe\ as dependent
  variable.  Notice the small scatter, with a fitting rms of
  $\sim$0.025~dex in \afe.  The black dashed curves, bracketing the
  best-fit line correspond to the 16th and 84th percentiles of the
  \afe\ distribution in bins of \afep.}
%\contcaption{.}
\label{fig:proxy}
\end{figure}
%%%%%%%%%%%%%%%%%%%%%%%%%%%%%%%%%%%%%%%%%%%%%%%%%%%%%%%%%%%%%%%%%%%%%%

{ We point out that all line indices used for the present analysis
  are measured at the nominal resolution (i.e.  SDSS plus $\sigma_0$)
  of each stack.  Since line indices are also sensitive to velocity
  dispersion -- as the broadening of lines changes their contribution
  to the fluxes measured both in the central and sidebands -- we
  always extract information (e.g.  IMF slope) by comparing indices to
  expectations of stellar population models smoothed to match the
  resolution of each individual stack.  This makes the derived trends
  (of IMF slope as well as elemental abundances) with $\sigma_0$
  independent of resolution. However, and for illustration purposes
  only, in the figures showing trends of line indices among different
  stacks, we correct the indices to a reference broadening of
  $\sigma_{\rm 0,ref}=200$~\kms. The correction is done by estimating the
  variation in each index when smoothing the best-fitting stellar
  population model to the $\sigma_0$ of a given stack and to
  $\sigma_{\rm 0,ref}$, respectively. We note that $200$~\kms corresponds
  to the central value of the $\sigma_0$ range of our sample, hence
  this choice of $\sigma_{\rm 0,ref}$ minimizes the correction for all
  stacks in the figures.  Notice that measuring indices at the
  original resolution of the stack, rather than degrading all stacks
  to the same maximum $\sigma_0$ value of $\sim 300$~\kms, has the
  advantage of maximizing the information extracted from the data,
  avoiding the effect of correlated variations among adjacent pixels
  when smoothing the data.  However, to test the impact of resolution
  on our results, as well as to explore a wider parameter space in the
  analysis, we smooth both data and models to the same $\sigma_0$ of
  $300$\,\kms, when performing the hybrid approach
  (Sec.~\ref{subsec:hybrid}).}

\subsection{Selection of spectral features}
\label{sec:sel_features}

The list of  spectral indices, adopted in this  study is assembled as
follows.

\begin{description}
 
\item[{\it  IMF-sensitive   spectral  indices}.]   We   have  visually
  inspected  the flux ratio  of two  bimodal SEDs,  within contiguous,
  $300$\,\AA\  width,  spectral  windows,  looking  for  IMF-sensitive
  wavelengths (as, e.g., in  figure~9 of~MIUSCAT-I).  The two SEDs are
  chosen  to have the same  age ($10$~Gyr)  and (solar)  metallicity, with
  $\Gamma=0.3$  and   $3.3$,  respectively.   Matching   the  list  of
  potentially useful wavelengths to existing lists of spectral indices
  resulted in  a list of  five IMF-sensitive features~\footnote{Notice
    that  the   \mgb\  index  was   also  first  selected   among  the
    IMF-sensitive indices. However,  because of its strong sensitivity
    to  \afe, we  have  decided not  to  include it  in the  analysis.
    Moreover, as discussed  by CvD12, the sensitivity of  \mgb\ to the
    IMF is significantly  reduced by the fact that a  $\rm Cr I$ blend
    partially overlaps  with the red  pseudo-continuum.}, namely \mgf,
  \tioi,  \tioiio, \naii,  and \cat.   Here, the  \cat\ is  defined as
  CaT$=$\cai+\caii+\caiii, following the definition of
%  \cai, \caii, and \caiii\ as in 
\citet{Cenarro2001}.   The definition  of spectral  indices  -- mostly
based on  previous works  -- is summarized  in Tab.~\ref{tab:indices}.
Two  exceptions  are  the  \naii\  and \tioiio\  features,  which  are
slightly different  versions of the  $\rm NaI8200A$ index,  defined in
MIUSCAT-I,  and  the  well-known,  Lick-based,  $\rm  TiO2$  molecular
band~\citep{Trager98},  respectively.  In  practice, we  have modified
the  blue (red)  pseudo-continua of  $\rm NaI8200A$  ($\rm  TiO2$), in
order  to minimize  systematic discrepancies  between models  and data
(independent of the IMF).  Further  details on this issue are given in
App.~\ref{sec:new_indices}.
\item[{\it Abundance-sensitive spectral indices}.]  The list of
  IMF-sensitive spectral indices is complemented with the \nad\ and
  \cahk\ spectral features. These indices, being much more sensitive
  to Na and Ca abundances than IMF, can be useful tools to break the
  degeneracy between IMF and abundance ratios (see e.g.~CvD12). For
  instance, by analyzing the diagram \nad\ vs.~NaI (their NaI index
  being similar to our \naii), STK12 concluded that both a
  dwarf-enriched IMF and overabundant $\rm [Na/Fe]$ are required to
  match the equivalent widths of a small set of ETGs from the SDSS. A
  powerful Ca abundance diagnostic is also given by the
  \caf\ index~\citep{Trager98}.  In fact, the \caf\ has been recently
  used by~\citet{JTM12} to conclude in favour of an underabundance of
  $[{\rm Ca/Fe}]$ in ETGs.  Unfortunately, we have found that none of
  our fitting schemes (including the modelling of abundance patterns)
  is able to produce an acceptable fit to \caf.  For this reason, we
  have excluded this index from the analysis, but for reference, we
  present a comparison between models and data for this index in
  Sec.~\ref{subsubsec:caf}.  The definition of abundance-sensitive
  spectral indices is summarized, together with that of IMF-sensitive
  indices, in Tab.~\ref{tab:indices}.
\item[{\it Age-metallicity sensitive  spectral indices}.]  In order to
  account  for the well-known  variation of  age and  metallicity with
  galaxy  properties (e.g.~\citealt{Thomas:05}),  we also  include the
  \hbo, \hgf, and  \mgfep\ spectral indices.  The \hbo\  is a modified
  \hb\    index,   optimized    to   minimize    the   age-metallicity
  degeneracy~\citep{CV09}.   The  \hgf\  is  also a  very  useful  age
  indicator, due to  its small sensitivity to \afe,  similar to \hb, in
  constrast to  higher-order Balmer lines~\citep{Thomas:04}.  Finally,
  we  adopt  the  total metallicity  estimator  \mgfep\ of \citet{TMB:03},
  whose    definition    removes    the   residual    dependence    of
  the~\citet{GON93} \mgfe\ spectral index on \afe.
\end{description}

{  Fig.~\ref{fig:sens_indices} plots  the  sensitivity of  selected
  spectral indices  to age, metallicity,  and IMF.}
%We do not  include \hbo, \hgf, and \mgfep, as  the behaviour of these
%  indices has been already extensively characterized elsewhere.
For  each index,  we show  three  panels, plotting  line strengths  of
extended MILES  (MIUSCAT) SSPs as a  function of (from  left to right)
age, metallicity, and (bimodal) IMF slope, $\Gamma$.  Each panel shows
the model predictions for a small set of simple stellar populations.
%(i) (bimodal) $\Gamma$ (top), for solar metallicity, and two age
%values of $6$ and $12$~Gyr; (ii) metallicity (middle), for a bimodal
%IMF with $\Gamma=1.3$, and ages of $2$, $6$, and $12$~Gyr; (iii) age
%(bottom), for a bimodal IMF with $\Gamma=1.3$, and $[Z/H]= -0.4, 0$
%and $0.2$.
The  extended MILES  (MIUSCAT) SSPs  have been  smoothed to  match the
wavelength  dependent resolution  of SDSS  (see Sec.~\ref{sec:stacks})
plus a  $\sigma$ of $200$\,\kms.   { Both \hbo\ and  \hgf\ decrease
  significantly  with   increasing  age.   However,   while  \hbo\  is
  essentially   independent  of   metallicity,  consistent   with  its
  definition~\citep{CV09}, \hgf\ tends  to decrease with $[Z/H]$. Both
  indices (and in particular \hgf) are independent of IMF.  \mgfep\ is
  also  independent of  $\Gamma$, increasing  with  total metallicity,
  and,  to a  lesser extent,  with age.   }  As  expected,  all \tioi,
\tioiio,  \naii, \cat,  and  \mgf\ show  a  significant dependence  on
$\Gamma$.   The \tioi,  \tioiio, \naii,  and \mgf\  increase  with IMF
slope,  while \cat\ exhibits  the opposite  behaviour.  The  $\rm TiO$
indices, as well as \mgf,  are insensitive to metallicity, but tend to
increase with  age. For  \tioiio (\tioi), the  latter trend  is weaker
(similar)  to that  with $\Gamma$.   Notice also  that \tioiio\  has a
dependence  on  age,  metallicity,   and  IMF,  very  similar  to  the
Lick-based  \tioii\ index  (dashed curves  in the  Figure).  Regarding
\naii, it strongly increases with  IMF slope, with a weaker dependence
on age and metallicity.  The  \nad\ shows a decent sensitivity to IMF,
but  it strongly  increases with  both  age and  metallicity.  On  the
contrary,  for  age  older  than  $\sim 2$~Gyr  and  either  solar  or
super-solar metallicity, the \cahk, i.e. the other abundance-sensitive
index included in this study,  is essentially independent of both age,
metallicity, and  IMF.  The  \cat\ index is  constant with  respect to
age, for SSPs older than a  few Gyr, while it depends significantly on
metallicity and decreases significantly with increasing IMF slope.

The different sensitivity of
all selected spectral indices to age, metallicity, and IMF, makes them
a powerful  tool to constrain all these  parameters simultaneously, as
shown in Sec.~\ref{sec:results}.

An important remark  concerns the effect of internal  dust on spectral
indices.   Being a  resonant line,  \nad\  can be  highly affected  by
interstellar  absorption.  This  has  hampered, so  far,  its use,  in
combination with \naii, to constrain the relative contribution of $\rm
[Na/Fe]$  and  IMF  to  gravity-sensitive features.   The  same  issue
applies to \tioi\ (see e.g.~TMJ11), whose blue pseudo-continuum can be
affected  by the Na  interstellar absorption,  leading to  overly weak
strengths.  Notice that  our work is virtually unaffected  by dust, as
the stacked spectra are constructed by removing, {\it ab initio}, ETGs
with significant amount of reddening.  Also, plotting \nad\ as well as
\tioi\  strengths as  a  function  of $E(B-V)$,  for  spectra in  each
$\sigma_0$ bin, does not show any significant correlation.

Another remark  is that all  our spectral indices are  measured within
the central $3''$ diameter  aperture of the SDSS fibres, corresponding
to a  smaller fraction  of the effective  radius for the  more massive
galaxies.   Radial trends  of  spectral indices  might  thus bias  our
conclusions   on   the   IMF   variation   with   galaxy   mass.    In
App.~\ref{sec:aper}  we  present  the  trends  of  some  IMF-sensitive
indices with  effective radius, showing that the  aperture effect does
not affect our conclusions.

\begin{table*}
\centering
\small
%\begin{minipage}{140mm}
 \caption{Targeted spectral indices to constrain the low-mass end of the IMF. }
  \begin{tabular}{c|c|c|c|c|c}
   \hline
 Index & Units & Blue Pseudo-continuum & Central feature & Red Pseudo-continuum & source \\
       &       & [\AA] & [\AA] & [\AA] &  \\
   \hline
\cahk   & \AA & $3806.5$--$3833.8$ & $3899.5$--$4003.5$ & $4020.7$--$4052.4$ & Serven+05 \\
\mgf    & \AA & $4738.9$--$4757.3$ & $4760.8$--$4798.8$ & $4819.8$--$4835.5$ & Serven+05 \\
%\nad    & \AA & $5862.2$--$5877.2$ & $5878.5$--$5911.0$ & $5923.7$--$5949.7$ & \\
\nad    & \AA & $5860.625$--$5875.625$ & $5876.875$--$5909.375$ & $5922.125$--$5948.125$ & Trager+98 \\
\tioi   & mag & $5816.625$--$5849.125$ & $5936.625$--$5994.125$ & $6038.625$--$6103.625$& Trager+98 \\
\tioiio & mag & $6066.625$--$6141.625$& $6189.625$--$6272.125$ & $6422.0$--$6455.0$& $\rm This \, work$ \\
%\tiocah & \AA & $6520.0$--$6545.0$ & $6600.0$--$6817.0$ & $7035.0$--$7050.0$ & Chen+12\\
\naii   & \AA & $8143.0$--$8153.0$&$8180.0$--$8200.0$ & $8233.0$--$8244.0$  & This work \\
\cai    & \AA & $8474.0$--$8484.0$& $8484.0$--$8513.0$ & $8563.0$--$8577.0$ & Cenarro+01 \\
\caii   & \AA & $8474.0$--$8484.0$& $8522.0$--$8562.0$ & $8563.0$--$8577.0$ & Cenarro+01 \\
\caiii  & \AA & $8619.0$--$8642.0$& $8642.0$--$8682.0$ & $8700.0$--$8725.0$ & Cenarro+01 \\
   \hline
  \end{tabular}
\label{tab:indices}
%\end{minipage}
\end{table*}

%%%%%%%%%%%%%%%%%%%%%%%%%%%%%%%%%%%%%%%%%%%%%%%%%%%%%%%%%%%%%%%%%%%%%%%%
\section{The effect of \afe}
\label{sec:afe}

{ There is a well-known correlation between velocity dispersion
and \afe\ abundance ratio \citep[see, e.g.,][]{Tra:00}. This correlation
needs to be addressed in our study, to confirm that our trends in
the IMF-sensitive spectral indices are not  simply caused by the
change in \afe. In order to study this issue,}
%In order to  study the impact of element  abundance ratios on spectral
%indices,  
we split  our sample  of ETGs within each  $\sigma_0$  bin (see
Sec.~\ref{sec:sample}), according  to \afe.  This  is done by
estimating a  proxy of \afe\  for each galaxy, as  detailed in
Sec.~\ref{subsec:aFe_stacks}.   In Sec.~\ref{sec:afe_corr}  we present
the correlations of indices vs.  proxy.

\subsection{\afe\  stacks at fixed $\sigma_0$}
\label{subsec:aFe_stacks}

For  each  ETG,  we  estimate  a  proxy of  \afe,  based  on  (nearly)
solar-scaled extended  MILES (MIUSCAT) models.  First,  we measure the
galaxy  luminosity-weighted  age by  spectral  fitting the  wavelength
range from  3900 to 7350\,\AA~\footnote{The lower  limit of $3900$\,\AA\
  avoids  the  bluest  part  of  the SDSS  spectral  range,  which  is
  potentially    more    affected    by   small-scale    flat-fielding
  uncertainties,  while the upper  limit corresponds  approximately to
  the  upper limit  of the  MILES  spectral library.},  with the  {\tt
  STARLIGHT} fitting code~\citep{CID05}.  We feed {\tt STARLIGHT} with
an  input  list  of  $132$   SSPs,  covering  a  wide  range  of  ages
($1$--$17.78$~Gyr) and metallicities  (from $[Z/H]=-0.71$ to $+0.22$),
assuming a Kroupa IMF.  For the given age, we estimate two independent
metallicities, $Z_{Mg}$ and $Z_{Fe}$,  from the spectral indices $Mgb$
and  ${\rm   Fe}3\equiv({\rm  Fe}4383+{\rm  Fe}5270+{\rm  Fe}5335)/3$,
respectively.       This     procedure      is      illustrated     in
Fig.~\ref{fig:proxy_estimate},      for      galaxies      in      the
$\sigma_0=200-210$\,\kms\  stack (grey dots).   The Figure  also plots
$Age$-Mg$b$ and  $Age$-Fe$3$ grids  for MILES extended  (MIUSCAT) SSPs
(Kroupa IMF).  While at solar scale both $[Z/H]_{Mg}$ and $[Z/H]_{Fe}$
should coincide, for  an $\alpha$-enhanced population the $[Z/H]_{Mg}$
is  larger than  $[Z/H]_{Fe}$, as  it is  actually the  case  for most
data-points.  Since the $[Z/H]_{Mg}$  is often larger than the maximum
metallicity  of  the  base  SSPs  ($[Z/H]=+0.22$), for  each  age,  we
extrapolate the  model Mg$b$ up  to a metallicity of  $[Z/H]=+0.5$, as
shown  by  the  green  portion  of  the grid  in  the  left  panel  of
Fig.~\ref{fig:proxy_estimate}. We  define as solar proxy  of \afe, the
difference  \afep$\equiv   [Z/H]_{Mg}-[Z/H]_{Fe}$.   Notice  that  for
single SDSS spectra estimating ages from spectral fitting, rather than
single spectral features (e.g.  \hb),  gives a more robust (i.e.  less
uncertain) age,  and thus metallicity,  estimate.  In fact,  the $S/N$
ratio of  single (with  respect to stacked)  SDSS spectra  prevents us
from obtaining accurate age estimates from \hb\ alone (with the median
\hbo\ equivalent width uncertainty  being $\sim 0.3$\,\AA).  Large age
uncertainties would  spread a significant  fraction of points  at both
high and  low ages.  As  seen from the  shape of the  $Age$-Mg$b$ grid
(Fig.~\ref{fig:proxy_estimate}), at fixed  $Mgb$, young populations do
require a larger extrapolation in  $[Z/H]$ than older ones, making the
proxy estimate more  uncertain when the errors in  the age are larger.
For the stacked  spectra, thanks to the high  $S/N$ ratio, we estimate
the  \afep\ from \hbo-based  (rather than  spectral fitting)  ages, as
shown  in Fig.~\ref{fig:proxy_estimate}.  In  practice, we  derive the
SSP-equivalent age by fitting  \hbo\ and \mgfep\ equivalent lines with
MILES  SSPs (Kroupa  IMF). Varying  the IMF  (according to  the $1SSP$
fitting  results  of Sec.~\ref{subsec:resone})  and/or  the method  to
estimate the age  ({\tt STARLIGHT} rather than \hbo),  does not change
the \afep\ more than by a few percent~\footnote{On the other hand, the
  age  estimates  differ  when  derived with  different  methods.  For
  instance,  STARLIGHT luminosity-weighted  ages  are about  $20$~$\%$
  older  than  \hbo-based,  SSP  equivalent,  ages.   This  difference
  implies a vertical offset between the blue-through-red curve and the
  center     of    the    distribution     of    grey     points    in
  Fig.~\ref{fig:proxy_estimate}.     }.     The    \afep\    increases
systematically  for  our  stacks,  from  $\sim  0.18$  in  the  lowest
$\sigma_0$  bin, up  to $\sim  0.47$ at  highest $\sigma_0$.   This is
consistent with the well-known result that the \afe\ of ETGs increases
with $\sigma_0$ \citep[see, e.g.,][]{Tra:00,Thomas:05}.

{ Since we aim to stack spectra with respect to \afep, an important
  issue is  to establish whether  reliable estimates of \afep\  can be
  obtained  for  individual  galaxies  in our  sample.   The  relevant
  parameter is  the median $S/N$ (per  \AA) over the  passbands of the
  \mgb\ and \fet\ spectral  features (hereafter $S/N_{MgFe}$). We find
  that for  individual spectra the $S/N_{MgFe}$ ranges  from $\sim 15$
  to  $30$  --  quoting  the   10-th  and  90-th  percentiles  of  the
  distribution -- with a median  value of $20$.  Notice that the $S/N$
  is larger for the Mg and  Fe than for the \naii\ feature, because of
  the rapid  decline of the  SDSS spectrograph throughput  above $\sim
  8000$\,\AA.   We estimate  the  uncertainty on  \afep\ by  comparing
  estimates of this  quantity for $2,313$ galaxies of  our sample with
  repeated observations  in SDSS (see  Paper I and Paper  IV), finding
  that the error  on \afep\ varies approximately in  a linear way with
  $S/N$ (in the range $12 \le  S/N_{MgFe} \le 32$), from $\sim 0.2$ at
  $S/N_{MgFe} = 15$, to $\sim  0.1$ at $S/N_{MgFe} = 30$. By selecting
  repeated observations  whose spectra  have large differences  in S/N
  ($\Delta S/N_{MgFe}>10$),  we also find no  significant variation in
  their   average  \afep\   ($\sim  0.03   \pm  0.04$),   implying  no
  differential bias  in \afep\  with $S/N$, and  thus with  respect to
  $\sigma_0$ (i.e. among different stacks). }

%%%%%%%%%%%%%%%%%%%%%%%%%%%%%%%%%%%%%%%%%%%%%%%%%%%%%%%%%%%%%%%%%%%%%%
\begin{figure*}
\begin{center}
\leavevmode
\includegraphics[width=15cm]{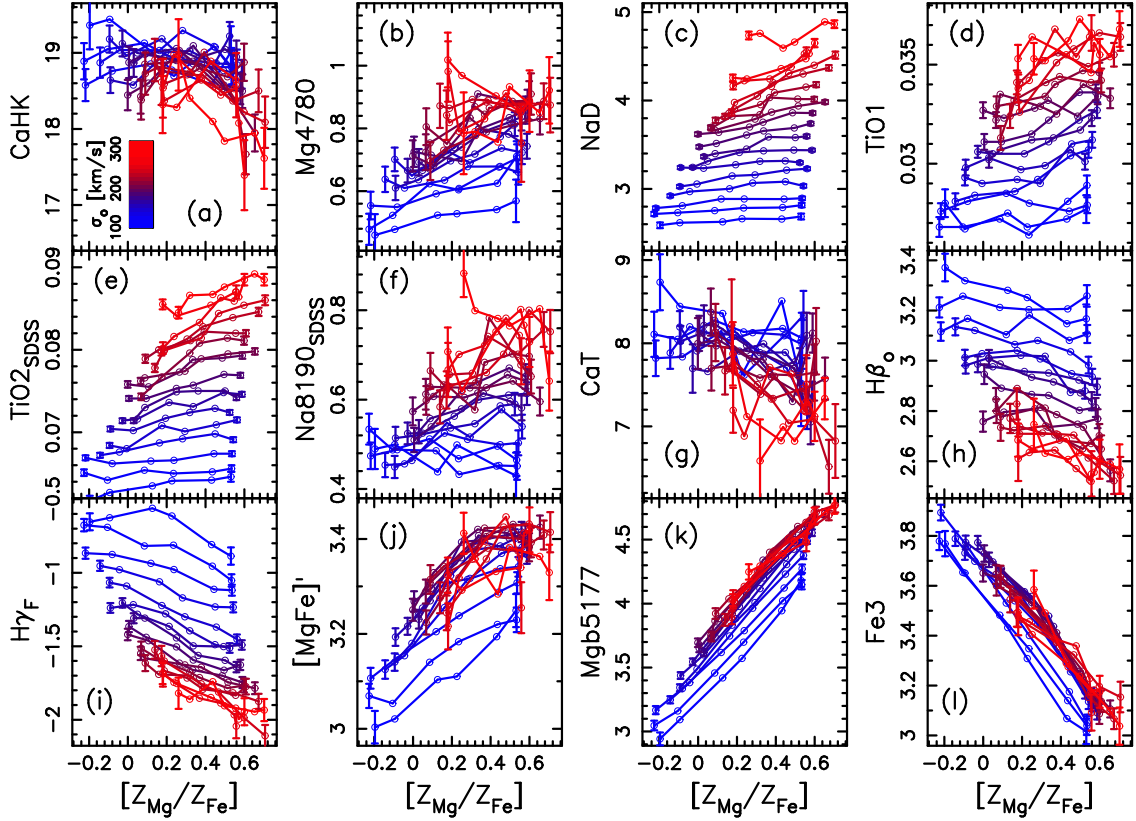}
\end{center}
\caption{Dependence   of   spectral  indices   on   solar  proxy   for
  $\alpha$-enhancement, \afep.   All indices refer to  stacks at their
  nominal   resolution,  i.e.    SDSS  instrumental   resolution  plus
  $\sigma_0$.   Different  colours  correspond to  different  velocity
  dispersion bins, as shown by  the colour bar in the top--left panel.
  Panels  (a--g),  as  labelled,  plot  IMF-  and  abundance-sensitive
  indices  (see the text),  while the  remaining panels  correspond to
  indices sensitive to age and metallicity (h--j), and \afe\ (panels k
  and l).   Error bars are  1~$\sigma$ measurement errors on  the line
  strengths. Notice the strong  increase (decrease) of \mgb\ (\fet) as
  a  function of  \afep, reflecting  the variation  of \afe\  at fixed
  $\sigma_0$.  The  dependence of \hbo\  and \mgfep\ on  \afep\ likely
  reflects  a  variation  of  age  and metallicity  in  each  velocity
  dispersion bin.  }
%\contcaption{.}
\label{fig:indices_fixed_sigma_nomod}
\end{figure*}
%%%%%%%%%%%%%%%%%%%%%%%%%%%%%%%%%%%%%%%%%%%%%%%%%%%%%%%%%%%%%%%%%%%%%%

Fig.~\ref{fig:proxy}       compares       the      ``presumably-true''
\afe~\footnote{The  expression  ``true''  refers  to  \afe\  estimates
  obtained  from  stellar population  models  taking abundance  ratios
  explicitly into account.} with \afep\  for our sample
of $24,781$  ETGs (see Sec.~\ref{sec:sample}).  The  ``true'' \afe\ is
estimated by comparing the \mgfep, \mgb, and \fet\ spectral indices of
each  galaxy  to  the prescriptions  of~\citet[hereafter  TMJ11]{TMJ11}
$\alpha$-enhanced   SSP  models   (covering   a  range   of  age   and
metallicity).  To this  effect, we fix the age of  TMJ11 models to the
luminosity-weighted  age  measured  from  {\tt STARLIGHT}  with  MILES
models,  as  described  above.    Hence,  both  \afe\  and  \afep\  in
Fig.~\ref{fig:proxy}  are  computed  for   the  same  age,  but  using
different models~\footnote{Although the two sets of models differ, one
  should  notice  that TMJ11  models  are  actually  based on  fitting
  functions  from  the  MILES  stellar library.}.   Notice  the  tight
relation between \afe\  and solar proxy, well described  by the linear
best-fit line in  the Figure, with a fitting  rms of $\sim 0.025$~dex.
This small scatter  proves that indeed, one can  study abundance ratio
effects by relying completely on the MILES-based (nearly solar-scaled)
\afep\ proxy. Notice that the  relation between \afe\ and \afep\ has a
slope of  about $0.5$.   Hence the \afe\  of our 18  $\sigma_0$ stacks
varies  from $\sim  0.09$  at  lowest $\sigma_0$,  to  $\sim 0.23$  at
highest $\sigma_0$ (see the \afep\ values reported above).

We stack galaxies according to  \afep, considering six \afep\ bins for
each $\sigma_0$  interval.  The \afep\  bins are defined  by selecting
ETGs (1) below  the 10-th percentile, (2) below  the 25-th percentile,
(3) between the 25-th and 50-th percentiles, (4) between the 50-th and
75-th percentiles, (5)  above the 75-th percentile, and  (6) above the
90-th  percentile  of the  \afep\  distribution.   For  each case,  we
median-combine the spectra, as described in Sec.~\ref{sec:stacks}.  At
fixed $\sigma_0$,  the \afep\ stacks span a  range of $\delta($\afep$)
\sim 0.6$ ($\delta($\afe$)\sim0.3$). We  use these stacks to study the
dependence of our targeted spectral indices on element abundance
ratios.

%Given  the redshift  range of  the SPIDER  sample, the  extracted SDSS
%spectra   extends   over    the   central   $\sim$3--5\,kpc   of   all
%galaxies\footnote{The  SDSS spectrograph  collects light  from sources
%  via  an  optical fibre  with  a projected  diameter  in  the sky  of
%  3~arcsec.}.  

%%%%%%%%%%%%%%%%%%%%%%%%%%%%%%%%%%%%%%%%%%%%%%%%%%%%%%%%%%%%%%%%%%%%%%%%
\subsection{Correlations with \afe}
\label{sec:afe_corr}

Fig.~\ref{fig:indices_fixed_sigma_nomod}  plots the  variation  of the
spectral   indices   as   a   function   of  the   solar   proxy   for
$\alpha$-enhancement  (\afep) for  all stacked  spectra within  the 18
bins in $\sigma_0$.   In addition to the indices  selected for the IMF
study  (see above),  we  also plot  \mgb\  and \fet,  which enter  the
definition  of  \afep\  (see Sec.~\ref{subsec:aFe_stacks}).   {  We
  point out that all indices are measured at the nominal resolution of
  the stacks (SDSS plus $\sigma_0$). Hence, while the variation of the
  indices  at fixed  $\sigma_0$  is not  affected  by resolution,  the
  variation of indices among  different $\sigma_0$ bins is also partly
  driven  by their different  velocity dispersion,  an effect  that we
  take into account in the analysis (see below).}

Most of the targeted spectral indices exhibit clear trends with \afep.
This might be  due to the dependence of the indices  on \afe, but also
due to a variation of age, metallicity, IMF, and/or individual element
abundances  within each  bin. The  \mgb\ increases  with  \afep, while
\fet\  shows  the  opposite  behaviour.  Notice  that  this  different
behaviour  is not  a  mere consequence  of  the fact  that we  stacked
spectra with  respect to \afep, as  \afep\ { can  change because of
  either  a variation  of \mgb\  or  \fet.  The  tight correlation  in
  panels (k)  and (i) of  Fig.~\ref{fig:indices_fixed_sigma_nomod},
  rather, reflect  the fact} that  \mgb\ (\fet) anticorrelates  with Fe
(Mg) abundance  (see, e.g.,  figure~1 of ~\citealt{TJM11}).   For what
concerns the IMF,  one can notice that, at  fixed \afep, the dynamical
range of all IMF-sensitive indices  with respect to $\sigma_0$ is much
larger than that  with \afep\ at fixed $\sigma_0$.   Hence, the IMF is
likely not driving the \afep\ trends. { This result also holds true
  when  removing the effect  of age  and metallicity,  as well  as the
  dependence    of   line    indices   on    $\sigma_0$,    from   the
  \afep\ correlations  (see below)}.  The panels with  \hbo, \hgf, and
\mgfep\  show that  age and  metallicity vary  significantly  at fixed
$\sigma_0$, in the sense  of galaxies with higher $\alpha$-enhancement
having  older ages  (i.e.  weaker  Balmer  lines), as  well as  higher
metallicities (i.e.  larger \mgfep).   This conclusion is based on the
fact that, { at fixed  total metallicity ($[Z/H]$)}, both \hbo\ and
\hgf,   as   well   as   \mgfep,  are   essentially   independent   of
\afe~\footnote{According  to~\citet{Thomas:04},  \hb\  and  \hgf\  are
  expected     to    slightly     increase     with    \afe,     while
  Fig.~\ref{fig:indices_fixed_sigma_nomod}  shows   a  trend  of  both
  indices  to {\sl  decrease} with  \afep, reflecting  a  variation in
  age. }.

%%%%%%%%%%%%%%%%%%%%%%%%%%%%%%%%%%%%%%%%%%%%%%%%%%%%%%%%%%%%%%%%%%%%%%
\begin{figure*}
\begin{center}
\leavevmode
\includegraphics[width=15cm]{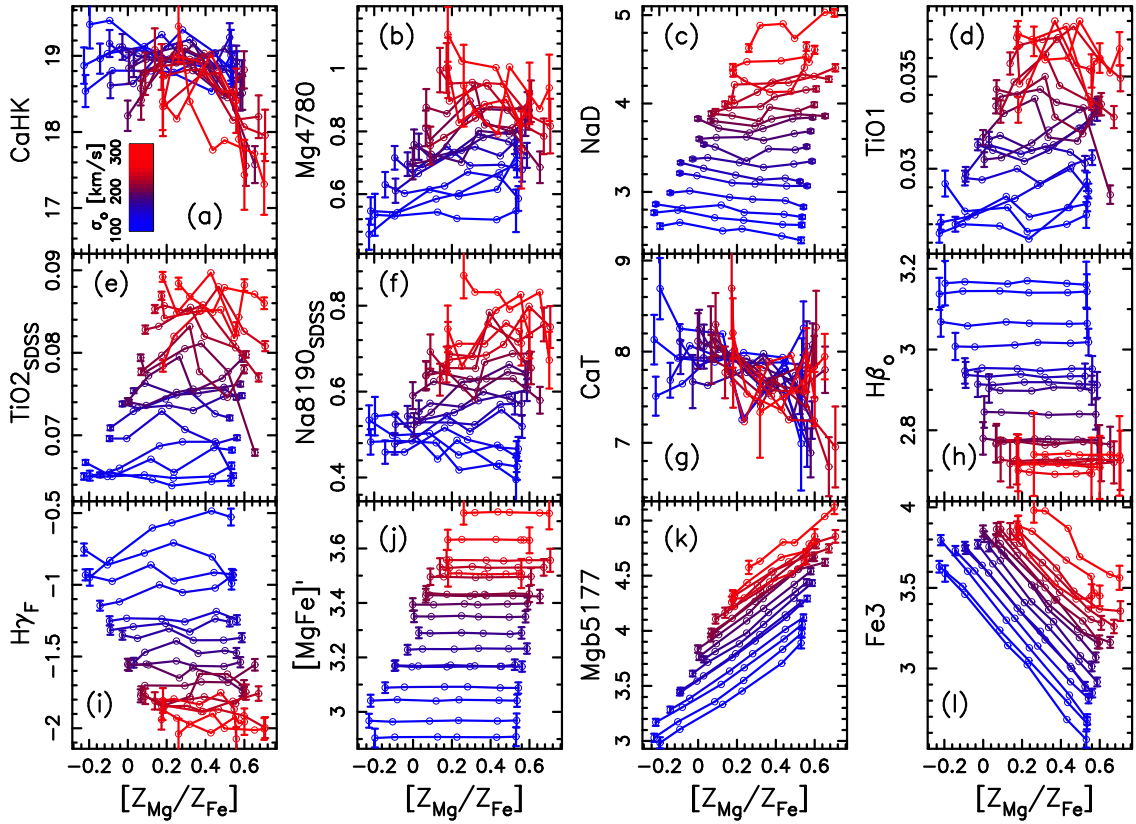}
\end{center}
\caption{Same as Fig.~\ref{fig:indices_fixed_sigma_nomod} but after
  removing the trends of \hbo\ and \mgfep\ in each $\sigma_0$ bin,
  using extended MILES (MIUSCAT) SSPs with a Kroupa IMF (see the
  text). }
%\contcaption{.}
\label{fig:indices_fixed_sigma}
\end{figure*}
%%%%%%%%%%%%%%%%%%%%%%%%%%%%%%%%%%%%%%%%%%%%%%%%%%%%%%%%%%%%%%%%%%%%%%

%%%%%%%%%%%%%%%%%%%%%%%%%%%%%%%%%%%%%%%%%%%%%%%%%%%%%%%%%%%%%%%%%%%%%%
\begin{figure*}
\begin{center}
\leavevmode
\includegraphics[width=15cm]{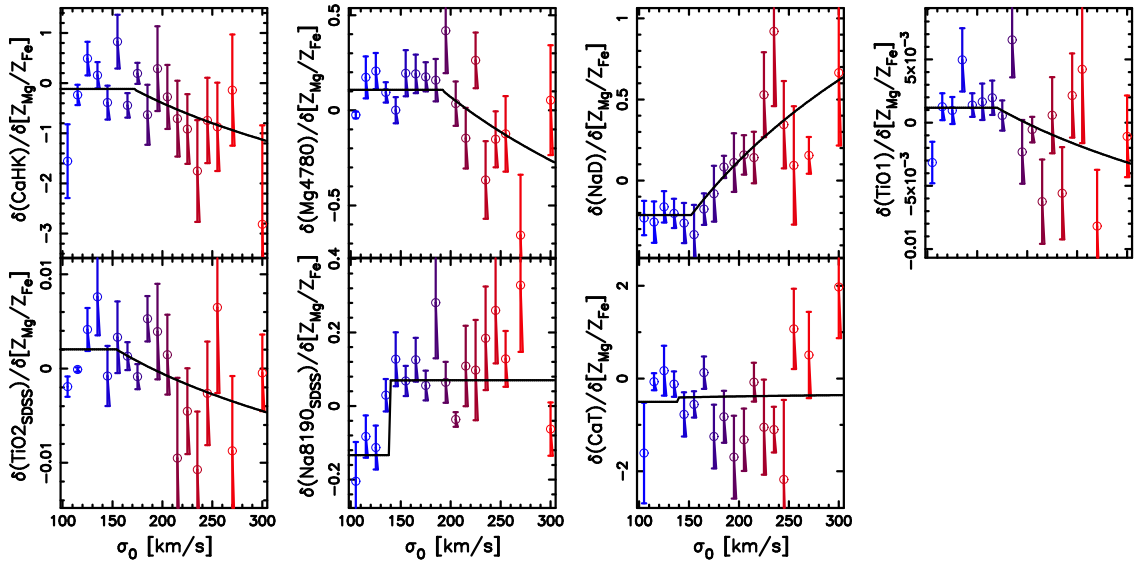}
\end{center}
\caption{Slopes of the trends of spectral indices vs. \afep\ as a
  function of velocity dispersion.  Colours correspond to different
  velocity dispersion values, as in
  Fig.~\ref{fig:indices_fixed_sigma_nomod}.  Error bars are 1\,$\sigma$
  uncertainties of the slope.  Black curves trace the best-fitting
  analytic approximations of the observed trends (see the text).  }
%\contcaption{.}
\label{fig:indices_slopes_sigma}
\end{figure*}
%%%%%%%%%%%%%%%%%%%%%%%%%%%%%%%%%%%%%%%%%%%%%%%%%%%%%%%%%%%%%%%%%%%%%%

{ We correct  the spectral indices for their  dependence on age and
  metallicity  at fixed  $\sigma_0$, as  well as  their  dependence on
  resolution among stacks  with different $\sigma_0$}.  The correction
procedure estimates age and metallicity from the \hbo--\mgfep\ diagram
for each stack,  and then uses those age  and metallicity estimates to
remove the age/metallicity dependence of each index, for each stack in
a given $\sigma_0$ interval.  More explicitly, we adopt the following
equation:
\begin{equation}
\rm
I_{corr} = I_{obs}-I_{mod}(H\beta_o; MgFe'; Kroupa)+\langle I_{mod}\rangle_{200},
\label{eq:afe_corr}
\end{equation}
where  $\rm  I_{obs}$   is  the  index  line  strength   for  a  given
\afep\ stacked  spectrum; $I_{corr}$ is the index  after correction is
applied; ${\rm  I_{mod}(H\beta_o; MgFe'; Kroupa)}$ is  the index value
expected for the same stack by  fitting its \hbo\ and \mgfep\ EWs with
extended  MILES  (MIUSCAT)  SSPs  having a  Kroupa  IMF;  ${\rm\langle
  I_{mod}\rangle_{200}}$ is the index  value predicted for an SSP with
a    {   Kroupa    IMF,   $\sigma_{\rm    0,ref}=200$\,\kms\   (see
  Sec.~\ref{sec:indices})},  and age  and metallicity  fixed  to their
average values within the  $\sigma_0$ bin.  Notice that the quantities
${\rm  I_{mod}(H\beta_o;  MgFe';   Kroupa)}$  are  computed  by  first
correcting \hbo\ for  nebular emission (see App.~\ref{sec:hb_emission}
for details).   { The  ${\rm I_{mod}(H\beta_o; MgFe';  Kroupa)}$ is
  computed at  the same  resolution as $\rm  I_{obs}$, i.e.   the SDSS
  resolution  plus   the  $\sigma_0$  of  the   given  stack}.   Since
${\rm\langle   I_{mod}\rangle_{200}}$  is  computed   at  $\sigma_{\rm
  0,ref}=200$\,\kms,  Eq.~\ref{eq:afe_corr}   does  also  correct  the
trends   of  Fig.~\ref{fig:indices_fixed_sigma_nomod}   to   the  same
spectral resolution.  We point out that the correction does not assume
any  variation of  IMF among  stacks, which  is the  hypothesis  to be
tested   in   this   paper.    {   Furthermore,   as   noticed   in
  Sec.~\ref{sec:indices},   the   reference   value  of   $\sigma_{\rm
    0,ref}=200$\,\kms is the central  value of the $\sigma_0$ range of
  our  stacks,  hence  minimizing  the resolution  correction  to  the
  indices  for all  bins.}  Fig.~\ref{fig:indices_fixed_sigma}  is the
same  as Fig.~\ref{fig:indices_fixed_sigma_nomod}  but  after applying
Eq.~\ref{eq:afe_corr},  i.e. it  shows the  genuine dependence  of the
indices on \afep.  Most of  the trends are weaker than those presented
in  Fig.~\ref{fig:indices_fixed_sigma_nomod}.  { In  particular, by
  construction,  \hbo\ and  \mgfep\  do not  show  any variation  with
  \afep, while  \hgf\ does not  show any significant  correlation with
  \afep, aside from a few  bins (e.g. for the lowest $\sigma_0$; where
  the correction  might have  been overestimated).  For  what concerns
  IMF-sensitive  indices, as  already noticed  above,  their variation
  with $\sigma_0$ is  similar, or even larger, at  low \afep, implying
  that $\sigma_0$ is  the main driver of IMF  variations among stacks.
  Since   the    error   on   \afep\   is    about   $0.15$~dex   (see
  Sec.~\ref{fig:proxy_estimate}), i.e.  significantly smaller than the
  \afep\ range of  $\sim 0.8$~dex, our finding is  likely not affected
  by the errors on \afep.  In addition, by stacking only galaxies with
  a higher  $S/N_{MgFe}$ ($>25$), we  find consistent trends  to those
  presented in Fig.~\ref{fig:indices_fixed_sigma}.  }

%%%%%%%%%%%%%%%%%%%%%%%%%%%%%%%%%%%%%%%%%%%%%%%%%%%%%%%%%%%%%%%%%%%%%%
\begin{figure}
\begin{center}
\leavevmode
\includegraphics[width=8cm]{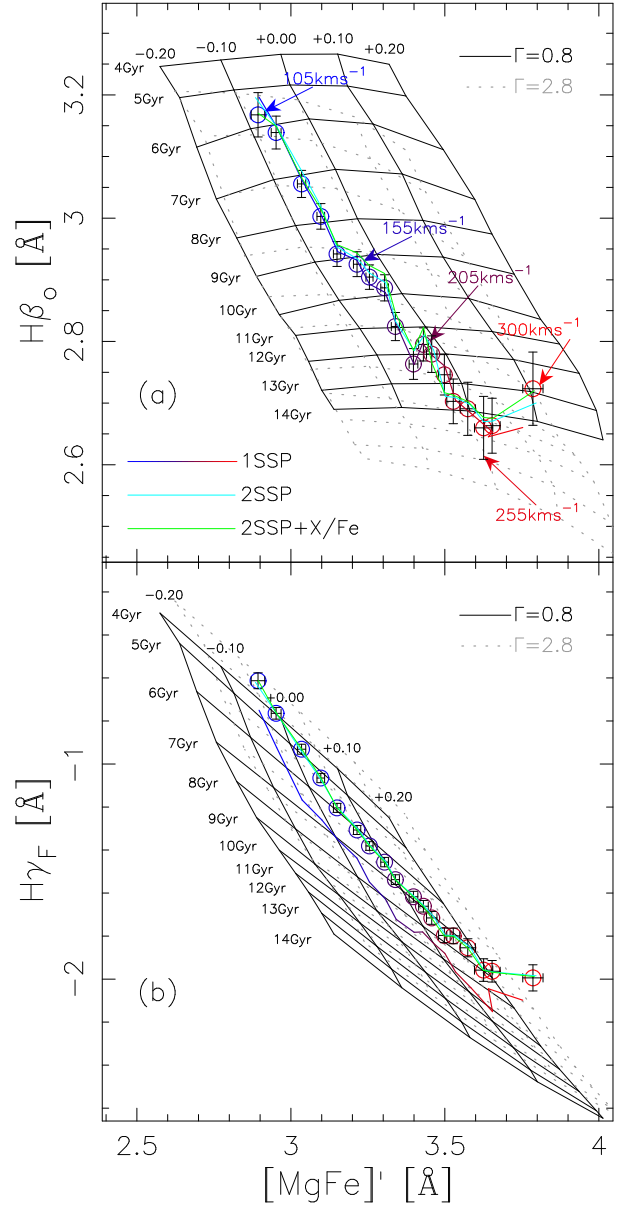}
\end{center}
\caption{Results of fitting spectral indices with $1SSP$, $2SSP$, and
  $2SSP+X/Fe$, models. The age-sensitive spectral indices \hbo\ ({\sl
    top}) and \hgf\ ({\sl bottom}) are shown as a function of the
  total metallicity proxy, \mgfep.  The grids correspond to 1SSP
  extended MILES (MIUSCAT) models, smoothed to a resolution of
  $200$\,\kms, with different age and metallicity, as labelled, for a
  top-heavy ($\Gamma=0.8$; black) and bottom-heavy ($\Gamma=2.8$;
  grey) bimodal IMF. Open circles plot line strengths of the 18
  $\sigma_0$ stacks corrected to a common resolution of $200$\,\kms.  Error
  bars are given at the 2\,$\sigma$ level (including measurement
  errors and \afep\ correction uncertainties).  Circles are plotted
  with different colours, as in Fig.~\ref{fig:indices_slopes_sigma},
  depending on the $\sigma_0$ bin.  Solid curves show the best-fitting
  solutions for $1SSP$ (blue-through-red curve), $2SSP$ (cyan), and
  $2SSP+X/Fe$ (green).  }
%\contcaption{.}
\label{fig:fit_indices_1}
\end{figure}
%%%%%%%%%%%%%%%%%%%%%%%%%%%%%%%%%%%%%%%%%%%%%%%%%%%%%%%%%%%%%%%%%%%%%%

In order to characterize the trends of spectral indices with \afep, we
fit them  with linear relations, treating index  values as independent
variables.  In Fig.~\ref{fig:indices_slopes_sigma}, we plot the slopes
of the  best-fitting relations, $\delta({\rm index})/\delta($\afep$)$,
as a function  of $\sigma_0$.  For each index  line strength, $I$, and
each $\sigma_0$ bin,  we also compute the rms  of the residuals around
the best fit  $(s_I)$. The quantity $s_I$ provides  an estimate of the
uncertainty in the use  of linear relations to interpolate/extrapolate
index values to solar abundances (\afep$=0$), and is included into the
error  budget   when  fitting   stellar  population  models   to  data
(Sec.~\ref{sec:fitting}).   The figure shows  that all  indices except
\cat, exhibit two regimes  with different slopes, featuring a constant
slope  at  low  $\sigma_0$  ($\simlt 130-150$\,\kms),  followed  by  a
correlation  with  $\sigma_0$  at  higher velocity  dispersion.   This
behaviour  is   traced  by   the  black  curves   in  each   panel  of
Fig.~\ref{fig:indices_slopes_sigma}.    Each  curve  is   obtained  by
connecting the median slope value  at low $\sigma_0$ to a best-fitting
power-law relation at higher $\sigma_0$.  For \cat, we notice that our
stacking procedure  involves fewer spectra  (i.e. only those  at lower
redshift,   see  Sec.~\ref{sec:stacks}),   making   the  corresponding
\afep\     trends     --     hence     the     slope     values     in
Fig.~\ref{fig:indices_slopes_sigma}  -- not  as robust  as  with other
indices.  For instance, given the  large uncertainty on the slopes for
the three  highest $\sigma_0$  bins, we cannot  exclude that  \cat\ is
insensitive to  \afep\ at lowest  $\sigma_0$, with the  slope becoming
negative at $\sigma_0> 150$\,\kms, as for \cahk.  The main features of
the \afep\ trends can be summarized as follows.

\begin{table*}
\centering
\small
%\begin{minipage}{150mm}
 \caption{Sensitivity of the \nad\ index to \afe\ according to different stellar population models.  
}
  \begin{tabular}{c|c|c}
   \hline
model$^\dag$ &  $\delta(\rm NaD)/\delta($\afe$)$ & model parameters \\
(1) & (2) & (3) \\
   \hline
TMJ11 & $+0.2$ & Salpeter IMF;  $[Z/H]=0.0$; Age$=10$~Gyr; $0 \le$\afe$\le 0.5$\\
CvD12 & $-2.0$ & Chabrier IMF;$[Fe/H]=0.0$; Age$=13.5$~Gyr; $0 \le$\afe$\le 0.3$\\
COE07 & $-1.3$ & Chabrier IMF; $[Fe/H]=0.0$; Age$=10$~Gyr; $0 \le$\afe$\le 0.4$\\
CER07 & $-2.5$ & Kroupa IMF;   $[Z/H]=0.0$; Age$=10$~Gyr; $0 \le$\afe$\le 0.4$\\
   \hline
  \end{tabular}
\label{tab:nad}

\noindent
$\dag$ TMJ11:  Thomas, Maraston \& Johansson (2011);  CvD12: Conroy \&
van Dokkum (2012a);  COE07: Coelho et al. (2007);  CER07: Cervantes et
al. (2007).
%\end{minipage}
\end{table*}

\begin{description} 
\item[{\it Na indices --}] { At low velocity dispersion, the median
  slopes are  negative, i.e.  both  \nad\ and \naii\ tend  to decrease
  with \afep. On the contrary, at high velocity dispersion, the slopes
  are  positive,  i.e.  the  Na  indices  increase  with \afep.   }The
  low-$\sigma_0$  behaviour   is  qualitatively  consistent   with  the
  predictions of  the CvD12 stellar population models,  where \afe\ is
  expected to  affect the  continua of both  \nad\ and  \naii, causing
  these indices to decrease  with \afe.  However, CvD12 models predict
  a      slope      of     $\delta(\rm      Na8190)/\delta($\afe$)\sim
  -0.5$~\footnote{This  value refers to  CvD12 models  with an  age of
    $13.5$~Gyr      and       Chabrier      IMF,      smoothed      to
    $\sigma=200$\,\kms. Notice that CvD12 models are computed for $\rm
    [Fe/H]=0$.    Hence,   higher  \afe\   SSPs   have  higher   total
    metallicity, and  the variation of  a given index with  \afe\ does
    also  reflect  a  metallicity  variation. Since  \naii\  tends  to
    increase  with $[Z/H]$,  one might  expect an  even  steeper slope
    (i.e. more negative) than  that reported above ($-0.5$).  }, which
  is significantly different from the value of $-0.15$ one obtains for
  the  lowest  three bins  of  $\sigma_0$  (after  accounting for  the
  relation  between \afep\ and  \afe, see  Fig.~\ref{fig:proxy}).  The
  same  discrepancy  exists for  \nad\  where  we measure  $\delta(\rm
  NaD)/\delta($\afe$)\sim  -0.4$ (at  lowest $\sigma_0$),  while CvD12
  models  predict $\sim  -2$.   Even stronger  differences exist  when
  comparing different models, as shown in Tab.~\ref{tab:nad}, where we
  report  the  \afe\  sensitivity   of  \nad\  for  different  stellar
  population models,  namely CvD12, TMJ11, \citet[CER07]{Cervantes07},
  and  \citet[COE07]{Coelho:07}.   We   find  large  differences  when
  comparing  different   models~\footnote{  Notice  that   while  some
    difference  between  models  can  be  expected  because  different
    conventions for metallicity are adopted (i.e.  either $\rm [Fe/H]$
    or $[Z/H]$), large differences  also exist between models adopting
    the  same definition of  metallicity (i.e.  TMJ11 vs.   CER07, and
    CvD12  vs. COE07).}.   Moreover,  while CvD12,  COE07, and  CER07,
  agree qualitatively,  in that the \nad\ weakens  as \afe\ increases,
  TMJ11  models  predict  a  mild  {\sl  increase}  with  enhancement.
  Unfortunately,  the same  model comparison  cannot be  performed for
  \naii,   as    only   CvD12   models    can   be   used    to   make
  predictions~\footnote{Neither  TMJ11  nor   CER07  models  have  the
    wavelength coverage  required to compute  \naii.  This is  not the
    case  for  COE07  synthetic  models,  but  for  these  models  the
    equivalent width of  \naii\ turns out to be  negative, in contrast
    to observations.  }  at $\lambda \sim 8200$\,\AA.
\item[{\it TiO indices -- }] The slopes change similarly for both
  indices, from slightly positive at low $\sigma_0$ to negative (on
  average) for $\sigma_0\simgt 200$\,\kms.  Even in this case, large
  differences exist when comparing different models, where the two TiO
  indices are expected to be either almost independent (TMJ11);
  strongly increasing (CER07); mildly increasing (CvD12); or mildly
  decreasing (COE07) with \afe.
\item[{\it  Ca  indices  --  }]  Both \cahk\  and  \cat\  have  slopes
  consistent with zero at  low $\sigma_0$.  The slopes become negative
  at  high $\sigma_0$ for  \cahk, while  for \cat\  the trend  is less
  clear.  Notice  that CvD12 models  predict both indices  to increase
  with \afe,  the same trend being  also predicted by  COE07 and CER07
  models for  the \cahk\ index, in  contrast to our  findings (at both
  low and high $\sigma_0$).  For \cahk, this discrepancy might also be
  due to the fact that this  index is expected to be affected, for low
  mass  stars,  by   chromospheric  emission  fill-in,  hampering  its
  modelling.
\item[{$Mg4780$ -- }] This index mildly increases with \afep\ at the
  lowest $\sigma_0$, consistent with the expected sensitivity to $\rm
  [Mg/Fe]$ \citep[see][]{Serven05}, while it decreases with \afep\ at high
  $\sigma_0$.  This double-regime behaviour is similar to that of the
  other indices shown, and indicates that $Mg4780$ might be sensitive to
  other elements~\footnote{For instance, using CvD12 models, one finds
    that the equivalent width of \mgf\ decreases as $[C/Fe]$
    increases, showing an opposite behaviour to $\rm [Mg/Fe]$. } besides Mg.
\end{description}

The fact  that the response of  spectral indices to  \afep\ depends on
$\sigma_0$  implies  that \afep\  traces  different element  abundance
ratios  at different  galaxy  mass scales.   At  high $\sigma_0$,  the
positive  slope of  the \afep\  trends of  \nad\ and  \naii\  might be
reflecting  a  stronger increase  in  Na  abundance  with \afep,  with
respect to other  $\alpha$ elements. { Would this  be the case, one
  might expect a  negative slope of Ca indices  at high $\sigma_0$, as
  the abundance of Na, at fixed Ca abundance, is expected to influence
  the Ca  line strengths  through its effect  on electron  pressure in
  stellar  atmospheres (see e.g.   CvD12).  While  this is  indeed the
  case  for  \cahk,  the  situation  for  \cat\  is  less  clear  (see
  above)\footnote{  The  fact that  we  do  not  see an  opposite
    behaviour  among \afep\  slopes of  \cat\ and  Na indices  at high
    $\sigma_0$  argues  against a  picture  whereby abundance  ratios,
    rather than IMF,  drive variations of Na and  Ca line strengths.}.
  In  practice,   a  variation  in  $[Na/Ca]$  with   \afe,  at  fixed
  $\sigma_0$,  might   dominate  the  effect   of  electron  pressure,
  eventually leading to  no correlation among the slopes  of Ca and Na
  indices. }  On the other hand, at low (relative to high) $\sigma_0$,
the trends with \afep\ might be reflecting instead the true dependence
of  the spectral  indices with  \afe.  Regardless  of what  causes the
variation of the slopes, the main  point for the present study is that
one can use these slopes, for each $\sigma_0$ bin, in order to correct
the observed indices to solar abundance ratio (\afep$=0$), and compare
them  with the  predictions from  (nearly solar-scale)  MILES extended
(MIUSCAT) models.

%%%%%%%%%%%%%%%%%%%%%%%%%%%%%%%%%%%%%%%%%%%%%%%%%%%%%%%%%%%%%%%%%%%%%%
\begin{figure*}
\begin{center}
\leavevmode
\includegraphics[width=18cm]{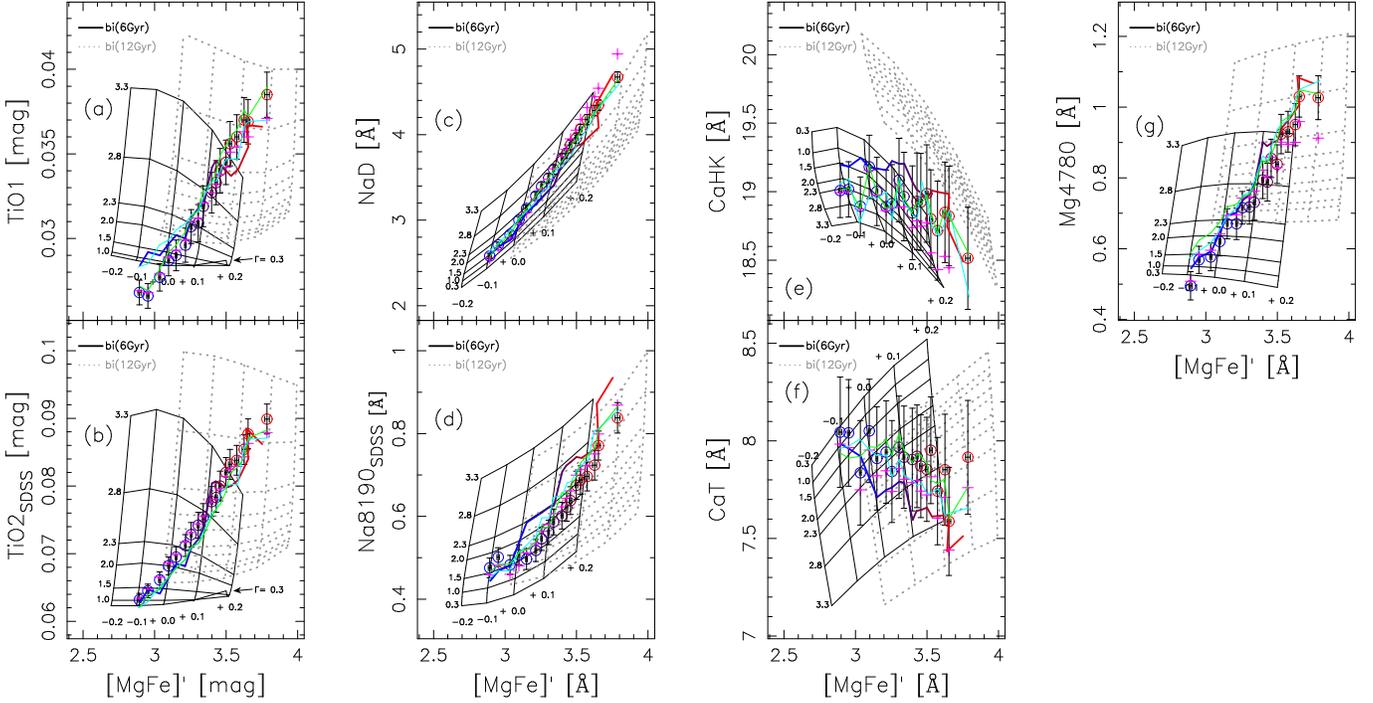}
\end{center}
\caption{Same as Fig.~\ref{fig:fit_indices_1}, but for IMF- and
  abundance-sensitive spectral indices. Black and grey grids show the
  effect of changing IMF slope and metallicity, for two SSPs, with
  ``young'' ($6$~Gyr) and ``old'' ($12$~Gyr) age, respectively (see
  upper-left corner of each panel).  Error bars are 1\,$\sigma$
  uncertainties (including measurement errors and \afep\ correction
  uncertainties).  Open circles plot the line strengths of the 18
  $\sigma_0$ stacks corrected to a common resolution of $200$\,\kms,
  and solar abundance (\afep=0). For reference, magenta crosses show
  line strengths when no abundance correction is applied.  }
%\contcaption{.}
\label{fig:fit_indices_2}
\end{figure*}
%%%%%%%%%%%%%%%%%%%%%%%%%%%%%%%%%%%%%%%%%%%%%%%%%%%%%%%%%%%%%%%%%%%%%%

%%%%%%%%%%%%%%%%%%%%%%%%%%%%%%%%%%%%%%%%%%%%%%%%%%%%%%%%%%%%%%%%%%%%%%%%
\section{Fitting procedure}
\label{sec:fitting}

For a given  IMF (either bimodal or unimodal),  we adopt a multi-index
fitting procedure, minimizing the following expression:
\begin{equation}
\chi^2(\Gamma) = \sum \frac{(I_{corr} - I_{MOD})^2}{\sigma_I^2+s_I^2},
\label{eq:chi2}
\end{equation}
where the sum extends over all the selected spectral indices (see
Sec.~\ref{sec:indices}), $I_{corr}$ and $s_I$ are the index line
strength corrected to \afep=0 and the correction uncertainty,
respectively (see Sec.~\ref{sec:afe_corr}), $\sigma_I$ is the
measurement uncertainty on $I$ (reflecting the statistical noise in a
stacked spectrum), and $I_{MOD}$ is the (extended MILES; MIUSCAT)
model line strength, with $MOD$ being either $1SSP$ or $2SSP$ (see
below).  The minimization is performed with respect to the model
parameters, and for each stack, both observed and model indices are
computed at the nominal resolution (SDSS plus $\sigma_0$) of a given
stacked spectrum. Three different models are considered, namely:
\begin{description}
\item[{\it 1 SSP -- }] The model parameters are the age and
  metallicity of a single extended MILES (MIUSCAT) SSP.  As described
  in Sec.~\ref{sec:models}, we consider SSPs covering a wide range of
  (200$\times$150) ages and metallicities.  Eq.~\ref{eq:chi2} is
  minimized over the grid for each value of $\Gamma$, and then, for a
  given choice of the IMF (i.e. bimodal or unimodal), the value of
  $\Gamma$ giving the minimum $\chi^2$ is selected.  The procedure is
  repeated by shifting indices according to their uncertainties
  ($\sigma_I$ and $s_I$), resulting in a PDF (probability distribution
  function) for $\Gamma$ (marginalized in age and metallicity).  For
  the 1SSP fits, we exclude \hgf\ from the fitting, for reasons
  explained in Sec.~\ref{subsec:resone}.
\item[{\it 2 SSPs -- }] The models consists of a linear combination of
  two extended MILES (MIUSCAT) SSPs, having the same IMF, but
  different ages and metallicities.  The five model parameters are the
  ages and metallicities of the 2 SSPs plus a mass fraction giving
  their relative contribution to the linear combination.  The
  minimization is performed by varying the ages and metallicities of
  the two populations independently, over the same grid as for 1SSP.
  Since the 2SSP fitting procedure involves a much larger volume of
  parameter space, we do not iterate the procedure to estimate
  uncertainties.  Instead, we weight each linear combination by
  $\exp(-\chi^2/2)$, generating a PDF for $\Gamma$ (marginalized over
  all other fitting parameters).
\item[{\it 2 SSPs$+X/Fe$  -- }] The models consist  of 2SSP (five free
  parameters), plus three free parameters describing the abundances of
  calcium,  sodium, and  titanium ($\rm  [Ca/Fe]$, $\rm  [Na/Fe]$, and
  $\rm [Ti/Fe]$, respectively).  In  practice, we replace $I_{MOD}$ in
  Eq.~\ref{eq:chi2}  with  $I_{2SSP+X/Fe}=I_{2SSP}  -  \delta_X  \cdot
  [X/Fe]$,    where    $\delta_X=\delta(I)/\delta([X/Fe])$   is    the
  sensitivity  of a given  index to  a variation  in the  abundance of
  element $X$.  Uncertainties on each fitting parameter (and $\Gamma$)
  are computed  as for the 2SSP  fits, marginalizing the  PDF over all
  the other fitting parameters.  For  each index, we consider only the
  contribution from the  dominant element the index is  expected to be
  sensitive to, i.e.  $\rm [Ca/Fe]$ for \cahk\ and \cat, $\rm [Na/Fe]$
  for \nad\ and \naii, and $\rm [Ti/Fe]$ for \tioi\ and \tioii.  For a
  given  stacked  spectrum  (i.e.   $\sigma_0$ bin),  we  compute  the
  corresponding  $\delta_X$'s by  using the  publicly  available CvD12
  models  with  varying  abundance  ratios.  These  models,  having  a
  Chabrier  IMF,  $\rm  [Fe/H]=0$,  and  an age  of  $13.5$\,Gyr,  are
  smoothed to match $\sigma=200$\,\kms (i.e. the average $\sigma_0$ of
  our  stacks)  plus  the   SDSS  resolution.   Varying  the  velocity
  dispersion  in  the  smoothing  does not  change  significantly  the
  $\delta_X$  estimates. In  Sec.~\ref{subsec:abundances},  we compare
  the $\delta_X$ estimates from  CvD12 models with estimates from some
  simple SSPs we create  by using different synthesis ingredients from
  CvD12  (see  App.~\ref{sec:synth_ssps}),  finding good  consistency.
  Using  CvD12 models, we  have explored  the sensitivity  of spectral
  indices to the abundance of  several individual elements (Ca, C, Mg,
  Na,  N, Si, Ti),  by estimating  the variation  of each  index (with
  respect to the solar scale) caused  by a factor of two change in the
  abundance of each element.  All  selected indices are found to be at
  least three  times more sensitive  to the dominant, than  any other,
  element.  For \tioi, \tioii, and \nad, the choice of a given element
  as the  dominant one  is also consistent  with the response  of Lick
  indices   to   single   element   abundance  ratios,   as   reported
  by~\citet[see their fig.~1]{JTM12}. Notice that the dominant element
  of \mgf\  is Mg.   Since the \afep\  should reflect directly  the Mg
  abundance, and the indices of our stacks are corrected to \afep$=0$,
  we do not consider the  effect of a residual $\rm [Mg/Fe]$ variation
  in the 2SSP$+X/Fe$ fits.
\end{description}

%%%%%%%%%%%%%%%%%%%%%%%%%%%%%%%%%%%%%%%%%%%%%%%%%%%%%%%%%%%%%%%%%%%%%%%%
\section{Fitting results}
\label{sec:results}

Figs.~\ref{fig:fit_indices_1} and~\ref{fig:fit_indices_2} compare
observed and best-fitting spectral indices for different fitting
methods, i.e.  $1SSP$, $2SSP$, and $2SSP+X/Fe$ models
(blue-through-red, cyan, and green curves, respectively).
Fig.~\ref{fig:fit_indices_1} plots the main age- and
metallicity-sensitive indicators, \hbo, \hgf, and \mgfep, while
Fig.~\ref{fig:fit_indices_2} shows abundance- and IMF-sensitive
indices.  The measured spectral indices, corrected to \afep$=0$ and
reported to a common resolution~\footnote{The correction to
  $\sigma_0=200$\,\kms\ is performed by computing best-fitting $1SSP$
  indices at the actual resolution of a given stack and the reference
  resolution of $\sigma_0=200$\,\kms.  The correction is given by the
  corresponding difference of line strength values.}  of
$\sigma_0=200$\,\kms, are plotted as blue-through-red circles.  All
indices are plotted as a function of \mgfep\ (proxy of total
metallicity).  For each panel in the Figures, the grids correspond to
single SSP models, with varying age, metallicity, and bimodal IMF
slope, as labelled.  The effect of \afep\ corrections is illustrated
by the magenta crosses, showing line strengths when no correction is
applied.  The corrections are generally mild, shifting data-points
(i.e.  blue-through-red relative to magenta symbols) within error
bars, with the main exception of \nad\ (see
Fig.~\ref{fig:fit_indices_2}), where the shift at high $\sigma_0$,
towards lower equivalent widths, is significant at more than the
3\,$\sigma$ level.

To compare the fitting quality among different $\sigma_0$ bins and
different fitting methods, we report in Tab.~\ref{tab:chi_fit}, for
each fit, the reduced chi-squared statistics, $\chi_\nu^2$, and the
probability, $P_{\ge \chi_{\nu}^2}$, for the $\chi_{\nu}^2$ to be
larger than the observed value, given the number of degrees of freedom
($\nu$, with $\nu=7$, $5$, and $2$, for $1SSP$, $2SSP$, and
$2SSP+X/Fe$, respectively).  By definition, lower values of $P_{\ge
  \chi_{\nu}^2}$ imply less likely best-fitting solutions, with
$P_{\ge \chi_{\nu}^2} <5\%$ representing a rejection at more than the
2\,$\sigma$ level for a random normal deviate.  In practice, because
the models may present small systematic deviations with respect to the
data, independent of the relevant stellar population properties (age,
metallicity, and IMF), we mainly use the $P_{\ge \chi^2}$ in a
relative sense here, i.e.  to compare different $\sigma_0$ bins and
different methods, as detailed below.  We notice that in addition to
constraining the IMF slope, different fitting models also give
information about the stellar population content (i.e.  age and
metallicity) of ETGs as a function of $\sigma_0$. Since our main focus
here is that of the stellar IMF, in the following sections we only
discuss some general features of the stellar population properties
inferred from different fitting schemes, postponing a more detailed
analysis to a future paper.

In Secs.~\ref{subsec:resone}, \ref{subsec:restwo},
and~\ref{subsec:abundances} we present the results obtained for
$1SSP$, $2SSP$, and $2SSP+X/Fe$ models, assuming bimodal IMF models.
The case of unimodal IMFs is considered in Sec.~\ref{subsec:unimod},
while Sec.~\ref{subsec:hybrid} presents results of the hybrid
approach.

\begin{figure*}
\begin{center}
\leavevmode
\includegraphics[width=12cm]{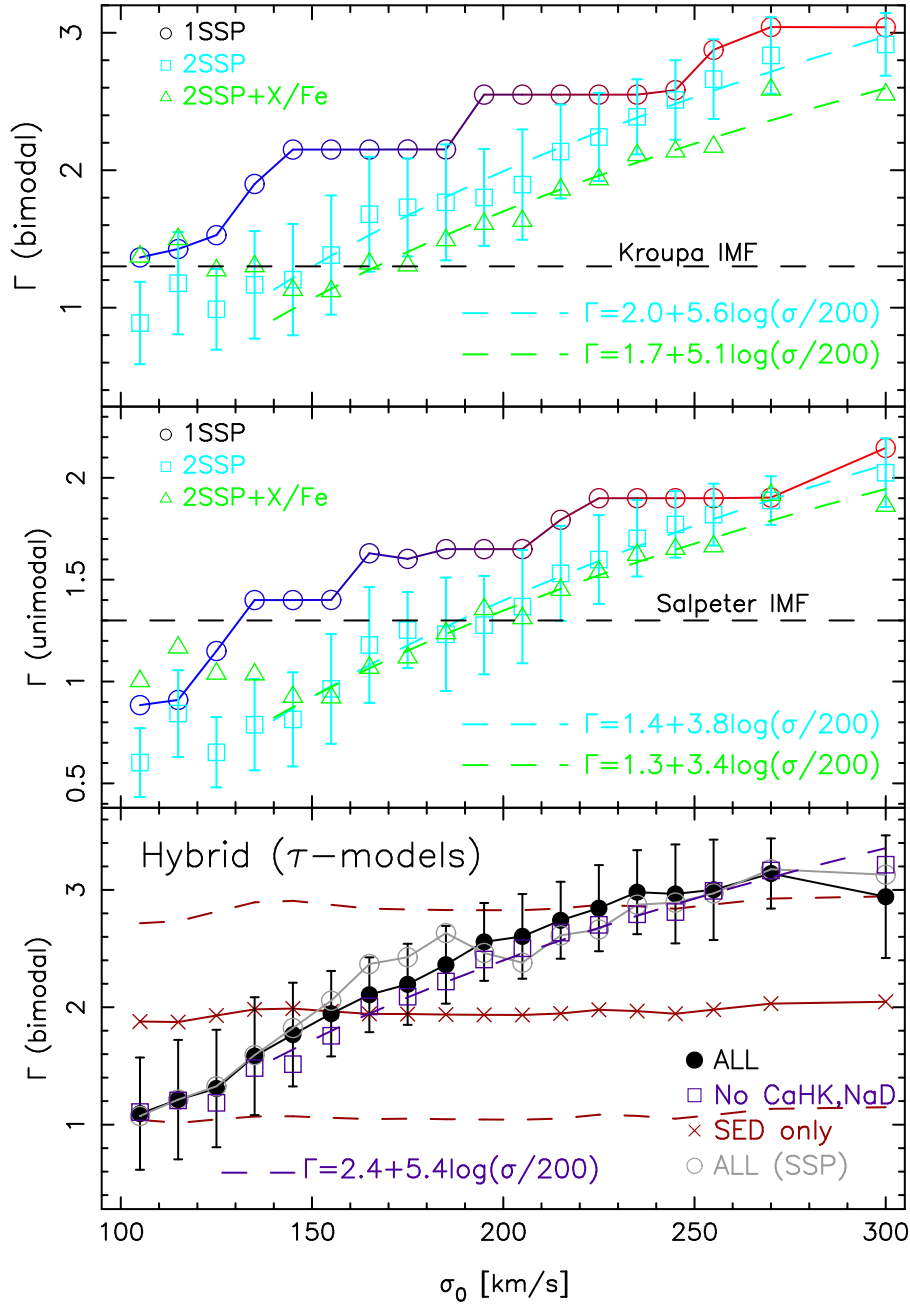}
\end{center}
\caption{Best-fit slope of the IMF, $\Gamma$, from different methods
  and models, as a function of central velocity dispersion
  ($\sigma_0$).  Top panel: best-fit slope of the bimodal IMF obtained
  by fitting spectral indices with $1SSP$, $2SSP$, and $2SSP+X/Fe$
  models.  Different symbols and colours correspond to the different
  models, as labelled in the top--left corner of the plot. The colours
  follow the same scheme as in Figs.~\ref{fig:fit_indices_1}
  and~\ref{fig:fit_indices_2}.  The error bars are given at the
  1\,$\sigma$ level, and are only shown for the $2SSP$ models for
  clarity. The horizontal dashed line marks the value of $\Gamma=1.3$,
  corresponding to a Kroupa-like IMF. For $\sigma_0>140$\,\kms, where a
  significant variation of $\Gamma$ with $\sigma_0$ is detected, we
  also show the best-fit relations to the $\Gamma$--$\sigma_0$ trends
  as dashed curves, for $2SSP$ and $2SSP+X/Fe$ models, respectively.
  The best-fit relations are reported in the lower-right corner of the
  plot.  Middle panel: same as top panel but for unimodal, rather
  than bimodal, IMFs.  Notice that both bimodal and unimodal models
  lead to the similar conclusion of a systematic increase of IMF slope
  with central velocity dispersion ($\sigma_0$).  The horizontal
  dashed line marks the value of $\Gamma=1.3$, corresponding, for a
  unimodal IMF, to the Salpeter slope.  Bottom panel: marginalized
  values of the (bimodal) IMF slope, obtained by our hybrid approach,
  with respect to velocity dispersion, including $1\sigma$
  confidence levels.  The joint method (SED fitting plus line
  strengths) for a grid of $\tau$ models is shown as black solid dots.
  The plot also shows the case for simple stellar populations (grey
  open dots) and for the $\tau$ model, where only the strongest
  IMF-sensitive line strengths are considered
  (\mgf,\,\tioi,\,\tioiio,\,\naii,\,\cat).  In this case, the best-fit
  relation to the $\Gamma$--$\sigma_0$ trend is also shown as a
  magenta, dashed, curve, for $\sigma_0>140$~\kms.  For reference, the
  same analysis restricted to spectral fitting alone is given by the red
  crosses and dashed red lines ($1\sigma$ level).}
%\contcaption{.}
\label{fig:gamma_fits_all}
\end{figure*}

%\caption{Same as Fig.~\ref{fig:gamma_fits} but for unimodal, rather
%  than bimodal, IMFs. Notice that both bimodal and unimodal models
%  lead to the similar conclusion of a systematic increase of IMF slope
%  with central velocity dispersion ($\sigma_0$).  The horizontal
%  dashed line marks the value of $\Gamma=1.3$, corresponding, for a
%  unimodal IMF, to the Salpeter slope.
% Best-fitting  slope  of  bimodal  IMFs  as  a  function  of
%  $\sigma_0$.  Different symbols and  colours correspond to the results
%  of fitting  different stellar population  models, as labelled  in the
%  top--left corner of the plot. Colours  are the same coding as in each
%  panel of  Fig.~\ref{fig:fit_indices}. Error bars are  plots only for
%  $2SSP$  models, to  make  the  plot more  clear,  and correspond  to
%  1~$\sigma$ uncertainties on $\Gamma$ values. 

%%%%%%%%%%%%%%%%%%%%%%%%%%%%%%%%%%%%%%%%%%%%%%%%%%%%%%%%%%%%%%%%%%%%%%%%
\subsection{Bimodal IMFs -- $1SSP$ fits}
\label{subsec:resone}
Overall,  single SSPs  reproduce fairly  well the  relative  trends of
\hbo,  \mgfep,  IMF- and  abundance-sensitive  spectral  indices as  a
function  of $\sigma_0$.   As seen  from the  \hbo\ vs.   \mgfep\ plot
(panel  a of  Fig.~\ref{fig:fit_indices_1}), $1SSP$  fits  yield older
ages and higher metallicities  with increasing velocity dispersion, in
agreement with previous work (e.g.~\citealt{GALL:06}).  The IMF slope,
$\Gamma$,  is found to  increase, becoming  more bottom-heavy  at high
$\sigma_0$.   This  is  shown  in  Fig.~\ref{fig:gamma_fits_all}  (top
panel), where we plot the best-fitting $\Gamma$, for bi-modal IMFs, as
a function of  $\sigma_0$, with the same colour  coding, for different
$\sigma_0$  bins,  as  in  Fig.~\ref{fig:fit_indices_1}.   The  $1SSP$
best-fitting $\Gamma$  increases from $\sim 1.4$  (i.e.  a Kroupa-like
slope) at  $\sigma_0 \sim  100$\,\kms\ to $\sim  3$ at  $\sigma_0 \sim
300$\,\kms.  This  IMF trend  is consistent with  that derived  in our
previous work (see  fig.~4 of FLD13), { where  we adopted a smaller
  set of  spectral indices,  and a different  approach to  analyze the
  stacked  spectra  (see  Sec.~\ref{subsec:hybrid}): we  analyzed  the
  stacks  after  smoothing all  of  them  to  the same  $\sigma_0$  of
  $300$\,\kms (rather than keeping them at their original resolution),
  and no  abundance ratio corrections  were applied.  To  further test
  whether our results are affected  by the different resolution of the
  stacks (see Sec.~\ref{sec:indices}), we repeated the $1SSP$ analysis
  by measuring  and fitting line  indices after smoothing  all spectra
  and models  to $300$\,km/s, finding negligible  differences ($<10 \,
  \%$)  in  the  resulting  values   of  $\Gamma$  as  a  function  of
  $\sigma_0$.}  Notice  that, since  \hbo\ slightly decreases  with IMF
slope (panel  a of Fig.~\ref{fig:fit_indices_1}), a  steeper slope for
more massive ETGs implies them to have younger ages than expected from
a  Kroupa/Chabrier IMF.   In  particular, at  highest $\sigma_0$,  for
$\Gamma=2.8$, the highest $\sigma_0$ stacks turn out to have an age of
$\sim 11$~Gyr,  i.e.  smaller than the  age of the  Universe, which is
not the  case when  assuming a Kroupa/Chabrier  IMF~\footnote{The fact
  that stellar population models predict  ages larger than that of the
  Universe is known as the age zero-point issue.  Although the problem
  might be solved  with a bottom-heavy IMF for  high-mass galaxies, it
  persists  for globular  clusters,  where no  evidence  exists for  a
  steepening of the IMF.  Hence, the zero-point issue still remains as
  a  fundamental  aspect to  deal  with  in  the modelling  of  stellar
  population (e.g.~\citealt{Vazdekis:01, Schiavon:02})}.

A thorough inspection of single panels in Fig.~\ref{fig:fit_indices_2}
reveals   that   for  IMF-   and   abundance-sensitive  indices   some
discrepancies exist, at the level  of a few $\sigma$'s, between models
and observational data, namely
\begin{description}
\item[--]  at  low  (high)  $\sigma_0$,  the model  line  strength  of
  \tioi\  is higher  (lower) than  the data;  i.e., the  slope  of the
  \tioi\--\mgfep\  relation is  shallower  for models than data  (see
  panel a of Fig.~\ref{fig:fit_indices_2});
\item[--] the model trend of \tioiio\ (panel b) is slightly offset
  downwards (by $\sim 1-1.5\,\sigma$) with respect to the data;
\item[--] at  $\sigma_0 \sim 230$\,\kms\,  the $1SSP$ fits give  too weak
  \nad\ (panel c);
\item[--] the  \naii\ (panel  d) is larger  for models than  data, for
  most  $\sigma_0$  bins;  the  largest differences  amounting  to  $\sim
  3$\,$\sigma$, found  for stacks at  $\sigma_0 \sim  195$ and
  $270$\,\kms;
\item[--] the \cat\ (panel f) tends to be weaker in models than data,
  for all stacks with $\sigma_0 > 130$\,\kms; with the largest
  discrepancies at high $\sigma_0$, but comparable to the error bars.
\end{description}
Albeit  small,  all  these  discrepancies  result  into  probabilities
$P_{\ge \chi^2}$ smaller  than a few percent for  most (13) $\sigma_0$
stacks  (see  Tab.~\ref{tab:chi_fit}).    The  discrepancies  tend  to
disappear for $2SSP$ models, as discussed in Sec.~\ref{subsec:restwo}.
Regarding Balmer lines, we remark  that $1SSP$ fits include only \hbo,
i.e. we  do not include \hgf\ (see  Sec.~\ref{sec:fitting}). When only
using  \hbo\  in the  $1SSP$  fits,  the  predicted \hgf\  values  are
significantly smaller than  the observations, as seen in  panel (b) of
Fig.~\ref{fig:fit_indices_1}.   In   other  words,  the   $1SSP$  ages
inferred from \hbo\ are older  (by $\sim 1-2$\,Gyr) than those derived
from  \hgf\  alone.  The  fact  that  different  Balmer lines  provide
discrepant   SSP-equivalent  ages   is  a   well-known   issue,  whose
explanation  might  involve  the  stronger sensitivity  of  high-order
(relative  to  \hb)   Balmer  lines  to  (i)  {\afe}~\citep{Thomas:04,
  ThomasDavies:06}, (ii) single element abundance ratios~\citep{SW:07},
and  (iii)  fraction  of  young  stars~\citep{SCR:04,  SerraTrager:07,
  br10}.  A detailed  analysis of these issues is  beyond the scope of
the   present  paper,  but   we  notice   that  using   $2SSP$  models
(Sec.~\ref{subsec:restwo})  allows  us   to  consistently  match  both
\hbo\  and \hgf\  within error  bars, indicating  that a  $1SSP$ model
might be  too simplistic to describe  the data (although  we note that
the contribution from a second,  young, component is not large either,
see below).

%%%%%%%%%%%%%%%%%%%%%%%%%%%%%%%%%%%%%%%%%%%%%%%%%%%%%%%%%%%%%%%%%%%%%%%%
\subsection{Bimodal IMFs -- $2SSP$ fits}
\label{subsec:restwo}

For two SSP models, the fits improve significantly with respect to the
$1SSP$ case,  as seen by  comparing cyan ($2SSP$)  to blue-through-red
($1SSP$)         curves        in        Figs.~\ref{fig:fit_indices_1}
and~\ref{fig:fit_indices_2},  and by  the fact  that they  give larger
$P_{\ge     \chi^2}$     values     than    $1SSP$     models     (see
Tab.~\ref{tab:chi_fit}; with  the exception of  the lowest $\sigma_0$,
see below).  In particular, $2SSP$  models fit well the trends of both
\hbo\ and  \hgf\ with \mgfep, match  \tioi\ and \nad\  indices at high
$\sigma_0$,  give  a better  description  of  \naii\,  and on  average
reproduce better  the trends  of Ca indices  with $\sigma_0$.   On the
other hand, some small discrepancies persist:
\begin{description}
\item[--] at low $\sigma_0$ ($<150$\,\kms), the model predictions of
  \tioi\ are still too high.
\item[--] a small ($\sim 1\,\sigma$) average offset remains in
  \tioiio, with models having weaker line strength than the data, as
  in the $1SSP$ case;
\item[--] the model \naii\ is still too high with respect to the
  observations, but the average difference ($-0.9\,\sigma$) is
  significantly reduced with the use of two populations.
\end{description}
While part of the discrepancy in  \tioiio\ and that of \tioi\ at low
$\sigma_0$ might  be explained by a varying  $\rm [Ti/Fe]$ abundance ratio
(see  Sec.~\ref{subsec:abundances}),  differences  between  model  and
observed  indices (e.g. \naii)  might also  have an  intrinsic origin,
reflecting   subpercent-level  uncertainties  in   stellar  population
models, and/or differences between the  true shape of the IMF and that
assumed in the present analysis  (either bimodal or unimodal). We come
back to the latter  point in Sec.~\ref{subsec:unimod}, when discussing
the  possibility of  discriminating  among different  IMF shapes.

\begin{figure*}
\begin{center}
\leavevmode
\includegraphics[width=16.2cm]{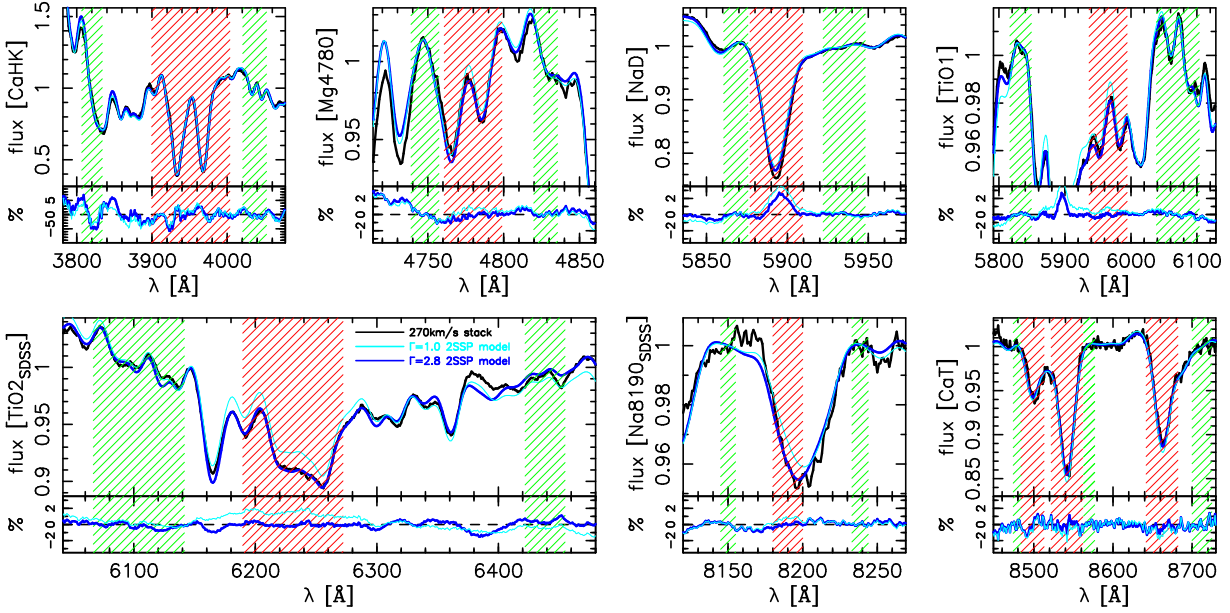}
\end{center}
\caption{Comparison of the $\sigma_0=260-280$\,\kms\ stacked spectrum
  to $2SSP$ best-fitting models with bimodal IMF $\Gamma=1$
  (Kroupa-like; cyan) and $\Gamma=2.8$ (bottom-heavy).  The six panels
  correspond to spectral regions of IMF- and abundance-sensitive
  spectral indices. Hatched green and red regions mark the sidebands
  and central features of spectral indices.  For each plot, the upper
  panel plots the observed spectrum and models, while the lower panel
  shows relative residuals (in percent) after subtracting the model
  from the data. For each spectral region, but \cat, models and data
  have been normalized by linear fitting the median fluxes in the
  sidebands. For \cat\, the normalization is obtained by fitting a
  second order polynomial to the pseudo-continua of the three Ca
  features. Notice that a bottom-heavy IMF clearly provides a better
  fit to the data for all selected spectral regions.}
%\contcaption{.}
\label{fig:obs_mod_2ssp}
\end{figure*}

For what  concerns stellar  population parameters, for  all $\sigma_0$
bins, the $2SSP$ best-fit models consist of an old ($11-14$\,Gyr), and
a younger population, the latter  contributing by less than $25\%$ (in
light) to the mixture~\footnote{We  notice that the young component is
  about  $2$\,Gyr old  only  for  the two  stacks  with $\sigma_0  \le
  120$\,\kms,  while  in  the   other  cases,  excluding  the  highest
  $\sigma_0$  bin, it  is older  than $\sim  3$\,Gyr.  At  the highest
  $\sigma_0$ ($\sim 300$\,\kms), the  young component is $1$\,Gyr old,
  but  this  contributes  negligibly,  by  less  than  $1\%$,  to  the
  mixture.}.  Both  $1SSP$  and  $2SSP$ fits  consistently  suggest  a
significant  variation of  IMF  slope with  velocity dispersion.   The
$2SSP$  best-fit  $\Gamma$  changes  from  $\sim  1$  at  $\sigma_0  <
150$\,\kms\ to $\sim 2.9$ at $\sigma_0 < 300$\,\kms\ (see top panel of
Fig.~\ref{fig:gamma_fits_all}).  The shape of the $\Gamma$--$\sigma_0$
relation  differs  significantly  between  $1SSP$ and  $2SSP$  fitting
schemes,  with  $1SSP$  models   exhibiting  a  faster  increase  with
$\sigma_0$ towards  a bottom-heavy IMF  for more massive  systems.  At
$\sigma_0 \sim 200$\,\kms, the $2SSP$ best-fit $\Gamma$ is $\sim 1.6$,
while $1SSP$  models give  $\Gamma \sim 2.6$.   On the other  hand, at
$\sigma_0>250$\,\kms,  both $2SSP$ and  $1SSP$ models  give compatible
values for  $\Gamma$ within error  bars.  At low  $\sigma_0$, spectral
indices are  less sensitive  to variations in  the IMF slope,  as seen
from the shape of the grids in Fig.~\ref{fig:fit_indices_2}, hence the
larger difference of model predictions for $\Gamma$ between $1SSP$ and
$2SSP$ fits.   The adoption  of different stellar  population mixtures
will therefore  affect significantly  the inferred $\Gamma$  values in
this regime. In contrast, at high $\Gamma$ (i.e. $\sigma_0$), spectral
indices are strongly sensitive  to IMF variations, and the constraints
on  $\Gamma$ appear  more robust  with  respect to  the other  stellar
population ingredients.   In summary, for  the purpose of  the present
work, the  main conclusion here is  that $2SSP$ models  plus a varying
IMF  reproduce  well  the  variation  of  all  spectral  indices  with
$\sigma_0$, and the quality of  the fits does not change significantly
with $\sigma_0$, as proven  by comparing $P_{\ge \chi^2}$ values among
different  $\sigma_0$  bins  (see  Tab.~\ref{tab:chi_fit}),  with  the
exception of the three lowest $\sigma_0$ bins ($\le 130$\,\kms), where
a small variation of single  element abundances might be required (see
Sec.~\ref{subsec:abundances}).

Following our  previous work (FLD13),  we model the trend  of $\Gamma$
with $\sigma_0$ with the relation:
\begin{equation}
\Gamma = A + B \log (\sigma_0[{\rm km/s}]/200),
\label{eq:gamma_sigma}
\end{equation}
where $A$ and $B$ are obtained by a least-squares fitting procedure to
the data, with $\Gamma$ as dependent variable in the fit.  The
uncertainties on $A$ and $B$ are bootstrap errors, reflecting the
uncertainties on $\Gamma$. Notice that the fit is restricted to stacks
with $\sigma_0 >140$\,\kms, where a significant variation of IMF slope
with velocity dispersion is detected. The values of $A$ and $B$, for
$2SSP$ as well as for other fitting methods, are summarized in
Tab.~\ref{tab:gamma_sigma}.  Notice that, considering the quoted
errors, the $2SSP$ best-fitting value of $B$ ($5.4 \pm 0.9$) is larger
than zero at a $6\sigma$ level, reflecting the fact that our data strongly
support an increase of IMF slope with velocity dispersion in ETGs.

\begin{table}
\centering
\small
%\begin{minipage}{150mm}
 \caption{Best-fit coefficients, $A$ and $B$, of the relation
   between IMF slope ($\Gamma$) and $\log \sigma_0$
   (Eq.~\ref{eq:gamma_sigma}) for different methods used to constrain
   the $\Gamma$. Uncertainties are quoted at the $1\sigma$ level.}
  \begin{tabular}{c|c|c}
   \hline
method &  $A$ & $B$ \\
(1) & (2) & (3) \\
   \hline
bimodal, $2SSP$       &    $2.0 \pm 0.1$ & $5.6 \pm 0.9$ \\
bimodal, $2SSP+X/Fe$  &    $1.7 \pm 0.1$ & $5.1 \pm 0.9$ \\
unimodal, $2SSP$      &    $1.4 \pm 0.1$ & $3.8 \pm 0.6$ \\
unimodal, $2SSP+X/Fe$ &    $1.3 \pm 0.1$ & $3.4 \pm 0.6$ \\
bimodal, hybrid       &    $2.4 \pm 0.1$ & $5.4 \pm 0.9$ \\
   \hline
  \end{tabular}
\label{tab:gamma_sigma}
\end{table}

Fig.~\ref{fig:obs_mod_2ssp}   shows,   in   wavelength,  rather   than
index--index, space,  the main conclusion of the  present work, namely
that high-mass ETGs  are on average better fit by an  IMF that is more
bottom heavy than the  standard Kroupa-like case.  The Figure compares
the  stacked   spectrum  at   $\sigma_0  \sim  270$\,\kms\   with  the
best-fitting  $2SSP$  model (blue)  and  a  $2SSP$ best-fitting  model
obtained  by   assuming  a  $\Gamma\sim  1$  IMF,   the  latter  being
representative    of    low-mass    ETGs    (see    top    panel    of
Fig.~\ref{fig:gamma_fits_all}).  The comparison is shown for different
spectral windows  where IMF- and  abundance-sensitive spectral indices
are defined.  Notice that a  high $\Gamma$ ($\sim 2.8$) gives the best
fit to all relevant features~\footnote{Notice that, in contrast to the
  indices, the observed spectrum in Fig.~\ref{fig:obs_mod_2ssp} is not
  corrected   to   solar-scale,  which   might   explain  some   small
  discrepancies between  models and  data in the  Figure (e.g.,  in the
  though  of the  \nad\ line).  }, i.e.   Na and  \cat\ lines,  and in
particular  the  observed spectrum  in  the  \tioiio\ spectral  range,
making this feature an extremely  useful tool to constrain the IMF (in
agreement with STK12).

\begin{table}
\centering
\small
\begin{minipage}{65mm}
 \caption{Reduced   chi-squared  statistics,  $\chi_\nu^2$,   for  all
   stacked spectra  and the three fitting models  ($1SSP$, $2SSP$, and
   $2SSP+X/Fe$) used to fit spectral indices in this work. The $P_{\ge
     \chi^2}  [\%]$  bracket value  is  the  probability  to obtain  a
   $\chi_\nu^2$ larger than the measured one.
}
  \begin{tabular}{c|c|c|c}
   \hline
 $\sigma_0$ range & \multicolumn{3}{|c}{$\chi_\nu^2$($P_{\ge \chi^2} [\%]$)} \\
      $[{\rm km/s}]$    &  $1SSP$ & $2SSP$ & $2SSP+X/Fe$ \\ 
(1) & (2) & (3) & (4) \\
   \hline
$100$--$110$ & $ 2.1(8)$ & $ 3.1(1)$ & $ 5.1(1)$ \\
$110$--$120$ & $ 2.6(3)$ & $ 2.0(7)$ & $ 2.1(12)$ \\
$120$--$130$ & $ 2.0(11)$ & $ 2.0(8)$ & $ 3.4(3)$ \\
$130$--$140$ & $ 1.4(31)$ & $ 1.2(31)$ & $ 2.7(7)$ \\
$140$--$150$ & $ 2.8(2)$ & $ 1.1(38)$ & $ 2.2(11)$ \\
$150$--$160$ & $ 3.4(0)$ & $ 1.5(19)$ & $ 3.0(5)$ \\
$160$--$170$ & $ 2.9(1)$ & $ 1.4(23)$ & $ 1.8(16)$ \\
$170$--$180$ & $ 2.5(3)$ & $ 1.6(17)$ & $ 2.2(11)$ \\
$180$--$190$ & $ 2.7(2)$ & $ 1.3(25)$ & $ 2.5(8)$ \\
$190$--$200$ & $ 3.4(0)$ & $ 1.3(27)$ & $ 4.1(2)$ \\
$200$--$210$ & $ 2.8(2)$ & $ 1.5(20)$ & $ 2.2(11)$ \\
$210$--$220$ & $ 1.8(14)$ & $ 1.6(17)$ & $ 2.6(8)$ \\
$220$--$230$ & $ 1.7(18)$ & $ 1.4(21)$ & $ 3.3(4)$ \\
$230$--$240$ & $ 2.8(2)$ & $ 1.3(28)$ & $ 2.2(12)$ \\
$240$--$250$ & $ 3.3(1)$ & $ 1.7(12)$ & $ 2.6(8)$ \\
$250$--$260$ & $ 3.3(1)$ & $ 1.5(17)$ & $ 1.9(15)$ \\
$260$--$280$ & $ 1.9(12)$ & $ 0.5(78)$ & $ 1.6(21)$ \\
$280$--$320$ & $ 3.8(0)$ & $ 1.9(9)$ & $ 2.2(12)$ \\
   \hline
  \end{tabular}
\label{tab:chi_fit}
\end{minipage}
\end{table}

\begin{figure}
\begin{center}
\leavevmode
\includegraphics[width=8cm]{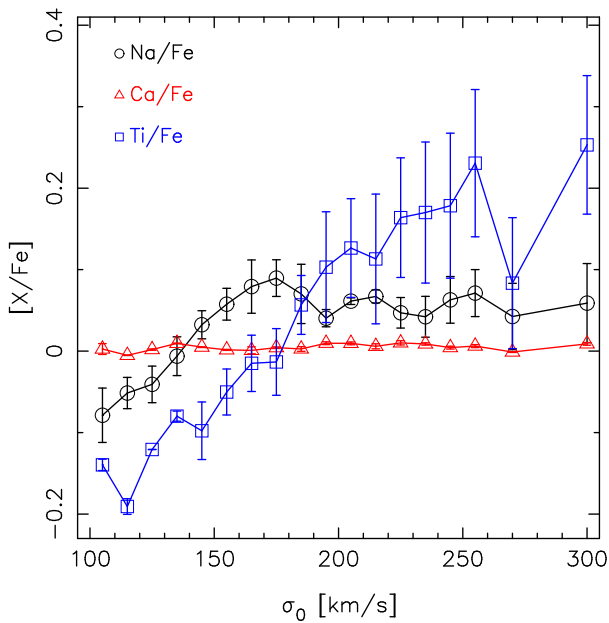}
\end{center}
\caption{Best-fit ``residual'' calcium, sodium, and titanium
  abundances ($\rm [X/Fe]$ with $\rm X=Ca$, $\rm Na$, and $\rm Ti$,
  respectively), as a function of central velocity dispersion
  $(\sigma_0)$. Residual abundances refer to the fact that we have
  already corrected all spectral indices to \afe$=0$ for each
  $\sigma_0$ bin.  The $\rm [X/Fe]$'s are obtained from the fitting of
  spectral indices with $2SSP$ extended MILES (MIUSCAT) synthetic
  populations, modelling the abundance sensitivity of different
  indices with the CvD12 models.  Since the indices are corrected to
  \afep$=0$, the $\rm [X/Fe]$'s represent residual abundance patterns, not
  accounted for by the \afep\ correction procedure.  Error bars are quoted
  at the 1\,$\sigma$ level.}
%\contcaption{.}
\label{fig:xfe_fits}
\end{figure}

%%%%%%%%%%%%%%%%%%%%%%%%%%%%%%%%%%%%%%%%%%%%%%%%%%%%%%%%%%%%%%%%%%%%%%%%
\subsection{The role of individual abundance patterns}
\label{subsec:abundances}

Comparing the  cyan and green  curves in Fig.~\ref{fig:fit_indices_2},
one  can see  that the  overall quality  of the  fits to  our targeted
spectral indices  does not improve significantly when  allowing for an
individual variation of  Ca, Na, and Ti abundance  patterns.  In fact,
for most $\sigma_0$ bins,  the probability $P_{\ge \chi^2}$ values are
{\sl    lower}    for   $2SSP+X/Fe$    than    $2SSP$   models    (see
Tab.~\ref{tab:chi_fit}). On  the other  hand, looking at  TiO features
alone, one can notice an  improvement with respect to the $2SSP$ case,
in  that  $2SSP+X/Fe$  models   match  better  \tioi\  at  the  lowest
$\sigma_0$ ($<150$\,\kms), as well  as \tioiio\ at high $\sigma_0$ (in
particular around  $250$\,\kms).  The  improved matching of  \tioi\ is
due  to  the  fact  that,  according  to  CvD12  models~\footnote{This
  different  sensitivity of \tioi\  and \tioiio\  to $\rm  [Ti/Fe]$ is
  qualitatively   consistent  with  the   predictions  of   the  TMJ11
  models.  }, \tioi\  is  slightly more  sensitive  to $\rm [Ti/Fe]$  than
\tioiio.   Hence,  by  lowering  $\rm [Ti/Fe]$, one  decreases  the  model
\tioi\ more than \tioiio, allowing both observed indices to be matched
simultaneously. This point  is illustrated in Fig.~\ref{fig:xfe_fits},
where the best-fit abundance ratios, $\rm [Ca/Fe]$, $\rm [Na/Fe]$, and
$\rm [Ti/Fe]$  are shown  as a function  of $\sigma_0$.  We  point out
that,  since spectral indices  have been  corrected to  \afep$=0$ (see
Sec.~\ref{sec:afe_corr}), the  $\rm [Ca/Fe]$, $\rm  [Na/Fe]$, and $\rm
[Ti/Fe]$  estimates  in Fig.~\ref{fig:xfe_fits}  do  not reflect  true
abundance ratios  in different $\sigma_0$  bins, but should  rather be
interpreted as  {\it residual} abundances,  i.e. not accounted  for by
the correlation of  single element abundances with \afe,  and the fact
that \afe\ increases with $\sigma_0$.   Another caveat is that we rely
on  a given  set of  {\it theoretical}  models (CvD12)  to  infer $\rm
[Ca/Fe]$, $\rm  [Na/Fe]$, and $\rm [Ti/Fe]$, although  we also explore
the   effect   of  using   our   own   simple   synthetic  SSPs   (see
App.~\ref{sec:synth_ssps})  partly based  on different  ingredients to
the CvD12 models  (see below).  For these reasons,  our conclusions on
individual  abundance ratios  should  be considered  on a  qualitative
basis.

\begin{description}
\item[{\it Ti abundances} --] The $\rm [Ti/Fe]$ is found to increase
  with $\sigma_0$, changing from negative ($\sim -0.2$~dex) at the
  lowest $\sigma_0$ to positive for $\sigma_0 > 180$\,\kms.  Notice
  that at $\sigma_0<150$\,\kms, the error bars for $\rm [Ti/Fe]$ are
  very small (less than a few percent), reflecting the fact that a
  negative value of $\rm [Ti/Fe]$ is required to reduce the
  discrepancy between model and observed \tioi\ strength for the
  lowest velocity dispersion stacks.  At high $\sigma_0$, the $\rm
  [Ti/Fe]$ becomes positive, as \tioiio\ increases, producing a better
  match to the observed \tioiio.  Notice, in fact, that $\rm [Ti/Fe]$
  peaks at $\sigma_0 \sim 250$\,\kms\ (Fig.~\ref{fig:xfe_fits}), where
  the $2SSP$ models differ the most from the data (see panel $b$ of
  Fig.~\ref{fig:fit_indices_2}). The error bars on $\rm [Ti/Fe]$ are
  large at $\sigma_0\simgt 200$\,\kms, meaning that a deviation
  of Ti abundance from solar scale is only marginally significant for
  massive systems.
\item[{\it Na abundances} --] The $\rm [Na/Fe]$ also increases with
  $\sigma_0$, varying by about $0.2$\,dex from $\sigma \sim 100$ to
  $300$\,\kms.  This amount of variation is far smaller than that
  recently reported by CvD12b, who found that $\rm [Na/Fe]$ can be as
  high as $\sim$1\,dex for ETGs in the SAURON sample.  Notice that
  this is not necessarily in disagreement with our findings.  In fact,
  the \nad\ feature is not covered by the observed spectra of CvD12b,
  with the $\rm [Na/Fe]$ being inferred indirectly by the authors because
  of its effect on the free electron pressure in stellar atmospheres.
  Also, as noticed above, our best-fit abundance ratios should be
  considered as residual (with respect to \afe), rather than absolute
  estimates.  The effect of varying $\rm [Na/Fe]$ on Na indices is
  illustrated in Fig.~\ref{fig:na_ca} (top panel), where \nad\ is
  plotted against \naii.  We also show the expected effect of varying
  $\rm [Na/Fe]$ by $+0.3$\,dex, when using (i) CvD12 models with a
  Chabrier IMF, solar metallicity, and an age of $13.5$\,Gyr (blue
  arrow); (ii) our synthetic SSP, with a Kroupa IMF, solar
  metallicity, and age of $12$\,Gyr (red arrow); (iii) the same SSP
  but for a Salpeter IMF (orange arrow).  We refer the reader to
  App.~\ref{sec:synth_ssps} for details on how our synthetic SSPs are
  created.  Notice the remarkable agreement between CvD12 and our-SSP
  predictions for a Chabrier/Kroupa IMF.  Increasing $\rm [Na/Fe]$
  makes both \nad\ and \naii\ to increase along a direction which is
  either steeper (Chabrier/Kroupa) or similar (Salpeter SSP) to the
  observed \nad--\naii\ trend.  This indicates that disentangling the
  effect of IMF and $\rm [Na/Fe]$ is in general very difficult, and
  one actually needs a multi-index approach, based on indices from
  different species, as in the present work.  We emphasize that a
  significant variation of $\rm [Na/Fe]$ with $\sigma_0$ alone (after
  correcting for the \afep\ trends) appears to be ruled out by our
  data, as the $2SSP$ fits reproduce well the \nad--\naii\ trend, as
  well as the other index-$\sigma_0$ trends.
\item[{\it Ca abundances} --] The best-fit $\rm [Ca/Fe]$ is very close
  to zero  for all stacks, with  small error bars  throughout the full
  $\sigma_0$ range.   The reason of  this behaviour is  illustrated in
  Fig.~\ref{fig:na_ca} (bottom  panel), where  \cahk\ is plotted  as a
  function  of \cat.   The  effect  of decreasing  Ca  abundance by  a
  ``small'' amount  ($-0.05$\,dex) is shown  by the blue  (CvD12), red
  (our-SSP, Kroupa  IMF), and  orange (our-SSP, Salpeter  IMF) arrows.
  Notice that in contrast to  $\rm [Na/Fe]$, Chabrier CvD12 models and
  our  simple  synthetic  SSP with  Kroupa  IMF  do  agree only  in  a
  qualitative way,  in that the $[Ca/Fe]$ variation  produces a strong
  change in  \cahk\, and only a  minor change to  \cat\ (especially in
  the  case  of  a   Salpeter  model).   However,  regardless  of  the
  prescription,  the arrows  show that  even  a small  change in  $\rm
  [Ca/Fe]$ would bring the model  indices away from the locus occupied
  by the  observations. Therefore, since the  extended MILES (MIUSCAT)
  base models match well the  \cahk\ and \cat\ grid, with \cahk\ being
  significantly sensitive  to $[Ca/Fe]$, there is not  much space left
  for a {\it residual} deviation of Ca abundance from solar scale.
%This  result does not exclude,  by itself, that  Ca might be
%  underabundant in  more massive galaxies, as  recently claimed, e.g.,
%  by~.
% 
%is underabundant in
%  the original stacks...
\end{description}

We point  out that the present  analysis considers only  the effect of
$\rm [Ca/Fe]$,  $\rm [Na/Fe]$, and $\rm [Ti/Fe]$  abundance ratios, as
these  are  expected to  dominate  the  contribution  to our  targeted
IMF-sensitive  spectral indices  (see  Sec.~\ref{sec:fitting}).  While
other  elements  may contribute  to  the line  strengths~\footnote{For
  instance, \citet{Worthey:98} raised the issue that TiO indices might
  also be sensitive to elements lighter  than Ti, like Sc and V.}, our
main  conclusion  here is  that,  after  removing  the effect  of  the
\afe\ increase  with velocity  dispersion, a variety  of IMF-sensitive
spectral indices from  different species (Ca, Na, Ti,  and Mg) can all
be  recovered  simultaneously with  (nearly)  solar-scaled models,  by
invoking a  variation of the  IMF to become  increasingly bottom-heavy
with  velocity  dispersion.   The  best-fitting  coefficients  of  the
$\Gamma$--$\sigma_0$    relation    (Eq.~\ref{eq:gamma_sigma}),    for
$2SSP+X/Fe$    models   with    bimodal   IMF,    are    reported   in
Tab.~\ref{eq:gamma_sigma}.   Notice  the  fair  agreement  of  fitting
coefficients between $2SSP$ and $2SSP+X/Fe$'s models.

\subsubsection{The case of $Ca4227$}
\label{subsubsec:caf}

\begin{figure}
\begin{center}
\leavevmode
\includegraphics[width=7.cm]{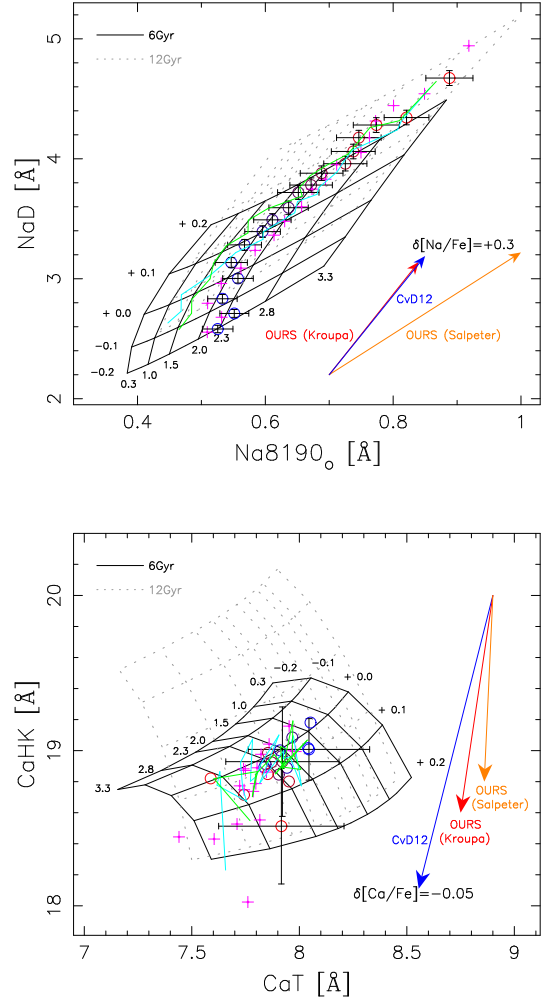}
\end{center}
\caption{Plots of abundance- vs. IMF-sensitive spectral indices. The
  top panel shows Na-sensitive indices, i.e. \nad\ vs. \naii, while
  the bottom panel shows Ca-sensitive indices, \cahk\ vs.  \cat. Only
  results from $2SSP$ and $2SSP+X/Fe$ fits are shown, as cyan and
  green curves, respectively.  Symbols are the same as in
  Figs.~\ref{fig:fit_indices_1} and~\ref{fig:fit_indices_2}.  The blue
  arrows show the effect of changing $\rm [Na/Fe]$ by $+0.3$\,dex
  (top) and $\rm [Ca/Fe]$ by $-0.05$\,dex according to CvD12 models.
  Error bars are shown only for stacks with $\sigma_0 \sim 100$,
  $200$, and $300$\,\kms, respectively.
%Notice that the  \nad\
%with \naii\  observed line strengths (blue-to-red circles) increase with $\sigma$, because of the 
}
%\contcaption{.}
\label{fig:na_ca}
\end{figure}

\begin{figure}
\begin{center}
\leavevmode
\includegraphics[width=7cm]{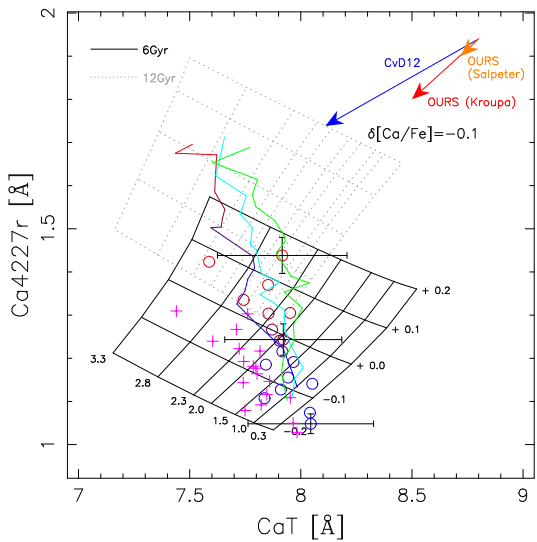}
\end{center}
\caption{Same  as  the lower  panel  of  Fig.~\ref{fig:na_ca} for  the
  \cafr\ index as a function of \cat.  The arrows show the effect of a
  change in $\rm [Ca/Fe]$ by $-0.1$\,dex, rather than $-0.05$\,dex, as
  in Fig.~\ref{fig:na_ca}, for displaying purposes.
%Notice that the  \nad\
%with \naii\  observed line strengths (blue-to-red circles) increase with $\sigma$, because of the 
}
%\contcaption{.}
\label{fig:caf}
\end{figure}

An  important  caveat related  to  Ca-sensitive  indices  is that,  as
anticipated  in  Sec.~\ref{sec:indices},  we  have not  been  able  to
include the \caf\ index in  the present analysis.  The reason for that
is  illustrated  in Fig.~\ref{fig:caf},  where  we  plot  \cafr\ as  a
function  of \cat.  The  \cafr\ is  a modified  version of  \caf, that
avoids the contamination of the blue sideband of \caf\ from the CN4216
molecular  band~\citep{PRS:05}.   Both  \cafr\  and  \cat\  have  been
corrected for  the index--\afep\  correlation at fixed  $\sigma_0$, as
discussed  in  Sec.~\ref{sec:afe_corr}.   The  \cafr\  decreases  with
\afep\, as seen by the fact that uncorrected data-points in the Figure
(magenta relative to blue-through-red symbols) have weaker \cafr.  All
models  ($1SSP$, $2SSP$, and  $2SSP+X/Fe$) give  \cafr\ EWs  too large
with  respect to  the  data,  the discrepancy  being  larger for  more
massive systems.  Including \cafr\  in the fitting procedure would not
solve the discrepancy, implying no significant change to the IMF trend
with  $\sigma_0$ (top  panel  of Fig.~\ref{fig:gamma_fits_all}).   The
\caf\ discrepancy is a well-known issue of stellar population studies,
the  origin  of  which   has  been  extensively  debated  (see,  e.g.,
\citealt{Vazdekis:1997,  TMB:03b,  Yamada:06}).   As recently  claimed
by~\citet{JTM12}, \caf\ EWs would be  explained by the fact that Ca is
underabundant in ETGs  with respect to Mg, following  more closely the
Fe abundance.  Using \cahk, \citet{worthey:11} concluded, indeed, that
both  $\rm [Ca/Fe]$  and  $\rm [Ca/Mg]$  systematically decrease  with
increasing elliptical galaxy mass.   While our \afe\ corrections to Ca
indices seem to  support this claim (as Ca  indices decrease, at fixed
$\sigma_0$,  with  \afep),  Fig.~\ref{fig:na_ca}  (bottom  panel)  and
Fig.~\ref{fig:caf} indicate  that the discrepancy  remains unresolved.
In  fact, changing  $[Ca/Fe]$ abundance  would shift  the models  in a
direction   orthogonal  to   the  sequence   of  data-points   in  the
\cafr\--\cat\  diagram  (see blue  arrow  in Fig.~\ref{fig:caf}),  and
would also not fit  the \cahk\--\cat\ diagram. Furthermore, our simple
SSPs  indicate that  for an  IMF more  bottom-heavy than  the standard
Kroupa/Chabrier (i.e. a Salpeter IMF;  see orange arrow in the Figure)
the effect of $\rm [Ca/Fe]$  abundance on \cafr\ becomes really minor,
further hampering  the explanation  of the discrepancy  between models
and data  for this  feature. We notice  that, while  understanding the
origin of  the \cafr\ discrepancy  goes certainly beyond the  scope of
the present study, to our knowledge,  this is the first time that such
a discrepancy is shown for spectral data with exceptionally high $S/N$
ratio, allowed by the stacking of a large sample of SDSS spectra.

%{ discuss plots with Na and Ca IMF- vs. abundance sensitive indices }

%{ discuss the puzzle of Ca4227}

%%%%%%%%%%%%%%%%%%%%%%%%%%%%%%%%%%%%%%%%%%%%%%%%%%%%%%%%%%%%%%%%%%%%%%%%
\subsection{Unimodal vs. bimodal IMFs}
\label{subsec:unimod}
\begin{figure}
\begin{center}
\leavevmode
\includegraphics[width=7cm]{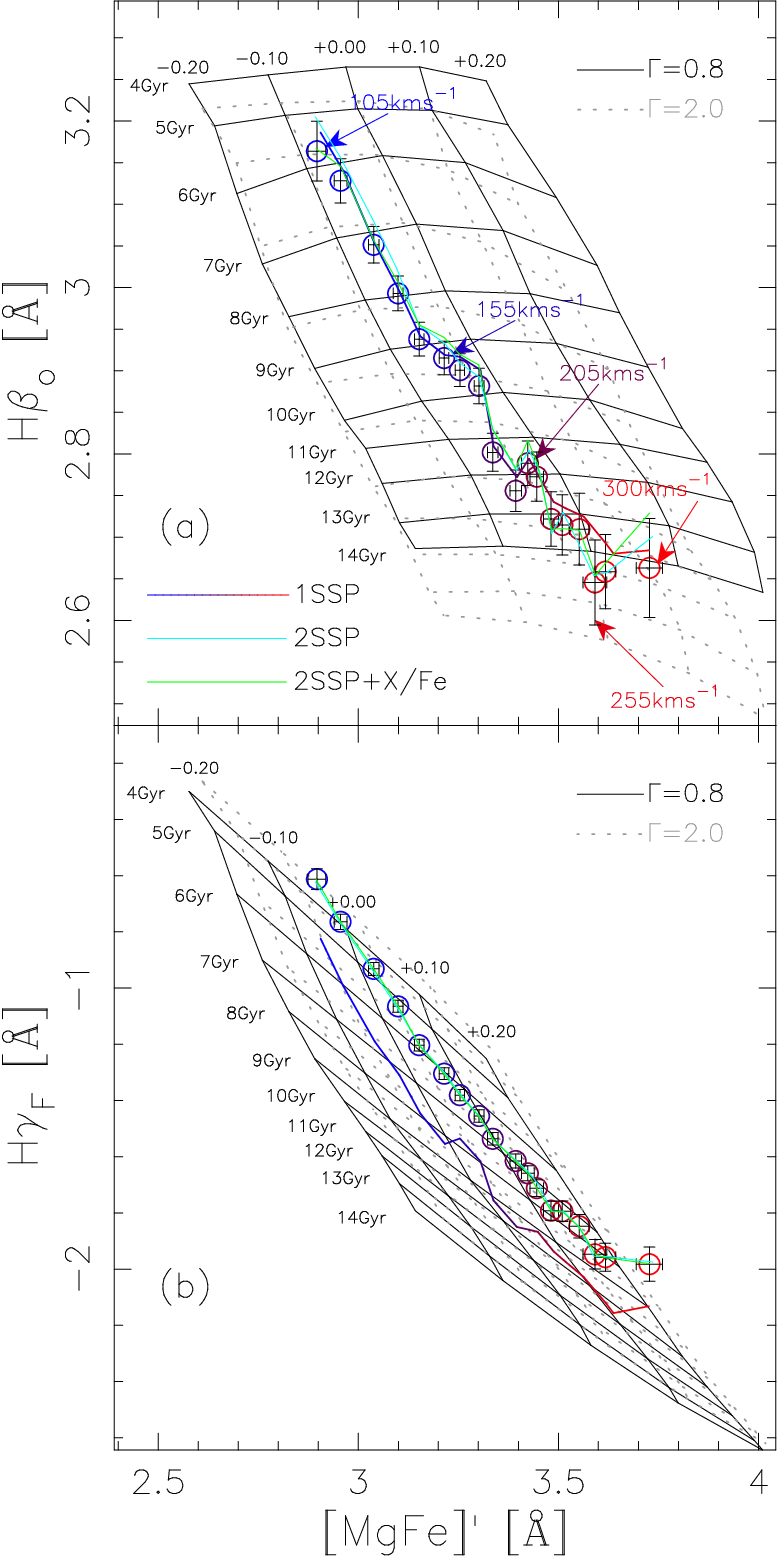}
\end{center}
\caption{Same as Fig.~\ref{fig:fit_indices_1}, for unimodal,
  rather than bimodal, IMF models. Notice that only models with
  unimodal $\Gamma \le 2.3$ are considered, as explained in
  Sec.~\ref{sec:models}.  }
%\contcaption{.}
\label{fig:fit_indices_un_1}
\end{figure}

\begin{figure*}
\begin{center}
\leavevmode
\includegraphics[width=18cm]{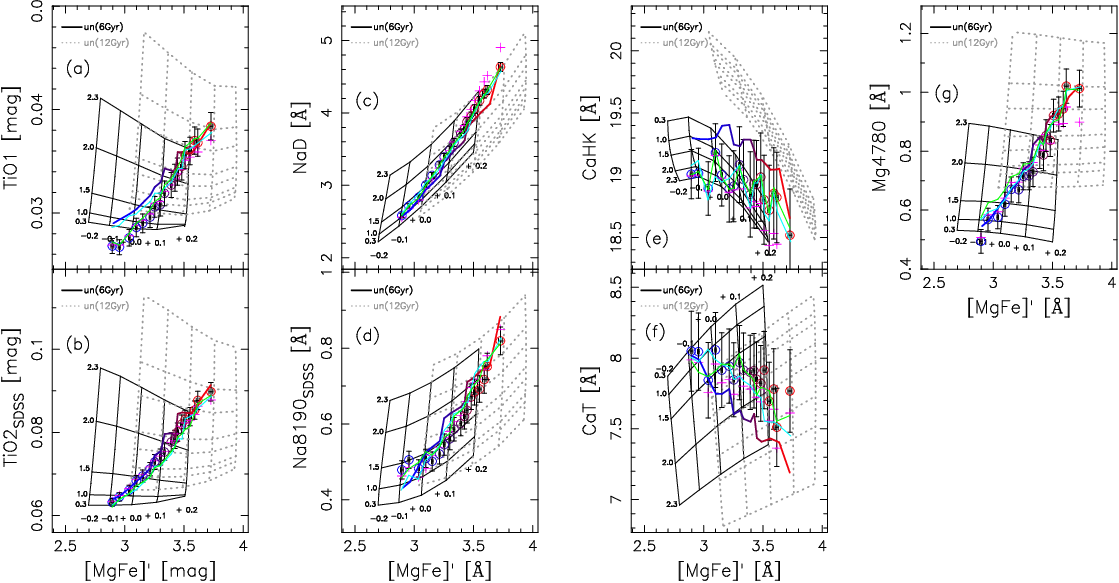}
\end{center}
\caption{Same as Fig.~\ref{fig:fit_indices_2}, but for unimodal,
  rather than bimodal, IMF models. Notice that only models with
  unimodal $\Gamma \le 2.3$ are considered, as explained in
  Sec.~\ref{sec:models}.  }
%\contcaption{.}
\label{fig:fit_indices_un_2}
\end{figure*}

Figs.~\ref{fig:fit_indices_un_1}  and~\ref{fig:fit_indices_un_2}  plot
the  results of fitting  spectral indices  with unimodal  (i.e. single
power law) IMF models.  The corresponding IMF slope--$\sigma_0$ trends
are  shown in Fig.~\ref{fig:gamma_fits_all}  (middle panel),  with the
best-fitting   coefficients  of   the   $\Gamma$--$\sigma_0$  relation
(Eq.~\ref{eq:gamma_sigma})  given  in Tab.~\ref{eq:gamma_sigma}.
As explained  in Sec.~\ref{sec:models}, we  have limited the  range of
unimodal slopes  to $\Gamma \le  2.3$. Indeed, the trends  of $\Gamma$
with $\sigma_0$ show  that higher values of $\Gamma$  are not required
to describe the data with single-slope IMFs.

Unimodal models  provide a  good fit to  spectral indices,  similar to
bimodal  IMFs. In  particular,  $2SSP$ fits  match  well all  selected
spectral indices.  At $\sigma_0\simlt  150$\,\kms, the models give too
high  values of  \tioi\ with  respect  to the  observations, both  for
$1SSP$ and  $2SSP$ models,  as in  the case with  a bimodal  IMF.  The
discrepancy  disappears when  including the  effect of  single element
abundance ratios, with $\rm [Ti/Fe]$ being more negative in lower mass
galaxies.  Indeed,  the individual trends  we find for  $\rm [Ca/Fe]$,
$\rm [Na/Fe]$, and  $\rm [Ti/Fe]$, vs. $\sigma_0$ are  very similar to
the bimodal case, with small deviations from solar scale.

\begin{figure}
\begin{center}
\leavevmode
\includegraphics[width=8cm]{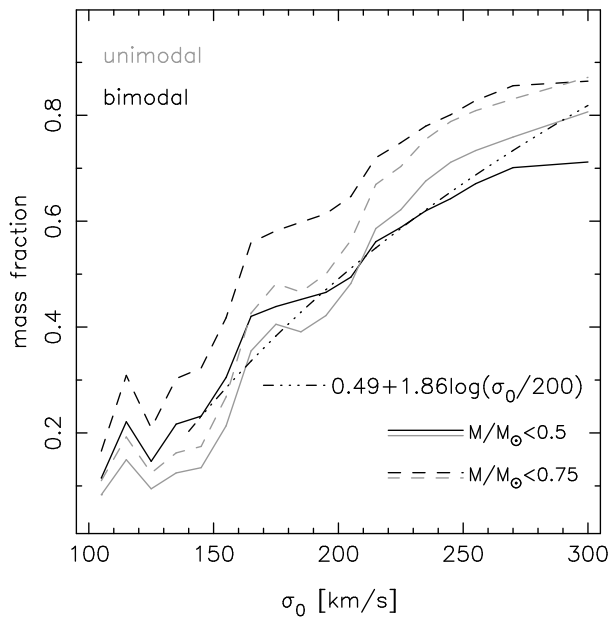}
\end{center}
\caption{Stellar  mass  fractions expected  from  unimodal (grey)  and
  bimodal   (black)   IMF   models,   as  a   function   of   velocity
  dispersion. Solid  and dashed lines  refer to the mass  fractions of
  stars   with  mass   smaller  than   $0.5$   and  $0.75$\,M$_\odot$,
  respectively.   The dot--dashed curve  shows the  best-fit relation,
  $0.49  + 1.83 \log(\sigma_0/200)$,  to the  mass fraction  trends of
  both   unimodal  and   bimodal   models,  for   mass  smaller   than
  $0.5$\,M$_\odot$, in  the range $\sigma_0>140$~\kms  (see the text).
}
%\contcaption{.}
\label{fig:mass_frac}
\end{figure}

Unimodal models  also feature  a clear tendency  for the IMF  slope to
increase   with  velocity  dispersion,   the  best-fitting   slope  in
Eq.~\ref{eq:gamma_sigma}  differing from zero  at more  than $6\sigma$
(see Tab.~\ref{tab:gamma_sigma}).  However, the amount of variation is
different from  the bimodal case:  $\Gamma$ ranges from $\sim  0.7$ to
$2$ for a  single-slope IMF, whereas best-fit bimodal  IMF slopes vary
from $\sim  1$ to  $3$. In fact,  for $2SSP$ models,  the best-fitting
slope of the $\Gamma$--$\sigma_0$ relation changes from $\sim 5.6$ for
bimodal to  $\sim 3.8$ for  unimodal IMF. This difference  arises from
the fact that -- at a  given $\Gamma$ -- unimodal and bimodal IMFs are
characterized by a different fraction  of high- to low-mass stars, and
thus the  $\Gamma$ has a  different physical meaning for  unimodal and
bimodal  models.   Fig.~\ref{fig:mass_frac}  illustrates  this  point,
where,  instead of $\Gamma$  we use  the mass  fraction in  stars with
masses  below some threshold,  as a  function of  velocity dispersion.
This  mass fraction  is given  at ``time  zero'', by  using  the given
functional form of  the IMF over the stellar mass  range from $0.1$ to
$100$\,M$_\odot$   (as   in~\citealt{Vazdekis:1996}).    At  a   given
$\sigma_0$,  we use  the $\Gamma$  value corresponding  to  the $2SSP$
best-fit  models  to  determine  the  mass fraction.   Two  cases  are
considered,    corresponding   to   a    threshold   of    $0.5$   and
$0.75$\,M$_\odot$, respectively.  The mass fractions agree pretty well
between  unimodal  and bimodal  IMFs.  In  particular, for  M/M$_\odot
<0.5$,  the  agreement is  better  than  $10  \%$, proving  that  mass
fractions  are  robustly constrained  regardless  of  the adopted  IMF
shape.  For M/M$_\odot <0.5$, the  trends in the initial mass fraction
for both unimodal and bimodal models can be described by the following
relation:
\begin{equation}
\begin{array}{ll}
{\rm Fraction}(<0.5M_\odot) & \equiv 
\frac{\int_{0.1M_\odot}^{0.5M_\odot} M\Phi(M)dM}
{\int_{0.1M_\odot}^{100M_\odot} M\Phi(M)dM} =\\
 & \\
 & = 0.49 + 1.86 \log(\sigma_0[{\rm km/s}]/200), 
\end{array}
\label{eq:mass_frac}
\end{equation}
where $\Phi(M)$ is  the IMF. The coefficients are  obtained by fitting
simultaneously both trends, for stacks with $\sigma_0>140$\,\kms (i.e.
the range where a significant variation of $\Gamma$ with $\sigma_0$ is
detected,  see Fig.~\ref{fig:gamma_fits_all}).  Eq.~\ref{eq:mass_frac}
is plotted  in Fig.~\ref{fig:mass_frac}  as a dot-dashed  curve.  From
Fig.~\ref{fig:mass_frac}  we conclude  that the  initial  stellar mass
fraction   in   low-mass   stars   (M/M$_\odot   <   0.5$)   increases
systematically   in  ETGs,   from  $\sim   20\%$  at   $\sigma_0  \sim
100$\,\kms\ to $\sim 80 \%$ at $\sigma_0 \sim 300$\,\kms.

%\begin{figure}
%\begin{center}
%\leavevmode
%\includegraphics[width=7cm]{FIGURES/GAMMA_ALL_noTiOCaH_Na8190_un.ps}
%\end{center}
%\caption{Same as Fig.~\ref{fig:gamma_fits} but for unimodal, rather
%  than bimodal, IMFs. Notice that both bimodal and unimodal models
%  lead to the similar conclusion of a systematic increase of IMF slope
%  with central velocity dispersion ($\sigma_0$).  The horizontal
%  dashed line marks the value of $\Gamma=1.3$, corresponding, for a
%  unimodal IMF, to the Salpeter slope.
% Best-fitting  slope  of  bimodal  IMFs  as  a  function  of
%  $\sigma_0$.  Different symbols and  colours correspond to the results
%  of fitting  different stellar population  models, as labelled  in the
%  top--left corner of the plot. Colours  are the same coding as in each
%  panel of  Fig.~\ref{fig:fit_indices}. Error bars are  plots only for
%  $2SSP$  models, to  make  the  plot more  clear,  and correspond  to
%  1~$\sigma$ uncertainties on $\Gamma$ values. 
%}
%\contcaption{.}
%\label{fig:gamma_fits_un}
%\end{figure}

%%%%%%%%%%%%%%%%%%%%%%%%%%%%%%%%%%%%%%%%%%%%%%%%%%%%%%%%%%%%%%%%%%%%%%%%
\subsection{Hybrid approach}
\label{subsec:hybrid}

So far, the analysis is based on equivalent widths alone, to constrain
{\sl both}  the IMF slope,  $\Gamma$, and the distribution  of stellar
ages  and   metallicities.   As  an  additional  test,   we  define  a
complementary  probability  distribution  function based  on  spectral
fitting,   and  we   combine  this   information  with   our  targeted
IMF-sensitive line strengths, following  a method analogous to the one
described in FLD13. Furthermore, we consider here the effect of a more
complex distribution of stellar ages and metallicities, by adopting
a  standard  exponentially  decaying  star  formation  rate  (i.e.   a
$\tau$-model),  where a star  formation history  is described  by four
parameters: formation epoch,  exponential timescale, metallicity (kept
fixed for  each individual  model), and IMF  slope, where  the bimodal
function  is  chosen.    Table~\ref{tab:hybrid}  shows  the  range  in
parameters. For  each stack, we  explore a grid of  $32\times 32\times
10\times 6$ models, defining two independent PDFs from the constraints
on either spectral fitting or  line strengths (we use in this analysis
\cahk,\,\mgf,\,\nad,\,\tioi,\,\tioiio,\,\naii,\, and \cat).  We apply the
same corrections for \afe\ as in the rest of the paper.  The final PDF
is given by the product of the individual PDFs, enabling us to combine
spectral fitting,  to constrain the age  and metallicity distribution,
and individual line strengths, to target the IMF slope.

\begin{table}
  \caption{Parameters used in the hybrid method ($\tau$-model)}
  \begin{tabular}{l|rcl}
   \hline
   Parameter & Steps & Range & Description\\
   \hline
   $t_{\rm FOR}$ & $32$ & $0.1\cdots tU^\star$ & Formation Epoch\\
   $\tau$\,(Gyr) & $32$ & $-1\cdots +0.7$ & Exponential timescale\\
   $[Z/H]$   & $6$  & $-1.7\cdots +0.22$ & Metallicity\\
   $\Gamma$     & $10$ & $0.3\cdots 3.3$ & IMF slope (bimodal)\\
   \hline
  \end{tabular}
\label{tab:hybrid}
$\star$ $t_U$ is the current ($z=0$) age of the Universe,
i.e. 13.5\,Gyr for a vanilla flavoured ($\Omega=0.3$;
$H_0=70$\,km\,s$^{-1}$\,Mpc$^{-1}$) $\Lambda$CDM cosmology.
\end{table}

The      spectral       fit      is      performed       over      the
$3900-5400$\,\AA\         range~\footnote{In        contrast        to
  Sec.~\ref{subsec:aFe_stacks},  where  we  adopt  an upper  limit  of
  $7350$~\AA, the value of $5400$~\AA\ is chosen here to avoid several
  features, like TiO molecular bands and \nad, which are already taken
  into account  as line  strength constraints.}, after  convolving the
synthetic spectra,  from the  original $2.51$\,\AA\ resolution  of the
extended  MILES   (MIUSCAT)  models,  to  a   velocity  dispersion  of
$300$\kms\ (i.e. the maximum value of $\sigma_0$ for stacked spectra),
plus the SDSS spectral resolution.   We avoid extending the fit over a
wider range of wavelengths to  minimise the effect of flux calibration
systematics.   Our results do  not change  significantly if  a smaller
spectral window is used, but it is important to include in the fitting
range  the  region  around  the  4000\,\AA\  break  for  an  effective
constraint    on    the     stellar    ages. The bottom    panel    of
Fig.~\ref{fig:gamma_fits_all}  shows the  best fit  values of  the IMF
bimodal   slope  with  respect   to  velocity   dispersion,  including
$1\,\sigma$ error bars.  In addition to the general model (black solid
dots), we include, for reference,  the constraints when only using the
SED fit (red crosses),  illustrating the expected complete degeneracy.
The magenta open squares give the results when only using the targeted
IMF-sensitive line  strengths (i.e.  \mgf,\,\tioi,\,\tioiio,\,\naii,\,
and  \cat ).   Finally, the  full hybrid  analysis is  also  shown for
models comprising  only simple stellar  populations (i.e a  single age
and  metallicity), as  open grey  dots. The  figure confirms  that our
previous results are robust with respect to more complex distributions
of   stellar   ages  and   metallicities.    In  particular,   fitting
Eq.~\ref{eq:gamma_sigma} to the  $\Gamma$--$\sigma_0$ trend, when only
using the targeted IMF-sensitive  line strengths (magenta dashed curve
in  the  bottom   panel  of  Fig.~\ref{fig:gamma_fits_all}),  we  find
consistent results to the $2SSP$ fits of line strengths alone (see the
values         of         best-fitting         coefficients         in
Tab.~\ref{tab:gamma_sigma}). Also, the  results of the hybrid approach
can be compared  to those of our previous work  (FLD13), where we used
the same hybrid approach to constrain the $\Gamma$, but with a smaller
set of  spectral indices, no  correction for abundance ratio  at fixed
$\sigma_0$,  and we  did not  optimize the  definition of  TiO  and Na
indices.   In  FLD13,  we  reported a  $\Gamma$--$\sigma_0$  slope  of
$B=7.2$,   consistent  with  that   of  $5.4   \pm  0.9$   found  here
(Tab.~\ref{tab:gamma_sigma}).  On  the other  hand, the offset  of the
$\Gamma$--$\sigma$  relation reported  by  FLD13 ($B  \sim 1.85$),  is
shallower  than that of  $2.4 \pm  0.1$ derived  here from  the hybrid
approach, but  fully consistent  with that we  derive from  the $2SSP$
analysis ($B=2.0 \pm 0.1$, see Tab.~\ref{tab:gamma_sigma}).

\begin{figure*}
\begin{center}
\leavevmode
\includegraphics[width=15cm]{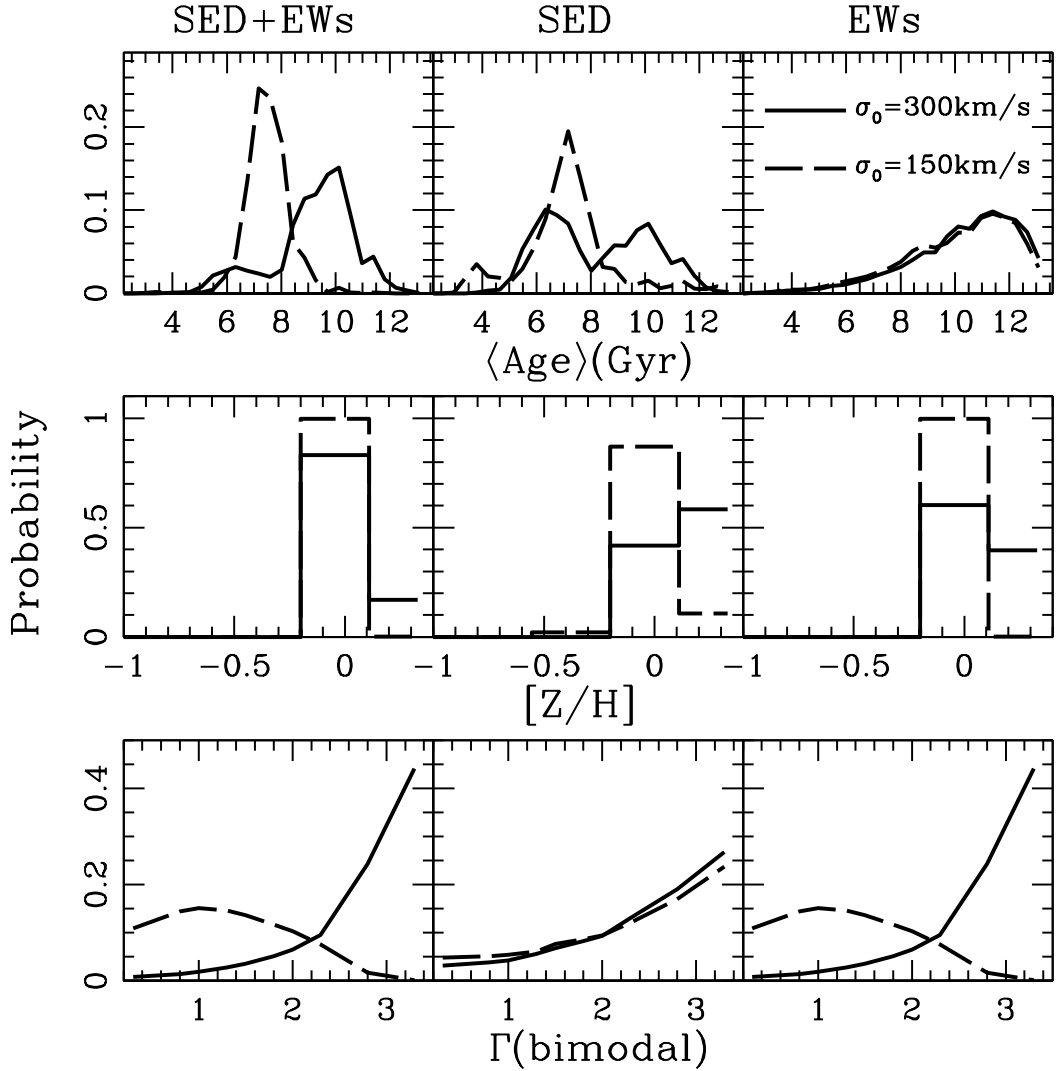}
\end{center}
\caption{Probability distribution functions (PDFs) with respect to
  average age (top), metallicity (middle) and IMF slope (bottom) for
  two stacks, with velocity dispersion $\sigma_0=150$\,\kms\ (dashed
  lines) and $300$\,\kms\ (solid lines). The hybrid method is
  considered (see \S \ref{subsec:hybrid}) for a grid of $\tau$ models
  (see text for details). In each case, the joint PDF is shown on the
  leftmost panels (SED+EWs), with the individual PDFs split between
  spectral fitting (SED, middle) and line strengths (EWs, right).  }
%\contcaption{.}
\label{fig:Hybrid_2}
\end{figure*}

Fig.~\ref{fig:Hybrid_2}   shows    jointly   and   independently   the
constraining power of spectral fitting (SED) and line strength fitting
(EWs).   From   top  to  bottom,  the  panels   give  the  probability
distribution function  with respect  to stellar age,  metallicity, and
IMF  slope, for  two  different stacks,  corresponding  to a  velocity
dispersion     of    $\sigma_0=150$\,\kms\    (dashed     line)    and
$300$\,\kms\  (solid line). The  figure shows  that an  analysis based
exclusively  on   line  strengths  cannot   discriminate  between  the
different  age  distributions  of  galaxies  with  different  velocity
dispersion. Notice in the hybrid  method we only use the IMF-sensitive
indices,  since   the  spectral  fitting  substitutes   the  age-  and
metallicity-sensitive indices: \hbo\ ,\hgf\ and \mgfep\ . In contrast,
differences  in IMF slope  cannot be  distinguished between  these two
stacks when using spectral fitting alone. It is the combination of the
two that enables us to obtain strong constraints on $\Gamma$, that are
complementary  to  our  previous  analysis,  only  based  on  spectral
lines. The  hybrid method confirms the strong  correlation between IMF
slope  and velocity dispersion,  even when  a complex  distribution of
ages and metallicities  is allowed for. The middle  panels reveal that
only  metallicities  $[Z/H]\geq  -0.1$\,dex  contribute  to  the
analysis,    in   agreement   with    our   previous    results   (see
e.g. Fig.~\ref{fig:fit_indices_1}).

%\begin{figure}               \begin{center}               \leavevmode
%\includegraphics[width=7cm]{FIGURES/Gamma.eps} \end{center} \caption{
%Marginalized  values  of the  IMF  slope  (bimodal)  with respect  to
%velocity  dispersion, including  $1\,\sigma$  confidence levels.  The
%joint method (SED  fitting plus line strengths) for  a grid of $\tau$
%models is shown  as black solid dots. The figure  also shows the case
%for simple  stellar populations (grey  open dots) and for  the $\tau$
%model,  where only  the  strongest IMF-sensitive  line strenghts  are
%considered  (\mgf,\,\tioi,\,\tioiio,\,\naii,\,\cat).   For reference,
%the same analysis restricted to  spectral fitting is given by the red
%crosses.  } \contcaption{.}  \label{fig:Hybrid_1} \end{figure}

%%%%%%%%%%%%%%%%%%%%%%%%%%%%%%%%%%%%%%%%%%%%%%%%%%%%%%%%%%%%%%%%%%%%%%%%
\section{Expected colours and $M/L$ ratios}
\label{sec:col_ml}

In addition  to gravity-sensitive features from  integrated light, the
stellar IMF can affect  indirectly other observable galaxy properties,
like broad-band colours (see~\citealt{GK:13}) and mass-to-light ratios
(hereafter  \ml; see,  e.g.,~\citealt{Cappellari:12a, FerreMateu:13}).
{ Hence,  it is interesting  to contrast the expectations  from our
  best-fit  stellar  population  models  with  those  from  broad-band
  photometry and (dynamical) $M/L$ estimates}.

\begin{table*}
\centering
\small
\begin{minipage}{100mm}
 \caption{Optical-NIR colours for  extended MILES (MIUSCAT) models with
   a bottom-heavy IMF, and our highest $\sigma_0$ stack.  }
  \begin{tabular}{c|c}
   \hline
 $g-K$ & Description \\
   \hline
$3.75$--$4.00$  & SSP, unimodal IMF, $\Gamma=2$, $ 0 \le [Z/H] \le 0.22$, Age\,$>12$\,Gyr \\
$3.65$--$3.97$  & SSP, bimodal IMF, $\Gamma=3.3$, $ 0 \le [Z/H] \le 0.22$, Age\,$>12$\,Gyr \\
$3.90 \pm 0.02$ & observed, $N_{\rm ETGs}=21$, $ 280 \le \sigma_0 \le 320$\,\kms \\
   \hline
  \end{tabular}
\label{tab:gK}
\end{minipage}
\end{table*}

%%%%%%%%%%%%%%%%%%%%%%%%%%%%%%%%%%%%%%%%%%%%%%%%%%%%%%%%%%%%%%%%%%%%%%%%
\subsection{Variation of colours}
\label{sec:colours}

{ Optical--NIR  photometry provide  additional constraints  on the
  IMF, } as a large fraction of  M dwarfs can enhance the NIR light of
a  stellar population  for an  IMF heavier  than  Salpeter~(see, e.g.,
\citealt{PVJ90}), making optical--NIR colours significantly redder. In
fact,  an excess  of reddening  at optical--NIR  wavelengths  has been
often invoked as  an important argument against the  steepening of the
low   mass   end  of   the   IMF   with   galaxy  mass   (see,   e.g.,
~\citealt{worthey:11}).   In order  to address  this issue,  we focus
here  on the bin  at the  highest velocity  dispersion (i.e.  $280 \le
\sigma_0 \le 320$\,\kms), for which the extended MILES (MIUSCAT)-based
fits give the most bottom-heavy IMF, i.e.  $\Gamma \sim 2$ and $\Gamma
\sim  3$, for  the unimodal  and bimodal  cases, respectively.  Out of
$160$ ETGs in this bin (see Tab.~\ref{tab:stacks}), $21$ galaxies have
NIR  (K-band) photometry  available  from the  UKIDSS-LAS survey  (see
Paper    I   for    details).     The   median    $g-K$   for    these
galaxies~\footnote{Notice  that SDSS  g-band photometry  is in  the AB
  system, while UKIDSS data are Vega-calibrated.  As detailed in Paper
  I,  colours  are  measured  in  an adaptive  aperture  of  $3  \cdot
  r_{Kron,i}$, where $r_{Kron,i}$ is the $i$-band Kron radius.  Median
  colours are corrected to the fibre aperture of radius $1.5''$, using
  $g-K$ colour gradient  estimates from~\citet{PaperIV} (Paper IV).  }
is  reported in Tab.~\ref{tab:gK},  and compared  to the  $g-K$ colour
range  of unimodal and  bimodal extended  MILES (MIUSCAT)  models with
bottom-heavy  IMF.   Model  colours  are estimated,  for  SSPs,  using
photometric  stellar libraries,  with the  code of~\citet{Vazdekis10}.
To  describe  the high-mass  systems,  we  consider  only colours  for
metallicities equal  or above solar,  and old ages ($>  12$~Gyr).  The
range  of model  colours in  the Table  is fully  consistent  with the
observed $g-K$, even for the  reddest unimodal models.  We remark here
that a  detailed comparison of  constraints from spectral  indices and
broad-band colours is  beyond the scope of the  present paper and will
be addressed in a forthcoming contribution.  Notice also that although
a  varying IMF  can affect  significantly some  optical  colours (e.g.
$r-i$), proper  modelling of  different optical colours  for quiescent
galaxies still  represents a challenge for  current stellar population
models, and the effect of  $\alpha$-enhancement on optical (as well as
NIR) colours remains to be fully understood (see MIUSCAT-II). However,
for the  purpose of the  present work, we emphasize  that optical--NIR
colours do not  pose any problem for the  hypothesis of a bottom-heavy
IMF in massive ETGs.

\subsection{Variation of M/L}
\label{sec:ml}

\begin{figure}
\begin{center}
\leavevmode
\includegraphics[width=7.5cm]{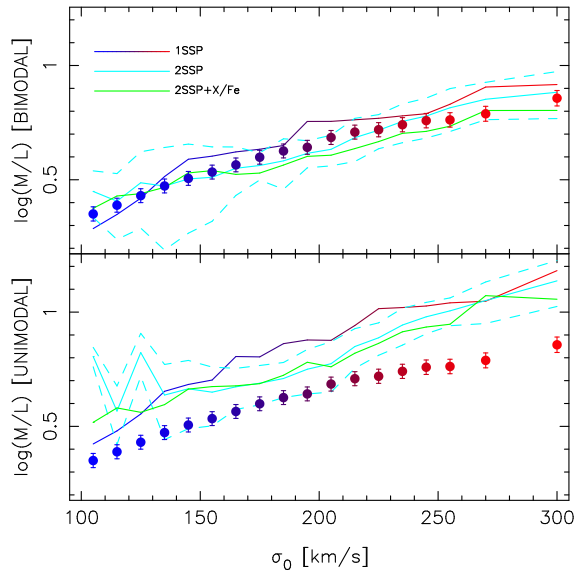}
\end{center}
\caption{Mass-to-light  ratios  as  a  function  of  central  velocity
  dispersion,  for bimodal  (top) and  unimodal  (bottom) best-fitting
  models  of  spectral  indices.   Dots  and  error  bars  are  median
  dynamical  mass-to-light  ratios, with  $1$~$\sigma$  error bars  on
  median values, while solid  curves give stellar mass-to-light ratios
  predicted  for  different  stellar  population  models  of  spectral
  indices,  including  the effect  of  a  bottom-heavier  IMF at  high
  (relative  to  low)  $\sigma_0$.   Dashed curves  mark  $2$~$\sigma$
  confidence intervals  on $2SSP$  model predictions. Notice  the good
  matching of bimodal models to the dynamical estimates.  }
%\contcaption{.}
\label{fig:logml}
\end{figure}

Fig.~\ref{fig:logml} plots the median  dynamical \ml\ as a function of
$\sigma_0$. The \ml\ is estimated as $5 \times R_e \sigma_e^2/(G \cdot
L)$ (see, e.g.,~\citealt{Cappellari:12}), where $R_e$ is the effective
radius, $\sigma_e$ is the  central velocity dispersion corrected to an
aperture of  $1\,R_e$, $G$ is  the gravitational constant, and  $L$ is
the total galaxy luminosity.  Both $R_e$ and $L$ are measured for each
galaxy  by  fitting  seeing-convolved  S\'ersic  models  to  the  SDSS
$r$-band  galaxy  images (see  SPIDER-I  for  details).  The  aperture
correction    to   $\sigma_0$    is   performed    by    using   eq.~1
of~\citet{Cappellari:06}.  The  above approximation for  \ml\ provides
an  estimate of the  \ml\ within  $1\,R_e$ with  an accuracy  of $\sim
0.03$~dex.   This  uncertainty  is   added  in  quadrature,  for  each
$\sigma_0$ bin, to  the error on the median \ml.   We refer the reader
to,  e.g.,   ~\citet{Tortora:12,  Cappellari:12,  Tortora:13}   for  a
discussion  on  the  impact  of  different model  assumptions  on  the
estimate of \ml.   For the purpose of the present work,  we do not aim
to discuss the details beyond  the computation of \ml, but instead, we
focus  on  the  comparison  between  the predictions  of  the  stellar
\ml\ (hereafter \mlstar) from  our best-fit population models with the
estimates of the dynamical \ml.  The key aspect is the constraint that
stellar mass-to-light  ratios cannot exceed  dynamical estimates, i.e.
\mlstar$\le$\ml.    The   expected   \mlstar's  are   overplotted   in
Fig.~\ref{fig:logml},  for  $1SSP$,  $2SSP$, and  $2SSP+X/Fe$  models,
respectively,   and  for  bimodal   (upper-)  and   unimodal  (lower-)
models~\footnote{{   The   \mlstar\   estimates,  as   plotted   in
    Fig.~\ref{fig:logml}, are made available on request to the authors
    in tabular  format, along with  the same estimates obtained  for a
    Kroupa Universal IMF.  }}.  The 1\,$\sigma$ confidence contours on
\mlstar\ are plotted only for  $2SSP$ models, for clarity.  The Figure
shows  that  a unimodal  IMF  gives  too  high, and  thus  unphysical,
\mlstar,  with respect  to  the dynamical  \ml.   Hence, although  the
unimodal models do  fit spectral indices as well  as the bimodal ones,
they seem  to be rejected based  on the \mlstar\  predictions.  On the
contrary, bimodal models give an \mlstar\ broadly consistent with \ml,
even  for  strongly bottom-heavy  models,  as  at highest  $\sigma_0$.
Considering error bars, our bimodal best-fits are also compatible with
stellar mass-to-light  ratios being  smaller than the  dynamical ones,
leaving room for  additional, non-stellar, matter (i.e.  dark-matter).
A similar conclusion has  been recently drawn by~\citet{CvD12b}, for a
subsample of $35$  ETGs from the SAURON survey.  {  We remark that the
  above conclusions  are based on  the assumption that  the expression
  for  the dynamical  \ml\ from~\citet{Cappellari:12}  applies  to the
  whole   population    of   ETGs,   represented    by   our   stacked
  spectra. Detailed dynamical and lensing studies for large samples of
  ETGs will help to address this point in the future.}

%\begin{figure}
%\begin{center}
%\leavevmode
%\includegraphics[width=7.5cm]{FIGURES/cols_gK.ps}
%\end{center}
%\caption{Optical-NIR  colours   as  a  function   of  central  velocity
%  dispersion, for bimodal (top)  and (unimodal) best-fitting models to
%  spectral indices.  Circles and error bars plot the median $g-K$, and
%  the corresponding  $1$~$\sigma$ error bar on median  colour, for each
%  bin  of velocity  dispersion, as  a function  of the  mid $\sigma_0$
%  value of  the bin. Solid and  dotted curves are  the expected colours
%  from  $1SSP$  and $2SSP$  models,  with  dashed  curves marking  the
%  $2$~$\sigma$ confidence  intervals on $1SSP$ model  colours.  A shift
%  $\delta(g-K)$ has been  applied to each model trend,  as reported in
%  each  panel, in  order to  match observed  and model  colours  at low
%  $\sigma_0$ (see the text).}
%%\contcaption{.}
%\label{fig:cols_gK}
%\end{figure}

%%%%%%%%%%%%%%%%%%%%%%%%%%%%%%%%%%%%%%%%%%%%%%%%%%%%%%%%%%%%%%%%%%%%%%%%
\section{Summary}
\label{sec:summary}

This paper explores in detail the recent claims of a systematic
variation of the low-mass end of the IMF in massive early-type
galaxies (ETGs). In order to set robust constraints, we need a large,
representative dataset of galaxy spectra at very high signal-to-noise
ratio. For this purpose, we stack a large, high-quality sample of SDSS
spectra, comprising $24,781$ low redshift ($z \sim 0.07$) ETGs from
the SPIDER survey~\citep{SpiderI}.  The stacking is performed in bins,
corresponding to their central velocity dispersion, and, at fixed
velocity dispersion, we also assemble a subsample split with respect
to $\alpha$-enhancement, by using a solar-scale proxy (\afep, see
Fig.~\ref{fig:proxy}).  We select a variety of IMF-sensitive spectral
features (\mgf, \tioi, \tioiio, \naii, \cat, see
Fig.~\ref{fig:sens_indices}), along with additional indices strongly
sensitive to the abundance of individual elements that could affect
their interpretation. The analysis includes standard age and
metallicity indicators (\mgfep, \hbo, \hgf). After being corrected to
solar-scale by means of semi-empirical correlations among line
strengths and \afep\ at fixed $\sigma_0$, the indices are fitted with
state-of-the-art population synthesis models. We use the extended
MILES (MIUSCAT) library \citep{Vazdekis:12}, covering a wide range of
age, metallicity, star-formation histories, and IMF.  The analysis is
complemented with an independent hybrid approach, where direct
spectral fitting in the optical is combined with constraints from
IMF-sensitive features \citep[following our previous
  work,][]{Ferreras:13}.  Two types of IMF are considered, a single
power-law distribution, generalizing the Salpeter law, and a bimodal
IMF, with a gradual turn-off at low mass, that generalizes the Kroupa
case.  The main results can be summarized as follows:

\begin{description}
\item[i.]  All selected spectral features consistently suggest a
  variation of IMF slope ($\Gamma$) in ETGs, with a trend from a
  Kroupa/Chabrier case at low central velocity dispersion, towards a
  more bottom-heavy IMF with increasing $\sigma_0$. For a bimodal IMF,
  the best-fit $\Gamma$ increases from about $1.3$ at $\sigma_0\simlt
  150$\,\kms\ to almost $3$ at $\sigma_0 \sim 300$\,\kms.  For
  unimodal models, the slope varies from $\Gamma\sim 1$ to $2$ over
  the same $\sigma_0$ range (Fig.~\ref{fig:gamma_fits_all}). The fits
  to the $\Gamma$--$\sigma_0$ correlation (eq.~\ref{eq:gamma_sigma})
  give consistent values between different methods
  (Tab.~\ref{tab:gamma_sigma}), although we note that the physical
  quantity being constrained is the mass fraction in low-mass stars
  (see iii. below)
\item[ii.]  Both unimodal and bimodal cases fit equally well all
  selected spectral indices, implying that we cannot distinguish
  between these two kinds of models based on the indices alone (see,
  e.g., Fig.~\ref{fig:fit_indices_2} and
  \ref{fig:fit_indices_un_2}). However, unimodal models seem to be
  clearly rejected by their overly high values of the stellar
  (relative to dynamical) mass-to-light (\mlstar) ratios
  (Fig.~\ref{fig:logml}).  On the other hand, the varying-IMF bimodal
  best-fits are in good agreement with independent constraints, i.e.
  dynamical $M/L$ estimates, and the optical-NIR broad-band colours of
  ETGs in our sample.
\item[iii.]  Although unimodal and bimodal models cannot be
  distinguished with the present analysis, they predict very
  consistent stellar mass fractions at birth in low-mass stars
  ($<0.5$\,M$_\odot$), varying from $\sim 20 \, \%$ at $\sigma_0 \sim
  100$\,\kms\ to $\sim 70 \%$ at $\sigma_0\sim 300$\,\kms
  (Fig.~\ref{fig:mass_frac}).  Such stellar low-mass fractions 
  (Eq.~\ref{eq:mass_frac}) represent the most robust constraint from
  the present analysis, and should be matched by theories aimed to
  explain the stellar IMF.
\item[iv.]  The fact that both solar-scale corrected IMF- and
  abundance-sensitive line strengths from different species, i.e.  Na,
  Ca, TiO, and Mg, can be all consistently explained by models with a
  varying IMF gives strong support to the claim that the IMF truly
  changes at the low-mass end in ETGs with high velocity dispersion,
  in constrast to a picture whereby single element adundances conspire
  to change simultaneously with $\sigma_0$.  Indeed, including a
  varying abundance of Na, Ca, and Ti in our fits (based on
  ~\citealt{CvD12a} stellar population models), we find no evidence
  for a significant variation of these abundances with $\sigma_0$
  besides the general \afe\ trend (Fig.~\ref{fig:xfe_fits}).
\item[v.] The range of variation of IMF-sensitive line strengths with
  $\sigma_0$ is essentially similar for both low- and
  high-\afe\ stacks, implying that velocity dispersion (i.e. possibly
  galaxy mass) is the main driver of IMF variation, regardless of
  changes in $\alpha$-enhancement.
\item[vi.]  Regarding the \afe, we find a remarkably tight correlation
  between the (nearly) solar-scale proxy, \afep, i.e. the difference
  between the two metallicities estimated from Mg and Fe lines with
  solar-scaled models, and the ``true'' \afe, i.e.  the one obtained
  by using $\alpha$-enhanced stellar population models
  (Fig~\ref{fig:proxy}).  This proves that one can study abundance
  ratio effects by relying entirely on solar-scaled models.
\item[vii.]  Besides \cat\ and \cahk, we have also tried to fit the
  well-known Lick-based \caf\ feature.  This index remains a puzzle
  for current stellar population models, as no fitting scheme is able
  to match its line strengths, especially at high-$\sigma_0$, where
  model indices are far too high with respect to the observations
  (Fig.~\ref{fig:caf}).  The $\rm [Ca/Fe]$ under-abundance explanation
  for this discrepancy seems to be excluded because the extended MILES
  (MIUSCAT) base models match well both the \cahk\ and \cat\ features
  (see, e.g., Fig.\ref{fig:fit_indices_2}).
\end{description}

From the modelling point of view, the suggested change in the IMF of
ETGs with velocity dispersion could be expected from the different
physical properties of the star forming clouds during the formation of
a massive galaxy \citep{Larson:05} -- such as a turbulent ISM with a
very high Mach number \citep{Hopkins:12}. Alternatively, variations in
the IMF properties convolved with the star formation history can leave
its imprint on the so-called integrated galactic IMF \citep[IGIMF,
  see, e.g.,][]{igimf}, which corresponds to the superposition of all
populations ever formed, and constitutes the observable we really map
with unresolved spectroscopic data. Regardless of whether the cause
lies in the microphysics of star formation, or the global formation
history, the robust constraints we impose here on the initial mass
fraction in low-mass stars should be met by any theory of star
formation.

%\subsection{ }
%\label{sec:stacks}

%\begin{figure}
%\begin{center}
%\includegraphics[height=120mm]{ours_tfg_re_n.ps}
%\end{center}
%\caption{Histograms   of   logarithmic   differences  between   z-band
%}
%\label{fig:comp_spars}
%\end{figure}

\section*{Acknowledgments}
{ We  would like  to thank the  anonymous referee for  his/her helpful
  report, which definitely helped improving parts of this manuscript.}
We also thank R. Smith  for helpful comments on this manuscript.  This
paper is  based on  data retrieved from  the Sloan Digital  Sky Survey
archives  ({\tt  http://www.sdss.org/collaboration/credits.html}).  We
have  also made  use of  the  4th data  release of  the UKIDSS  survey
\citep{Law07}, which is described in detail in \citet{War07}.  Funding
for the  SDSS and SDSS-II  has been provided  by the Alfred  P.  Sloan
Foundation,  the  Participating  Institutions,  the  National  Science
Foundation, the  U.S.  Department of Energy,  the National Aeronautics
and Space Administration, the  Japanese Monbukagakusho, the Max Planck
Society, and  the Higher Education  Funding Council for  England.  JFB
acknowledges support from the Ram\'on y Cajal programme by the Spanish
Ministry of  Economy and Competitiveness (MINECO). This  work has been
supported by the Programa  Nacional de Astronom\'ia y Astrof\'isica of
MINECO, under grants AYA2010-21322-C03-01 and AYA2010-21322-C03-02 and
by  the  Generalitat Valenciana  under  grant PROMETEO-2009-103..   MT
acknowledges the support of FAPESP, process no. 2012/05142-5.

\appendix

%%%%%%%%%%%%%%%%%%%%%%%%%%%%%%%%%%%%%%%%%%%%%%%%%%%%%%%%%%%%%%%%%%%%%%%%
\section{Definition of \naii\ and \tioiio\ spectral indices}
\label{sec:new_indices}

\begin{figure*}
\begin{center}
\leavevmode \includegraphics[width=15cm]{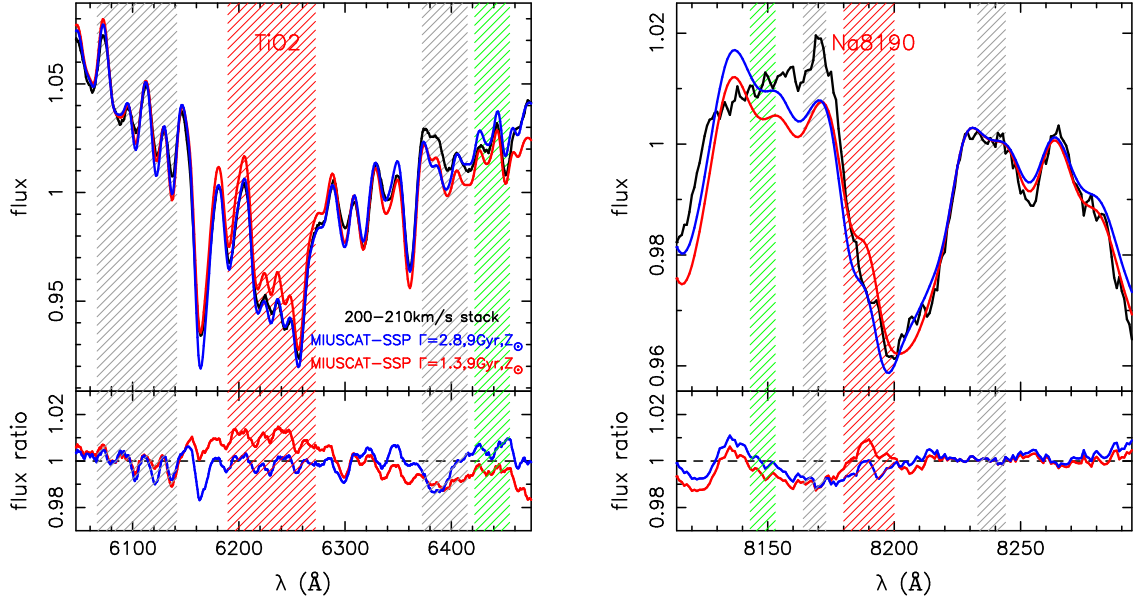}
\end{center}
\caption{Re-definition of \tioii\ (left) and Na8190 (right) spectral
  indices. The top panels show the SDSS stacked spectrum for $200 \le
  \sigma_0 \le 210$\,\kms\ (black), and two extended MILES (MIUSCAT)
  SSPs, with solar metallicity, age of $9$\,Gyr, and two bimodal IMFs,
  namely a Kroupa-like IMF ($\Gamma=1.3$; red) and a bottom-heavy
  model ($\Gamma=2.8$, blue). The SSPs were smoothed to match the
  spectral resolution plus $\sigma_0$ of the stack. Both data and
  models have been scaled by the median flux computed over the
  spectral regions encompassed by the TiO2 and NaI8200A definitions.
  The red hatched regions map the index central passbands, while grey
  hatched regions mark the index sidebands.  The bottom panels plot
  residuals between the two IMF models and data.  Notice the deviation
  of both Kroupa-like and bottom-heavy models with respect to the data
  in the red and blue sidebands of TiO2 and NaI8200A, respectively.
  To minimize this discrepancy, we have optimized the definition of
  red and blue sidebands of \tioii\ and NaI8200A (see text for
  details). The modified sidebands (green regions) define the spectral
  indices \tioiio\ and \naii, used in the present work. }
%\contcaption{.}
\label{fig:tio_na_redef}
\end{figure*}

In the present work, we adopt a modified version of \tioii\ and
NaI8200A spectral indices, previously defined by \citet{Trager98} and
\citet{Vazdekis:12}, respectively.  The NaI8200A index is, in turn, a
modified version of the NaI doublet index proposed
by~\citet{SchiavonNaD:97}. The reason for adopting a modified version
of these indices is illustrated in Fig.~\ref{fig:tio_na_redef}, where
we plot the stacked spectrum with $200 \le \sigma_0 \le 210$\,\kms\ in
the spectral range of \tioii\ ($6050 \le \lambda \le 6470$\,\AA; left)
and NaI ($8110 \le \lambda \le 8290$\,\AA; right).  Two extended MILES
(MIUSCAT) SSP models, with age $9$\,Gyr, solar metallicity
($[Z/H]=0$), and two different IMFs, i.e.  $\Gamma=1.3$ (Kroupa-like)
and $\Gamma=2.8$, are overplotted.  The SSPs have been smoothed to
match the spectral resolution and velocity dispersion of the stack.
The red hatched regions mark the central bandpasses of the indices,
for which we keep the original definition of \tioii\ and NaI8200A.
The grey hatched regions are the original sidebands of \tioii\ and
NaI8200A.  While there is good agreement between models and data in
the blue (red) pseudo-continuum of \tioii\ (NaI8200A), for both IMFs,
the models show a $\sim 1\%$ deviation from the observed spectrum in
the red (blue) sidebands of \tioii (NaI8200A).  The largest deviations
are seen at $\lambda \sim 6380$\,\AA\ and $\lambda \sim 8170$\,\AA,
and are also present in the other stacked spectra.  The discrepancies
are not removed by changing the age and metallicity (as well as IMF)
of the SSP models, for $Age \ge 3$\,Gyr and $[Z/H] \ge -0.4$. Hence, we
have re-defined the red (blue) sideband of \tioii\ (NaI8190A) in
order to minimize any deviation between models and data. This is done
as follows.
\begin{description}
\item[1.]  We adopt the original definition of \tioii\ and NaI8190A,
  and compute the $2SSP$ best-fit model to spectral indices (see
  Sec.~\ref{sec:fitting}). We assume here a Kroupa IMF in order to
  avoid any bias towards a $\sigma_0$-varying IMF (which is the
  hypothesis we want to test);
\item[2.] For \tioii, we vary the position (i.e. the lowest wavelength
  endpoint, $\lambda_1$) and width, $\delta \lambda$, of the red
  sideband, minimizing the absolute deviation of the index equivalent
  width between models and data. To this effect, we keep the central
  band and blue sideband definitions fixed to those of \tioii.  The
  absolute deviation is computed by summing up the EW absolute
  deviations for five stacks ($\sigma_0=100$, $150$, $200$, $250$, and
  $300$\,\kms), spanning the whole velocity dispersion range.  The
  deviation is minimized for $\lambda_1=6422$\,\AA\ and $\delta
  \lambda= 33$\,\AA\ (see the green hatched region in the left panel
  of Fig.~\ref{fig:tio_na_redef}). We adopt this definition throughout
  the present work, referring to the corresponding spectral index as
  \tioiio, to point out that the definition has been optimized for the
  SDSS spectra.
\item[3.]  For NaI8190A, we adopt a similar procedure to that at point
  2 for \tioii, but varying the position of the blue sideband of the
  index.  We impose the constraint $\delta \lambda \ge 10$\,\AA, i.e.
  the value adopted in \citet{Vazdekis:12}, in order to avoid overly
  reducing the region where the blue pseudo-continuum is measured.
  The minimum absolute deviation is obtained for $\lambda=8143$\,\AA,
  and $\delta \lambda = 10$\,\AA\ (see the green hatched region in the
  right panel of the Figure).  We refer to the corresponding spectral
  index as \naii.
\end{description} 
Repeating the entire procedure,  but replacing \tioii\ and NaI8190A at
step  1   with  \tioiio\  and  \naii\  gives   very  similar  sideband
definitions  as those  adopted  here, proving  that  the procedure  is
self-consistent.  We  also experimented  with changing the  Kroupa IMF
assumption   in   step~1  with   a   $\sigma_0$-varying  bimodal   IMF
(blue-through-red       curve        in       top       panel       of
Fig.~\ref{fig:gamma_fits_all}),  finding  the  optimum definitions  of
$\lambda_1$ and $\delta \lambda$ to change by only a few Angstr\"om.

\begin{figure*}
\begin{center}
\leavevmode
\includegraphics[width=15cm]{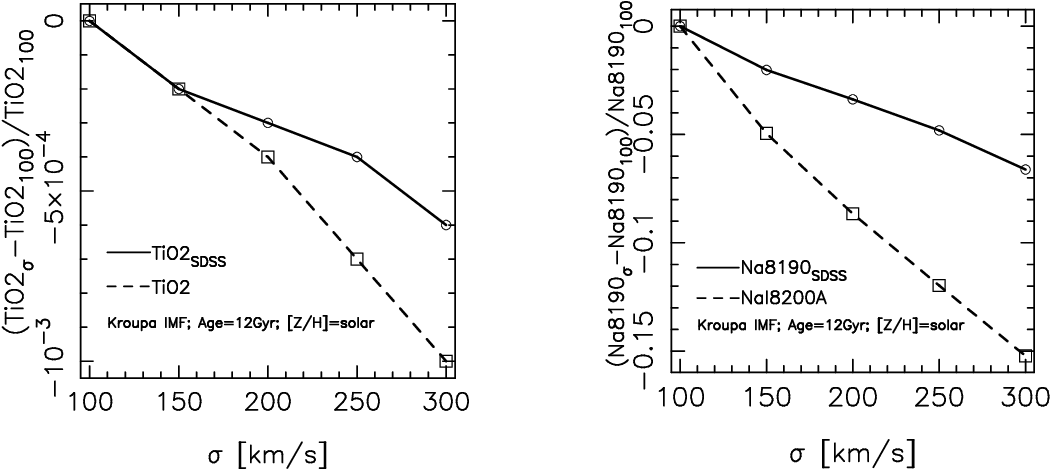}
\end{center}
\caption{Sensitivity of the \tioiio\ (left) and \naii\ (right)
  spectral indices to resolution.  The plot shows the relative
  variation of line strengths as a function of velocity dispersion, in
  the same range as our stacked spectra (namely, from $\sigma=100$ to
  $300$\,\kms). Solid lines refer to indices defined in previous
  works, i.e.  \tioii\ \citep{Trager98} and NaI8200A (MIUSCAT-I).  The
  dashed lines correspond to the indices adopted in the present study,
  optimized to better match the SDSS data with extended MILES
  (MIUSCAT) $2SSP$ models (independent of the IMF). The curves refer
  to an extended MILES (MIUSCAT) SSP model with Kroupa IMF, solar
  metallicity, and age of $12$\,Gyr, as labelled.  Notice that the
  optimized indices are less sensitive to velocity dispersion than
  those defined in previous works.}
%\contcaption{.}
\label{fig:tio_na_sigma}
\end{figure*}

In Fig.~\ref{fig:tio_na_sigma}, we show the dependence of \tioiio\ and
\naii\ EWs on velocity dispersion.  These indices have the further
advantage to be less sensitive to resolution than the previously
defined \tioii\ and NaI8200A. For \naii, the decreased sensitivity is
due to the fact that the blue sideband is measured in a more distant
region from the feature.  We also note that, as shown in
Sec.~\ref{sec:indices}, the new \tioiio\ and \naii\ indices have
similar sensitivity to age, metallicity, and IMF as \tioii\ and
NaI8200A.

%%%%%%%%%%%%%%%%%%%%%%%%%%%%%%%%%%%%%%%%%%%%%%%%%%%%%%%%%%%%%%%%%%%%%%%%
\section{Emission correction to \hb}
\label{sec:hb_emission}

\begin{figure}
\begin{center}
\leavevmode \includegraphics[width=7.5cm]{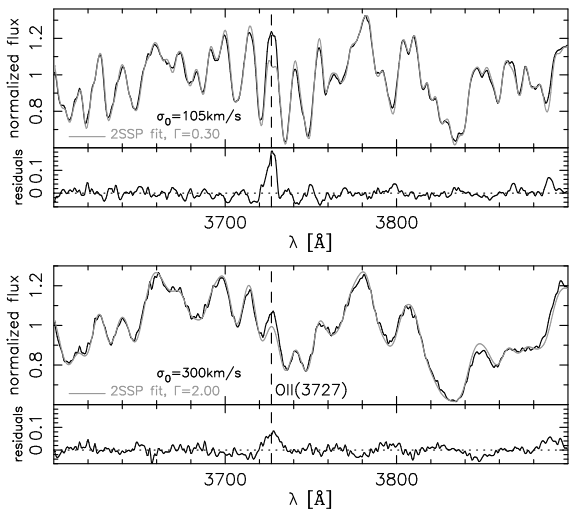}
\end{center}
\caption{{ Evidence for nebular emission in stacked spectra of ETGs
    at the extremes of velocity dispersion in our sample, namely $100
    \le \sigma_0 \le 110$\,\kms\ ({\sl top plot}) and $280 \le
    \sigma_0 \le 320$\,\kms\ ({\sl bottom plot}).  For each plot, the
    upper panel shows the observed spectrum (black) at its nominal
    resolution, and the MIUSCAT best-fitting model (grey), obtained by
    assuming a Kroupa IMF (see text for details).  The lower panel
    shows residuals, obtained by subtracting the model to the observed
    spectrum.  Vertical dashed lines mark the [OII] feature at
    $\lambda=3727\,$\AA, while horizontal dotted lines mark a residual
    value of zero. } }
%\contcaption{.}
\label{fig:OII}
\end{figure}

In order to fit the spectral indices with stellar population models,
we correct the \hbo\ spectral index for contamination from nebular
emission.  The presence of such contamination in our stacked spectra
of ETGs is shown in Fig.~\ref{fig:OII}, where we plot the spectral
region around the [OII] feature at $\lambda\! =\!3727$\,\AA\ for the
stacks at the endpoints of the velocity dispersion range, i.e.  $100
\le \sigma_0 \le 110$\,\kms\ (top) and $280 \le \sigma_0 \le
320$\,\kms\ (bottom), respectively.  Best-fitting extended MILES
(MIUSCAT) models, obtained by direct fitting of the spectral region of
interest are overplotted in grey (assuming a Kroupa IMF).  Notice the
flux excess at $\lambda \sim 3727$\,\AA, in particular at the lowest
$\sigma_0$, revealing the presence of nebular emission in the spectra.

\begin{figure}
\begin{center}
\leavevmode
\includegraphics[width=7.cm]{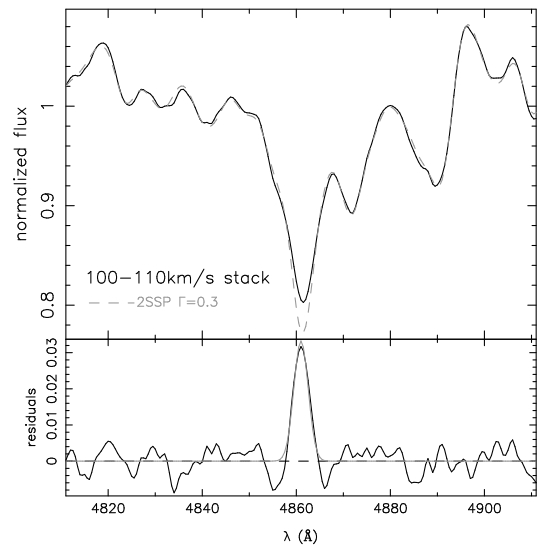}
\end{center}
\caption{Correction method for the \hbo\ equivalent width of stacked
  spectra from nebular emission. ({\sl Top}) The lowest $\sigma_0$
  stack (black) is shown along with the best-fit MIUSCAT model for a
  bimodal IMF with $\Gamma=0.3$ (grey). ({\sl Bottom}) The residuals
  of the fit (black) are fitted with a Gaussian function (grey), that
  determines the \hbo\ correction for nebular emission (see text for
  details). }
%\contcaption{.}
\label{fig:ecorr}
\end{figure}

The    correction    procedure   to    \hbo\    is   illustrated    in
Fig.~\ref{fig:ecorr}. For a given IMF (either bimodal or unimodal), we
fit  the \hb\  spectral region  (from  $4810$ to  $4910$\,\AA) with  a
linear combination  of two MIUSCAT  SSPs, excluding the trough  of the
line (i.e.  from  $4856$ to $4866$\,\AA).  To improve  the matching of
the continuum, a multiplicative  fourth degree polynomial is also used
in the fitting. The residuals  (bottom panel of the Figure) are fitted
with  a Gaussian  function. The  emission correction  is given  by the
difference  between  the  \hbo\  index   of  the  stack  and  the  one
re-measured  on the  same stack  after  adding the  Gaussian fit.   By
definition,  this  procedure will  depend  on  the  assumed IMF.   For
instance, for the $100 \le \sigma_0 \le 110$\,\kms\ ($280 \le \sigma_0
\le  320$\,\kms)  stack, the  emission  correction  varies from  $\sim
0.13$\,\AA\     ($0.09$\,\AA)    for     $\Gamma=0.3$,     to    $\sim
0.08$\,\AA\ ($0.05$\,\AA)  for $\Gamma=3.3$ (bimodal  IMFs).  Although
the $\Gamma$ dependence is small, following our previous work (FLD13),
we have implemented  an iterative procedure for the  $1SSP$ and $2SSP$
models (see Sec.~\ref{sec:fitting}).  First, we correct \hbo\ with the
median  correction (among  all IMFs),  and  then we  fit the  spectral
indices.   Once  the  best-fit   $\Gamma$  is  derived,  the  emission
correction  is  updated  accordingly,  and  the  fitting  is  repeated
again. In practice, we found that further iterations are not required,
as  the  fits do  not  change  appreciably  when performing  a  second
iteration. Notice that in FLD13 we applied a similar procedure to that
described  here. However,  rather  than fitting  the  stacks with  two
MIUSCAT SSPs in  the \hb\ spectral region, we  ran the {\tt STARLIGHT}
spectral fitting code over a  larger spectral range. We found that the
two SSP fits  provide more robust results than  our previous approach,
being less sensitive  to the adopted IMF. We  note that either varying
the degree  of the multiplicative  polynomial between two and  six, or
changing  the size  of the  line  trough, excluded  from the  fitting,
between $5$ and $15$\,\AA, does  not produce any significant change to
our  results.  { The  \hbo\ correction  turns  out to  vary
  smoothly  with  $\sigma_0$,  from  $\sim  0.12  \,  \AA$  at  lowest
  $\sigma_0$ to $\sim 0.04 \,  \AA$ for $\sigma_0 > 250$~\kms, further
  confirming the robustness of our approach. }

%%%%%%%%%%%%%%%%%%%%%%%%%%%%%%%%%%%%%%%%%%%%%%%%%%%%%%%%%%%%%%%%%%%%%%%%
\section{Aperture effects}
\label{sec:aper}

\begin{figure}
\begin{center}
\leavevmode
\includegraphics[width=8.5cm]{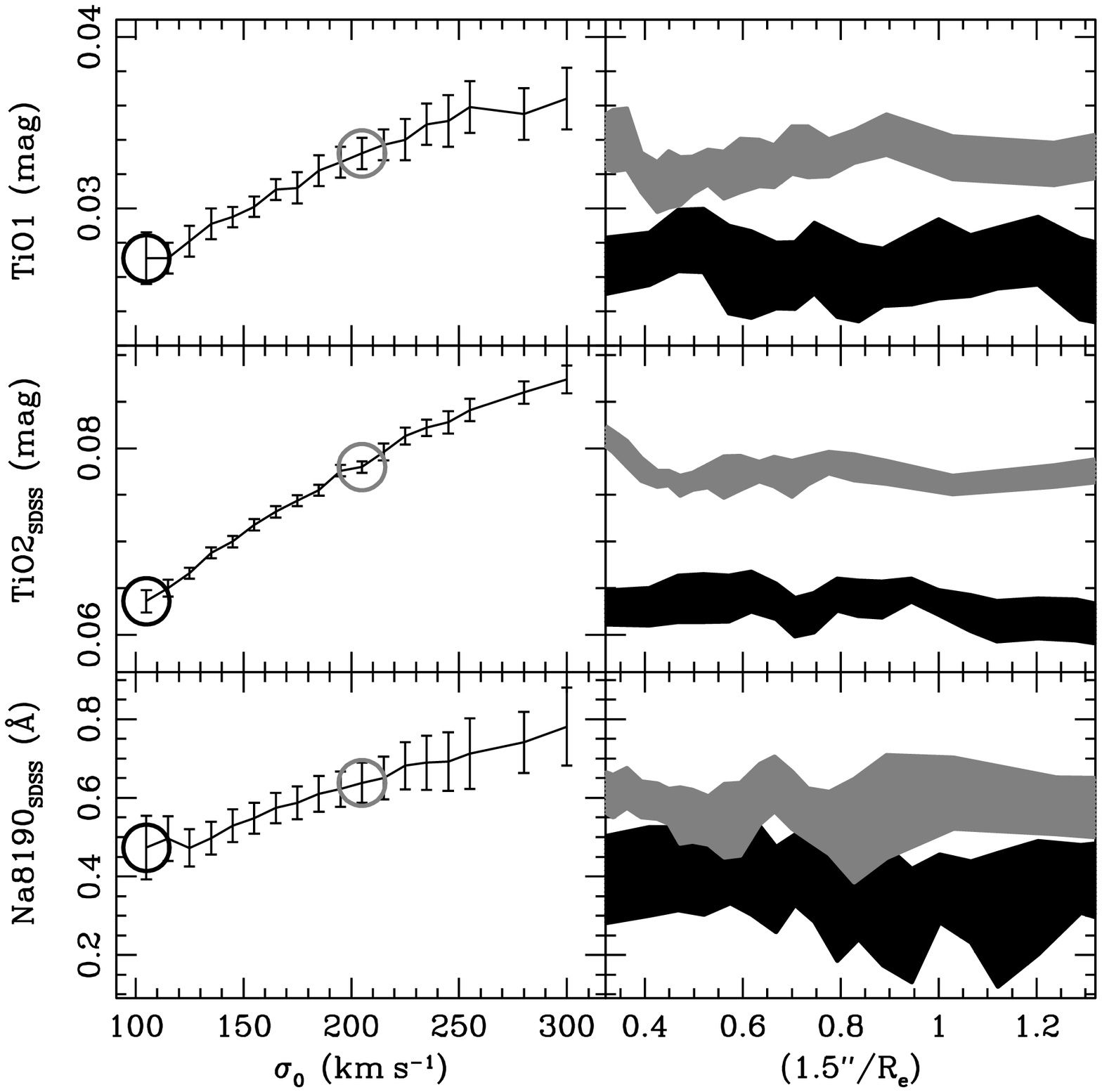}
\end{center}
\caption{ ({\sl Left}) Variation of three IMF-sensitive line strengths
  with respect to velocity dispersion. ({\sl Right}) Two stacks are
  considered (circled in the panels on the left), with velocity
  dispersion of $\sigma_0=100$ (black) and $200$ (grey) \kms. The
  shaded regions represent the variation in the corresponding line
  strengths with respect to effective radius, compared with the fixed
  $3\,$arcsec diameter fibre of the SDSS spectrograph. Error bars are
  shown at the $1\,\sigma$ confidence level.}
%\contcaption{.}
\label{fig:aper}
\end{figure}

Given that the spectra used in this paper correspond to the central
3\,arcsec region of each galaxy, a significant radial gradient of the
properties of the stellar populations could affect the results. In
Fig.~\ref{fig:aper} we explore this issue by showing, in the rightmost
panels the variation of three of the most IMF-sensitive line
strengths, with respect to effective radius, in comparison to the SDSS
3\,arcsec diameter fibre.  The shaded regions encompass the
$1\,\sigma$ confidence level of the $100$\,\kms\ (black) and
$200$\,\kms\ (grey) stacks.  For reference, the panels on the left
show the variation of the line strengths over the whole range in
velocity dispersion.  The figure confirms that our results are not
affected by variations in radial gradients of the stellar populations.

%%%%%%%%%%%%%%%%%%%%%%%%%%%%%%%%%%%%%%%%%%%%%%%%%%%%%%%%%%%%%%%%%%%%%%%%
\section{Synthetic SSPs with varying abundance ratios}
\label{sec:synth_ssps}
%In order  to create models for  the spectrum of  a stellar population,
%the  following ingredients are  required: stellar  evolutionary tracks
%(i.e., isochrones)  and a library  of stellar spectra.  The isochrones
%determine the  mass and  the atmospheric parameters  (surface gravity,
%$\log g$, and effective temperatures, $T_{eff}$) of stars belonging to
%an  SSP  with  a  given  age  and  chemical  composition  (metalicity,
%\afe). With the list of stars which should be included in the
%synthesis, the integrated  light spectrum of the SSP  can be generated
%by combining  the spectra  of individual stars,  according to  a given
%IMF.

%Our goal is to investigate how changes in the abundances of individual
%elements affect the spectral  indices. Empirical stellar libraries are
%not  well  suited for  this  purpose,  since  they are  restricted  to
%abundance  patterns  of stars  in  the  solar  neighborhood. 

In  addition to  the CvD12  stellar  population models,  we have  also
tested  the  effect  of  individual  variations  of  Ca,  Na,  and  Ti
abundances  on the  targeted spectral  indices  by creating  a set  of
simplified SSP models. Empirical stellar libraries are not well suited
for this purpose,  since they are restricted to  abundance patterns of
stars in the  solar neighbourhood. For this reason,  we rely completely
on synthetic  stellar libraries. To  generate stellar spectra,  we use
the Padova  evolutionary tracks~\citep{bertl08}, identifying  16 $\log
g$--$T_{\rm eff}$ pairs along the $12$\,Gyr isochrone, from the tip of
the RGB ($T_{\rm eff} = 3000$\,K and $\log g = -0.1$) down to the main
sequence  locus corresponding  to a  stellar mass  of 0.15\,M$_{\odot}$
($T_{\rm  eff}  =  3400$\,K and  $\log  g  =  5.2$).  For  each  $\log
g$--$T_{\rm eff}$  pair.  we  generate synthetic stellar  spectra with
the   {\tt   PFANT}   code  described   in   \citet{Cayrel.etal:1991},
\citet{Barbuy.etal:2003}   and   \citet{Coelho.etal:2005}.   Given   a
stellar model atmosphere and lists  of atomic and molecular lines, the
code  computes  a  synthetic  spectrum  assuming  local  thermodynamic
equilibrium (LTE).  We  have used a refined atomic  and molecular line
list,  calibrated   through  several  stellar   spectroscopic  studies
\citep[see,   e.g.,][]{Barbuy.etal:2003}.    For  stellar   atmosphere
models,    we   used    the   {\tt    MARCS   1D}    hydrostatic   LTE
models~\citep{Gustafsson.etal:2008}.    For   $\log   g  >   3$,   the
atmospheric model geometry is plane-parallel and mass independent; for
$\log  g < 3$  (giants with  non-negligible photospheric  depths), the
models are calculated  for a spherical geometry.  In  the latter case,
models for  1\,M$_{\odot}$ are  adopted. A microturbulent  velocity of
$2$\,\kms\  is  adopted  for   all  spectra.   Notice  that  the  flux
predictions of {\tt  PFANT} are less accurate in the  blue part of the
spectrum,  in particular  at  $\lambda <  5000$\,\AA,  because of  the
so-called  ``predicted  lines''  problem~\citep{Kurucz:92}.  For  this
reason, the use of our synthetic SSPs to model features at wavelengths
bluer      than     5000~$\AA$      (e.g.,     the      \cahk,     see
Sec.~\ref{subsec:abundances}) should be taken with some caution.

%Since we are
%mainly interested  in the study of  spectral indices, we  do not apply
%any flux correction to synthetic spectra.

\begin{figure*}
\centering
\includegraphics[width=15cm]{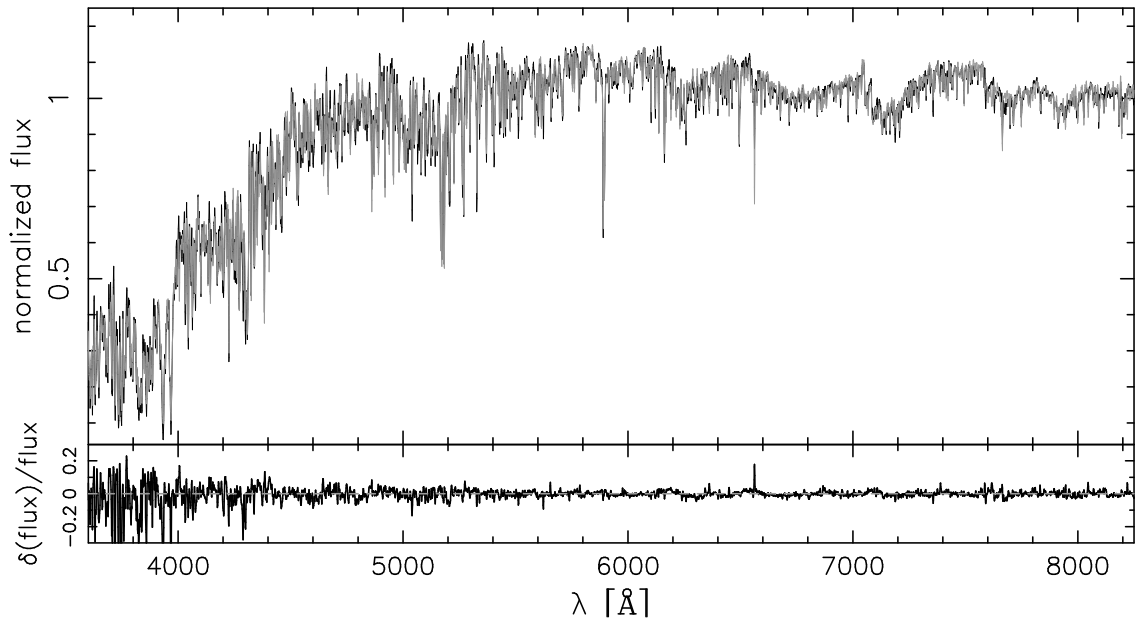}
\caption{{\sl (Top)} Comparison of one of our simplified, synthetic,
  SSPs, with solar abundances, to a MILES extended SSP, for an age of
  $12.5$\,Gyr, and solar metallicity.  The models have been
  continuum-matched, using a polynomial fitting of order twenty. Both
  the synthetic and MIUSCAT SSPs correspond to a Kroupa IMF. {\sl (Bottom)}
  relative residuals, between the synthetic and MILES SSPs. }
\label{fig:toy_ssp}
\end{figure*}

The SSP spectra, corresponding to an age of 12\,Gyr, are created using
the integral
 $$f(\lambda) = \int_ {m_1}^{m_2} s(\lambda, m) \phi(m) dm$$
\noindent where $s(\lambda, m)$ is  the spectrum of an individual star
with mass  $m$ at  a given $\lambda$,  and $\phi(m)$ is  the (assumed)
IMF. We adopt a mass interval from $m_1 = 0.15$\,M$_{\odot}$ to $m_2 =
1.01$\,M$_{\odot}$. We  consider Kroupa as well as  unimodal IMFs, the
latter  with  $\Gamma  =  0.3$,  $1.35$ (i.e.   Salpeter)  and  $2.3$,
respectively.  Notice  that stars with M$<  0.15$\,M$_{\odot}$ are not
included in the synthesis, as  there are no Padova evolutionary tracks
available below this  mass limit.  As discussed by  CvD12, even in the
case  of a strongly  bottom-heavy population  (e.g.  $\Gamma  = 3.3$),
this approximation  should have little impact on  the results, because
of  the   negligible  contribution   of  stars  with   0.08  $<   M  <
0.15$\,M$_{\odot}$ to the SSP integrated light. Fig.~\ref{fig:toy_ssp}
compares one  of our simple SSPs,  with solar composition,  to a MILES
extended (MIUSCAT)  SSP, with similar  age, metallicity, and  the same
(Kroupa) IMF.  Despite of the  simple approach to create the synthetic
models, the agreement  with the MILES extended models  is fairly good,
with an rms of $\sim 2\%$ at $\lambda >4500$\,\AA, and $\sim 10 \%$ in
the bluer part of the  spectrum ($\lambda <4500$\,\AA).  We remind the
reader  that the simplified  SSPs presented  here are  used only  in a
relative sense,  i.e.  to compute differences in  the spectral indices
when varying  individual element abundances.  We also  notice that our
synthetic  SSPs are  based on  rather different  ingredients  than the
CvD12  ones. In particular,  we use  the {\tt  MARCS} and  {\tt PFANT}
codes  for the  atmospheric models  and stellar  spectra,  while CvD12
adopt     the     {\tt     ATLAS12}~\citep{Kurucz:70}     and     {\tt
  SYNTHE}~\citep{Kurucz:81}  codes, respectively.  CvD12  adopt atomic
and  molecular line  lists with  oscillator strengths  computed either
theoretically or from lab  measurements, while our oscillator strengths
have   been   calibrated   through   several   stellar   spectroscopic
studies. The synthetic stellar spectra in both models are created down
to a stellar mass  of 0.15\,$M_\odot$.  However, CvD12 extrapolate the
library down  to 0.08\,$M_\odot$, while we  completely disregard these
low-mass  stars. Finally,  we  use Padova  evolutionary tracks,  while
CvD12 use Dartmouth isochrones~\citep{Dotter:08}.

%code  (Kurucz 1970, 1993).  Our synthetic  stellar spectra are
%created  with PFANT,  while CvD12  use  the SYNTHE  program (Kurucz  &
%Avrett  1981).   CvD12  adopt   the  atomic  and  molecular  linelists
%available on Kurucz’s webpage.   Oscillator strengths, $gf$, in Kurucz
%linelists are  computed theoretically  or from lab  measuremets, while
%$gf$s  used in  our  models were  calibrated  through several  stellar
%spectroscopic studies.   The synthetic stellar spectra  in both models
%were  created down  to a  stellar mass  of 0.15  Msun.  However, CvD12
%extrapolate the  library down to 0.08  Msun, while in  our model these
%low mass  stars are disregarded.   We use Padova  evolutionary tracks,
%while CvD12 use Dartmouth (Dotter et al. 2008) isochrones.

\label{lastpage}


\begin{thebibliography}{}


\bibitem[\protect\citeauthoryear{Abazajian et al.}{2009}]{SDSS:DR7} 
Abazajian, K.~N., et al., 2009, ApJS, 182, 543

\bibitem[\protect\citeauthoryear{Adelman-McCarthy  et al.}{2008}]{SDSS:DR6}  
Adelman-McCarthy, J.K., Ag\" ueros, M.A., Allam, S.S., et al., 2008,
ApJS, 175, 297
%\appendix
%\bibitem[\protect\citeauthoryear{Adelman-McCarthy et al.}{2008}]{ade08} 
%Adelman-McCarthy, J.~K., et al. 2008, ApJS, 175, 297

%\bibitem[\protect\citeauthoryear{Annibali et al.}{2007}]{Annibali:07}
%Annibali, F., Bressan, A., Rampazzo, R., Zeilinger, W. W., Danese, L.,
%2007, A\&A, 463, 455

%\bibitem[\protect\citeauthoryear{Aringer et al.}{2009}]{ba09}
%Aringer et al.\ 2009, A\&A, 503, 913

%\bibitem[Balcells \& Peletier(1994)]{BaP:94} 
% Balcells, M., \& Peletier, R.F.  1994, AJ, 107, 135

%\bibitem[\protect\citeauthoryear{Auger et al.}{2010}]{Auger:10}
%Auger, M.W., Treu, T., Gavazzi, R., Bolton, A.S., Koopmans, L.V.E.,
%Marshall, P.J., 2010, ApJ, 721, 163

%\bibitem[\protect\citeauthoryear{Beers et al.}{1990}]{Beers:90}
%Beers, T.C., Flynn, K., Gebhardt, K. 1990, AJ, 100, 32

%\bibitem[Bell et al.(2003)]{bell03} Bell, E.~F., McIntosh, D.~H.,
%Katz, N., Weinberg, M.~D.\ 2003, ApJS, 149, 289

%\bibitem[\protect\citeauthoryear{Bergval et al.}{2010}]{BZC:10}
%Bergval, N., Zaricksson, E., Caldwell, B., 2010, MNRAS, 405, 2697

%\bibitem[\protect\citeauthoryear{Berlind et al.}{2006}]{Berlind:06} 
%Berlind, A.~A., et al., 2006, ApJS, 167, 1

\bibitem[\protect\citeauthoryear{Auger et al.}{2010}]{Auger:10} 
Auger, M.W., Treu, T., Bolton, A.S., Gavazzi, R., Koopmans, L.V.E.,
Marshall, P.J., Moustakas, L.A., Burles, S., 2010, ApJ, 724, 511

\bibitem[\protect\citeauthoryear{{Barbuy}, {Perrin}, {Katz}, {Coelho},
  {Cayrel}, {Spite} \& {Van't Veer-Menneret}}{{Barbuy}
  et~al.}{2003}]{Barbuy.etal:2003}
{Barbuy} B., {Perrin} M.-N., {Katz} D., {Coelho} P., {Cayrel} R.,
{Spite} M., {Van't Veer-Menneret} C., 2003, A\&A, 404, 661

\bibitem[\protect\citeauthoryear{Barnab\'e et al.}{2011}]{Barnabe:11} 
Barnab\'e, M., Czoske, O., Koopmans, L.V.E., Treu, T., Bolton, A.S.,
2011, MNRAS, 415, 2215

\bibitem[\protect\citeauthoryear{Bastian et al}{2006}]{Bastian:06} 
Bastian, N., Saglia, R.P., Goudfrooij, P., Kissler-Patig, M.,
Maraston, C., Schweizer, F., Zoccali, M., 2006, A\&A, 448, 881

\bibitem[\protect\citeauthoryear{Bastian et al.}{2010}]{Bastian:10}
Bastian, N., Covey, K.~R. \& Meyer, M.~R.  2010, ARA\&A, 48, 339


\bibitem[\protect\citeauthoryear{Bernardi et al.}{2003a}]{Bernardi:03} 
Bernardi, M., et al., 2003a, AJ, 125, 1817

\bibitem[\protect\citeauthoryear{Bernardi et al.}{2003b}]{BER03} 
Bernardi, M., Sheth, R.K., Annis, J. 2003b, AJ, 125, 1849

\bibitem[\protect\citeauthoryear{Bernardi et al.}{2005}]{Bernardi:05} 
Bernardi, M., Sheth, R.~K., Nichol, R.~C., Schneider, D.~P. \&
Brinkmann, J.  2005, AJ, 129, 61

%\bibitem[\protect\citeauthoryear{Bernardi et al.}{2006}]{BERN:06}
%Bernardi, M., Nichol, R.~C., Sheth, R.~K., Miller, C.~J., Brinkmann,
%J., 2006, AJ, 131, 1288 (B06)

\bibitem[\protect\citeauthoryear{Bertelli et al.}{2008}]{bertl08}
Bertelli, G. et al.\ 2008, A\&A, 484, 815


%\bibitem[Blanton et al.(2003a)]{BL03a} Blanton, M.R., Lin, H.,
%Lupton, R.H., Maley, F.M., Young, N., Zehavi, I., Loveday, J. 2003,
%AJ, 125, 2276

%\bibitem[Blanton et al.(2005)]{Blanton:05} Blanton, M.R., et  al.,
%  2005, AJ, 129, 2562

%\bibitem[Brinchmann \& Ellis(2000)]{bri00} Brinchmann, J., \& Ellis,
%R.~S.\ 2000, ApJL, 536, L77

%\bibitem[\protect\citeauthoryear{Brewer et al.}{2012}]{Brewer:12}
%Brewer, B.~J, et al., 2012, MNRAS, in press ($1201.1677$)

%\bibitem[\protect\citeauthoryear{Bruzual}{2007}]{Br07}
%Bruzual, G.,\ 2007, in IAU Symposium 241
%"Stellar populations as building blocks of galaxies", eds. A.
%Vazdekis and R. Peletier, Cambridge: Cambridge University Press, 125

\bibitem[\protect\citeauthoryear{Bruzual \& Charlot}{2003}]{BrC03}
Bruzual, G., \& Charlot, S. 2003, MNRAS, 344, 1000 (BC03)

%\bibitem[Bundy et al.(2005)]{bun05} Bundy, K., Ellis, R.~S., \&
%Conselice, C.~J.\ 2005, ApJ, 625, 621

%\bibitem[Calzetti et al.(2000)]{cal00} Calzetti, D., Armus, L.,
%Bohlin, R.~C., Kinney, A.~L., Koornneef, J., \& Storchi-Bergmann,
%T.\ 2000, ApJ, 533, 682

%\bibitem[Cantiello et al.(2005)]{CAN05} Cantiello, M. et al. 2005,
%ApJ, 634, 239

%\bibitem[\protect\citeauthoryear{Caon et al.}{1993}]{CCD93}
%Caon, N., Capaccioli, M., \& D'Onofrio, M.\ 1993, MNRAS, 265, 1013

\bibitem[\protect\citeauthoryear{Cappellari et al.}{2006}]{Cappellari:06} 
Cappellari, M., et al., 2006, MNRAS, 366, 1126

\bibitem[\protect\citeauthoryear{Cappellari et al.}{2012a}]{Cappellari:12a}
Cappellari, M., et al., 2012, Nature, 484, 485

\bibitem[\protect\citeauthoryear{Cappellari et al.}{2012b}]{Cappellari:12} 
Cappellari, M., et al., 2012, MNRAS, submitted (arXiv:1208.3522)

\bibitem[\protect\citeauthoryear{Cappellari et al.}{2012c}]{Cappellari:12c}
Cappellari, M., et al., 2012c, MNRAS, submitted (arXiv:1208.3523)

%\bibitem[\protect\citeauthoryear{Cardelli et al.}{1989}]{Cardelli}
%Cardelli, J.A., Clayton, G.C., and Mathis, J.S., 1989, ApJ, 345, 245

\bibitem[\protect\citeauthoryear{Cardelli, Clayton \& Mathis}{1989}]{Cardelli} 
Cardelli, J.~A., Clayton, G.~C., Mathis, J.~S., 1989, ApJ, 345, 245

\bibitem[\protect\citeauthoryear{Carter et al.}{1986}]{Carter:86}  
Carter, D., Visvanathan, N., Pickles, A.J., 1986, ApJ, 311, 637

\bibitem[\protect\citeauthoryear{Cayrel et~al.}{1991}]{Cayrel.etal:1991}
Cayrel, R., Perrin, M.-N., Barbuy, B., Buser, R., 1991, A\&A, 247, 108

%\bibitem[\protect\citeauthoryear{Carollo et al.}{1993}]{cmc93}
%Carollo C.~M., Danziger I.~J., Buson L., 1993, MNRAS, 265, 553

%\bibitem[\protect\citeauthoryear{Carlberg}{1984}]{carlberg84} 
%Carlberg, R.~G., 1984, ApJ, 286, 403

\bibitem[\protect\citeauthoryear{Chabrier}{2003}]{Chabrier03}
Chabrier, G., PASP, 115, 763

%\bibitem[\protect\citeauthoryear{Charlot \& Bruzual}{2007}]{CBr07}
%Charlot, S., \& Bruzual, G.\ 2007, unpublished models distributed on demand (CB07)

%\bibitem[\protect\citeauthoryear{Charlot \& Bruzual}{2013}]{CBr13}
%Charlot, S., \& Bruzual, G. 2013, in preparation (CB13)


\bibitem[\protect\citeauthoryear{Cid  Fernandes  et al.}{2005}]{CID05}
  Cid Fernandes, R., Mateus, A., Sodr\'e, L., Stasinska, G., Gomes,
J. M., 2005, MNRAS, 358, 363

%\bibitem[\protect\citeauthoryear{Coccato, Gerhard \& Arnaboldi}{2010}]{CGA:10}
%Coccato, L., Gerhard, O., Arnaboldi, M., 2010, MNRAS, 407, L26 (CGA10)

\bibitem[\protect\citeauthoryear{Capaccioli, Caon \& D'Onofrio}{1992}]{capaccioli1992} 
Capaccioli, M., Caon, N., \& D'Onofrio, M. 1992, MNRAS, 259, 323

\bibitem[\protect\citeauthoryear{Cenarro et al.}{2001}]{Cenarro2001} 
Cenarro, A. J., Cardiel, N., Gorgas, J., Peletier, R. F., Vazdekis, A.,
Prada, F., 2001, MNRAS, 326, 959

\bibitem[Cenarro et al. (2003)]{Cenarro:03}  
Cenarro, A.~J., Gorgas, J., Vazdekis, A., Cardiel, N., Peletier,
R.~F., 2003, MNRAS, 339, L12

\bibitem[\protect\citeauthoryear{Cervantes et al.}{2007}]{Cervantes07} 
Cervantes, J. L., Coelho, P., Barbuy, B., and Vazdekis, A., 2007,
Proceedings IAU Symposium No. 241 'Stellar Populations as Building
Blocks of Galaxies', A. Vazdekis and R.F. Peletier, eds, 167

\bibitem[\protect\citeauthoryear{Cervantes \& Vazdekis}{2009}]{CV09} 
Cervantes, J. L., Vazdekis, A., 2009, MNRAS, 392, 691

\bibitem[\protect\citeauthoryear{Coelho et~al.}{2005}]{Coelho.etal:2005}
Coelho, P., Barbuy, B., Mel{\'e}ndez, J., Schiavon, R.~P., Castilho, B.~V.,  2005, A\&A, 443, 735

\bibitem[\protect\citeauthoryear{Coelho et al.}{2007}]{Coelho:07}
Coelho, P., Bruzual, G., Charlot, S., Weiss, A., Barbuy, B., Ferguson,
J.W., 2007, MNRAS, 382, 498

\bibitem[\protect\citeauthoryear{Cohen}{1978}]{Cohen:78} 
Cohen, J. G., 1978, ApJ, 221, 788

\bibitem[\protect\citeauthoryear{Conroy \& van Dokkum}{2012a}]{CvD12a} 
Conroy, C., van Dokkum, P., 2012a, ApJ, 747, 69

\bibitem[\protect\citeauthoryear{Conroy \& van Dokkum}{2012b}]{CvD12b} 
Conroy, C., van Dokkum, P., 2012b, ApJ, 760, 71

%\bibitem[Casali et al.(2007)]{Casali:07} Casali, M., Adamson, A.,
%Alves de Oliveira, C., et al. 2007, A\&A, 467, 777


%\bibitem[Cascone et al.(200[]{})]{CGP:02} Cascone, E., Grado, A.,
%Pavlov, M., Capasso, G. 2002, SPIE, 4848, 328

%\bibitem[Cervantes et al.(2007)]{Cervantes07} Cervantes, J.L.,
%Coelho, P., Barbuy, B., and Vazdekis, A. 2007, Proceedings IAU
%Symposium No. 241 'Stellar Populations as Building Blocks of
%Galaxies', A. Vazdekis and R.F. Peletier, eds, 167

%\bibitem[Chabrier(2003)]{cha03} Chabrier, G.\ 2003, PASP, 115, 763


%\bibitem[Choi et al.(2007)]{Choi:07} Choi, Y.Y., Park, C., Vogeley,
%M.S. 2007, ApJ, 658, 884

%\bibitem[Cid Fernandes et al.(2005)]{CID05} Cid Fernandes, R., Gonz\'
%alez Delgado, R.M., Storchi-Bergmann, T., Martins, L.P., Schmitt,
%H. 2005, MNRAS, 356, 270

%\bibitem[\protect\citeauthoryear{Clemens et al.}{2009}]{Clemens:09}
%Clemens, M.S., et al. 2009, MNRAS, 392, 35

%\bibitem[Coelho et al.(2005)]{Coelho05} Coelho, P., Barbuy, B.,
%Melendez, J., Schiavon, R.P., and Castilho, B.V. 2005, A\&A, 443, 735

%\bibitem[Cole et al.(2001)]{Cole01} Cole, S., et al.\ 2001, MNRAS,
%326, 255

%\bibitem[Coppola, La Barbera, Capaccioli(2009)]{CLB09} Coppola, G.,
%La Barbera, F., Capaccioli, M. 2009, PASP, 121, 437

%\bibitem[\protect\citeauthoryear{Cooper et al.}{2010}]{Cooper:10}
%Cooper, M.C., Gallazzi, A., Newman, J.A., Yan, R., 2010, MNRAS, 402,
%1942

%\bibitem[\protect\citeauthoryear{Daddi et al.}{2005}]{daddi05} 
%Daddi E., et al., 2005, ApJ, 626, 680 

%\bibitem[\protect\citeauthoryear{Davies et al.}{1987}]{davies87} 
%Davies, R.~L., Burstein, D., Dressler, A., Faber, S.~M., Lynden-Bell,
%D., Terlevich, R.~J., Wegner, G., 1984, ApJS, 64, 581

%\bibitem[\protect\citeauthoryear{Davies et al.}{1993}]{davies93} 
%Davies, R.~L., Sadler, E.~M., Peletier, R.~F., 1993, MNRAS, 262, 650

%\bibitem[Cowie et al.(1996)]{Cowie:96} Cowie, L.L., Songaila, A., Hu,
%E.M., Cohen, J.G. 1996, AJ, 112, 839

%\bibitem[De Lucia et al.(2006)]{deLucia:06} De Lucia, G., Springel,
%V.,, White, S.D.M., Croton, D., Kauffmann, G. 2006, MNRAS, 366, 499

%\bibitem[de Propris et al.(2005)]{dP05} de Propris, R., et al. 2005,
%MNRAS, 357, 590

%\bibitem[\protect\citeauthoryear{de Jong}{2008}]{deJong:08}
%de Jong, R.S., 2008, MNRAS, 388, 1521

%\bibitem[\protect\citeauthoryear{de La Rosa et al.}{2007}]{deLaRosa:07}
%de La Rosa, I.G., de Carvalho, R.~R., Vazdekis, A., Barbuy, B., 2007,
% AJ, 133, 330

%\bibitem[\protect\citeauthoryear{Dekel et~al.}{2009}]{dekel09}
%Dekel, A., Sari, R., Ceverino, D., 2009, ApJ, 703, 785

%\bibitem[\protect\citeauthoryear{de Vaucouleurs}{1961}]{devac61} 
%De~Vaucoulerus, G., 1961, ApJS, 5, 233

\bibitem[de la Rosa et al.(2012)]{delaRosa:12} de la Rosa, I.~G., la Barbera, F., Ferreras, I., de Carvalho, R.~R. 2012, MNRAS, 418, L74

\bibitem[Delisle \& Hardy(1992)]{Delisle:92} Delisle,  S.,  Hardy, E., 1992, AJ, 103, 711

\bibitem[Diaz, Terlevich, Terlevich(1989)]{Diaz:89} Diaz, A.I., Terlevich, E., Terlevich, R., 1989, MNRAS, 239, 325

\bibitem[Dotter et al.(2008)]{Dotter:08} Dotter, A., Chaboyer, B., Jevremovi\'c, D., Kostov, V., Baron, E., Ferguson, J.~W., 2008, ApJ, 178, 89


\bibitem[\protect\citeauthoryear{Dutton, Mendel \& Simard}{2012}]{Dutton:12}
Dutton, A. A., Mendel, J. T., Simard, L., 2012, MNRAS, 422, 33

%\bibitem[\protect\citeauthoryear{Eggen et al.}{1962}]{els62} 
%Eggen, O.~J., Lynden-Bell, D., Sandage, A.~R., 1962, ApJ, 136, 748

%\bibitem[D'Onofrio et al.(2008)]{DOF08} D'Onofrio, M., et al. 2008,
%ApJ, 685, 875

%\bibitem[\protect\citeauthoryear{Duc \& Renaud}{2011}]{Duc:11} 
%Duc P.-A., Renaud F., 2011, arXiv, arXiv:1112.1922

%\bibitem[\protect\citeauthoryear{Faucher-Gigu{\`e}re et al.}{2011}]{faucher11}
%Faucher-Gigu{\`e}re C.-A., Kere{\v s} D., Ma C.-P., 2011, MNRAS, 417,
%2982


%\bibitem[\protect\citeauthoryear{Ferreras et al.}{2009}]{ferr09}
%Ferreras, I., Lisker, T., Pasquali, A., Kaviraj, S., 2009, MNRAS, 395,
%554

\bibitem[\protect\citeauthoryear{Faber \& French}{1980}]{FaberFrench:80} 
Faber, S.M., French, H.B., 1980, ApJ, 235, 405

\bibitem[\protect\citeauthoryear{Falc\'on-Barroso et al.}{2011}]{Jesus:11} 
Falc\'on-Barroso, J., S\'anchez-Bl\'azquez, P., Vazdekis, A.,
Ricciardelli, E., Cardiel, N., Cenarro, A. J., Gorgas, J., Peletier,
R. F., 2011, A\&A, 532, 95

\bibitem[\protect\citeauthoryear{Ferreras et al.}{2005}]{FSW:05} 
Ferreras, I., Saha, P., Williams, L.~L.~R., 2005, ApJ, 623, 5

\bibitem[\protect\citeauthoryear{Ferreras et al.}{2008}]{FSB:08}
Ferreras, I., Saha, P., Burles, S., 2008, MNRAS, 383, 857

\bibitem[\protect\citeauthoryear{Ferreras et al.}{2010}]{ECross:10}
Ferreras, I., Saha, P., Leier, D., Courbin, F., Falco, E.~E. 2010,
MNRAS, 409, L30

\bibitem[\protect\citeauthoryear{Ferreras  et al.}{2013}]{Ferreras:13}
  Ferreras, I.,  La Barbera, F., de  la Rosa, I.~G.,  Vazdekis, A., de
  Carvalho,  R.~R.,  Falc\'on-Barroso,  J.,  Ricciardelli,  E.,  2013,
  MNRAS, 429, L15 (FLD13)


\bibitem[\protect\citeauthoryear{Ferr\'e-Mateu et al.}{2013}]{FerreMateu:13}
Ferr\'e-Mateu, A., Vazdekis, A., de la Rosa, I.~G., 2013, MNRAS, in
press (arXiv:1301.7066)

%\bibitem[\protect\citeauthoryear{F{\"o}rster Schreiber et al.}{2009}]{fsb09}
%F{\"o}rster Schreiber N.~M., et al., 2009, ApJ, 706, 1364

\bibitem[\protect\citeauthoryear{Gallazzi et al.}{2006}]{GALL:06}
Gallazzi, A., et al. 2006, MNRAS, 370, 1106

%\bibitem[Gallazzi et al.(2005)]{Gall:05} Gallazzi, A., Charlot, S., Brinchmann, J., White, S.D.M., Tremonti, C.A., 2005, MNRAS, 362, 41

%\bibitem[\protect\citeauthoryear{Gargiulo et al.}{2011}]{Gargiulo:11}
%Gargiulo, A., Saracco, P., Longhetti, M., 2011, MNRAS, 412, 1804

%\bibitem[Girardi et al.(2010)]{gir10} Girardi, L\'eo, et al., 2010, ApJ, 724, 1030

%\bibitem[G\'{o}mez et al.(2003)]{GOMEZ03} G\' omez, P.L., Nichol,
%R.C., Miller, C.J., et al. 2003, ApJ, 584, 210

\bibitem[\protect\citeauthoryear{Gonz\'alez}{1993}]{GON93} 
Gonz\'alez, J.~J. 1993, Ph.D. thesis, Univ. California

\bibitem[\protect\citeauthoryear{Goudfrooij \& Kruijssen}{2013}]{GK:13} 
Goudfrooij, P., Kruijssen, J.M.D., 2013, ApJ, 762, 107

%\bibitem[Goudfrooij et al.(1994)]{GOU94} Goudfrooij, P.  et al. 1994,
%A\&AS, 104, 179

%\bibitem[\protect\citeauthoryear{Gonz\'alez-P\'erez et al.}{2011}]{GP11}
%Gonz\'alez-P\'erez, V., Castander, F.J., Kauffmann, G., 2011, MNRAS,
%411, 1151

%\bibitem[\protect\citeauthoryear{Gorgas et al.}{1990}]{gorgas90}
%Gorgas, J., Efstathiou, G., Arag\'on-Salamanca, A., 1990, MNRAS, 245, 217

%\bibitem[\protect\citeauthoryear{Goudfrooij \& de Jong}{1995}]{GJON95}
%Goudfrooij, P.  \& de Jong, T. 1995, A\&AS, 298, 784

\bibitem[\protect\citeauthoryear{Graham \& Guzm\'an}{2003}]{graham&guzman2003} 
Graham, A.~W., Guzm{\'a}n, R., 2003, AJ, 125, 2936


\bibitem[\protect\citeauthoryear{Gustafsson et~al.}{2008}]{Gustafsson.etal:2008}
Gustafsson, B., Edvardsson, B., Eriksson, K., J{\o}rgensen, U.~G.,
 Nordlund, \AA.,  2008, A\&A, 486, 951

\bibitem[\protect\citeauthoryear{Hardy \& Couture}{1988}]{Hardy:88} 
Hardy, E., Couture, J., 1988, ApJ, 325, 29 

\bibitem[\protect\citeauthoryear{Hopkins}{2012}]{Hopkins:12} 
Hopkins, P., 2013, arXiv:1204.2835

%\bibitem[\protect\citeauthoryear{Greene et al.}{2012}]{Greene:12}
%Greene, J.~E., Murphy, J.~D., Comerford, J.~M., Gebhardt, K., Adams,
%J.~J., 2012, ApJ, in press ($arXiv1202.4464G$)

%\bibitem[\protect\citeauthoryear{Guo et al.}{2011}]{GUo:11}
%Guo, Y., et al., 2011, ApJ, 735, 18
%\bibitem[Guo  et  al.(2009)]{Guo:09} Guo,  Y.,  McIntosh,  D. H.,  Mo,
%  H.  J., Katz,  N., van  den Bosch,  F. C.,  Weinberg,  M., Weinmann,
%  S. M., Pasquali, A., Yang, X., 2009, MNRAS, 398, 1129

%\bibitem[Graves et al.(2009)]{Graves:09} Graves, G.J., Faber, S.M.,
%\& Schiavon, R.P. 2009, ApJ, 698, 1590

%\bibitem[Hambly et al.(2009)]{Hambly:09} Hambly, N.C., Collins, R.S.,
%Cross, N.J.G., et al. 2008, MNRAS, 384, 637

%\bibitem[Hewett et al.(2006]{Hewett:06} Hewett, P.C., Warren, S.J.,
%Leggett, S.K., Hodgkin, S.T. 2006, MNRAS, 367, 454

%\bibitem[\protect\citeauthoryear{Hinkley \& Im}{2001}]{HI01} 
%Hinkley S., Im M., 2001, ApJ, 560, L41

%\bibitem[Hogg et al.(2004)]{Hogg04} Hogg, D.W., Blanton, M.R.,
%Brinchmann, J., et al. 2004, ApJ, 601, 29

%\bibitem[\protect\citeauthoryear{Hyde \& Bernardi}{2009}]{HydeBernardi:09}
%Hyde, J.B., Bernardi, M., 2009, MNRAS, 394, 1978

%\bibitem[Ilbert et al.(2009)]{ilb09} Ilbert, O., et al.\ 2009, ApJ,
%690, 1236

%\bibitem[J\o rgensen et al.(1996)]{JFK96} J\o rgensen, I., Franx, M.,
%\& Kjaergaard, P. 1996, MNRAS, 280, 167

%\bibitem[J\o rgensen(1999)]{JORG:99} J\o rgensen, I. 1999, MNRAS,
%306, 607

\bibitem[\protect\citeauthoryear{Johansson et al.}{2012}]{JTM12} 
Johansson, J., Thomas, D., Maraston, C., 2012, MNRAS, 421, 1908

\bibitem[\protect\citeauthoryear{Kroupa}{2001}]{Kroupa01} 
Kroupa, P., 2001, MNRAS, 322, 231

\bibitem[\protect\citeauthoryear{Kroupa \& Weidner}{2003}]{igimf} 
Kroupa, P., Weidner, C. 2003, ApJ, 598, 1076

\bibitem[\protect\citeauthoryear{Kurucz}{1970}]{Kurucz:70} 
Kurucz, R.~L., 1970, SAOR, 309

\bibitem[\protect\citeauthoryear{Kurucz \& Avrett}{1981}]{Kurucz:81} 
Kurucz, R.~L., Avrett, E.~H., 1981, SAOR, 391

\bibitem[\protect\citeauthoryear{Kurucz}{1992}]{Kurucz:92} 
Kurucz, R.L., 1992, RMxAA, 23, 45

%\bibitem[Kauffmann \& Charlot(1998)]{kau98} Kauffmann, G., \&
%Charlot, S.\ 1998, MNRAS, 297, L23
%\bibitem[Ilbert et al.(2006)]{Ilbert:06} Ilbert, O., et al., 2006, A\&A, 457, 841

%\bibitem[\protect\citeauthoryear{Kere\v{s} et~al.}{2005}]{keres05} 
%Kere\v{s}, D., Katz, N., Weinberg, D.~H., \& Dav{\'e}, R. 2005, MNRAS,
%363, 2

%\bibitem[\protect\citeauthoryear{Kobayashi}{2004}]{koba04} 
%Kobayashi, C., 2004, MNRAS, 347, 740

%\bibitem[\protect\citeauthoryear{Komatsu et al.}{2011}]{wmap7} 
%Komatsu, E., et al. 2011, ApJS, 192, 18

%\bibitem[\protect\citeauthoryear{Kormendy et al.}{2009}]{Kormendy:09} 
%Kormendy, J., Fisher, D.B., Cornell, M.E., Bender, R., 2009, ApJS,
%182, 216

%\bibitem[\protect\citeauthoryear{Kriek et al.}{2010}]{kriek10}
%Kriek, M. et al.\ 2010, ApJ, 722L, 64

%\bibitem[\protect\citeauthoryear{Kuntschner et al.}{2010}]{Kuntschner:10}
%Kuntschner, H., et al. 2010, MNRAS, 408, 97

%\bibitem[\protect\citeauthoryear{Kuntschner et al.}{2002}]{Kunt:02}
%Kuntschner, H., Smith, R.J., Colless, M., Davies, R.L., Kaldare, R.,
%Vazdekis, A. 2002, MNRAS, 337, 172

%\bibitem[La Barbera et al.(2003)]{LBM03} La Barbera, F., et al. 2003,
%A\&A, 409, 21

%\bibitem[La Barbera et al.(2005)]{LdC:05} La Barbera, F., de
%Carvalho, R.~R., Gal, R.~R., Busarello, G., Merluzzi, P., Capaccioli,
%M., Djorgovski, S.G. 2005, ApJ, 626, 19

%\bibitem[\protect\citeauthoryear{La Barbera et al.}{2008}]{LBdC08}
%La Barbera, F., et al. 2008, PASP, 120, 681

%\bibitem[La Barbera et al.(2008b)]{LBM08} La Barbera, F., Busarello,
%G., Merluzzi, P., de la Rosa, I.G., Coppola, G., Haines, C.P. 2008,
%ApJ, 689, 913

%\bibitem[\protect\citeauthoryear{La Barbera \& de Carvalho}{2009}]{LdC:09}
%La Barbera, F. \& de Carvalho, R.~R. 2009, ApJ, 699, 76L

\bibitem[\protect\citeauthoryear{La Barbera et al.}{2010a}]{SpiderI} 
La Barbera, F., de Carvalho, R.~R., de la Rosa, I.G., Lopes, P.A.A.,
Kohl-Moreira, J.L., Capelato, H.V., 2010a, MNRAS, 408, 1313 (Paper I)

\bibitem[\protect\citeauthoryear{La Barbera et al.}{2010b}]{SpiderII} 
La Barbera, F., de Carvalho, R.R., de la Rosa, I.G., Lopes, P.A.A.,
2010, MNRAS, 408, 1335 (Paper II)

%\bibitem[La  Barbera et al.(2010b)]{PaperIII}
%La Barbera, F., Lopes, P.A.A., de Carvalho, R.~R., de La Rosa, I.G.,
%Berlind, A.A., 2010b, MNRAS, 408, 1361 (Paper III)

\bibitem[\protect\citeauthoryear{La Barbera et al.}{2010c}]{PaperIV} 
La Barbera, F., de Carvalho, R.~R., de la Rosa, I.G., Gal, R.~R.,
Swindle, R., Lopes, P.A.A., 2010c, AJ, 140, 1528 (Paper IV)

%\bibitem[\protect\citeauthoryear{La Barbera et al.}{2011}]{LF:11}
%La Barbera, F., Ferreras, I., de Carvalho, R.~R., Lopes, P.A.A.,
%Pasquali, A., de la Rosa, I.G., de Lucia, G., 2011, ApJ, 740, 41

%\bibitem[\protect\citeauthoryear{Lan\c con \& Mouhcine}{2002}]{lm02}
%Lan\c con, A. \& Mouhcine, M. \ 2002, A\&A, 393, 167

%\bibitem[\protect\citeauthoryear{Lanz \& Hubeny}{2003}]{lh03}
%Lanz, T. \& Hubeny, I.\ 2003, ApJS, 146, 417

%\bibitem[\protect\citeauthoryear{Lanz \& Hubeny}{2007}]{lh07}
%Lanz, T. \& Hubeny, I.\  2007, ApJS, 169, 83

%\bibitem[\protect\citeauthoryear{Le Borgne et al.}{2003}]{jfl03}
%Le Borgne, J.-F. et al.\ 2003, A\&A, 402, 433

%\bibitem[\protect\citeauthoryear{Larson}{1974}]{lar74} 
%Larson, R.~B., 1974, MNRAS, 166, 585

%\bibitem[\protect\citeauthoryear{Larson}{1975}]{lar75} 
%Larson, R.~B., 1975, MNRAS, 173, 671

\bibitem[\protect\citeauthoryear{Larson}{2005}]{Larson:05} 
Larson, R.~B., 2005, MNRAS, 359, 211

\bibitem[\protect\citeauthoryear{Lawrence et al.}{2007}]{Law07}
Lawrence, A., Warren, S.J., Almaini, O., et al. 2007, MNRAS, 379, 1599

%\bibitem[\protect\citeauthoryear{Lopes et al.}{2009a}]{lop09a} 
%Lopes, P.A.A., de Carvalho, R.~R., Kohl-Moreira, J.L., Jones, C., 2009,
%MNRAS, 392, 135

%\bibitem[\protect\citeauthoryear{McClure}{1969}]{McClure69}
%McClure, R,~D., 1969, AJ, 74, 50

%\bibitem[\protect\citeauthoryear{Marigo \& Girardi}{2007}]{MG07}
% Marigo, P., \& Girardi, L. 2007, A\&A, 469, 239

%\bibitem[\protect\citeauthoryear{Marigo et al.}{2008}]{MG08}
%Marigo, P. et al.\ 2008, A\&A, 482, 883

%\bibitem[Menanteau et al.(2001)]{MAE01} Menanteau, F., et al. 2001,
%MNRAS, 322, 1

%\bibitem[Mehlert et al.(2000)]{MEH00} Mehlert, D. et al. 2000, A\&AS,
%141, 449

%\bibitem[\protect\citeauthoryear{Mehlert et al.}{2003}]{mehlert03}
%Mehlert D., Thomas D., Saglia R.~P., Bender R., Wegner G., 2003, A\&A,
%407, 423

%\bibitem[\protect\citeauthoryear{Melbourne et al.}{2012}]{melb12}
%Melbourne, J. et al.\ 2012, ApJ, 748, 47

%\bibitem[Menanteau et al.(2004)]{MEN04} Menanteau, F., et al. 2004,
%ApJ, 612, 202

%\bibitem[\protect\citeauthoryear{Michard}{2005}]{MIC:05}
%Michard, R., 2005, A\& A, 441, 451

%\bibitem[Michard(1999)]{MIC99} Michard, R. 1999, A\&A, 137, 245

%\bibitem[\protect\citeauthoryear{Naab et al.}{2009}]{naab09} 
%Naab T., Johansson P.~H., Ostriker J.~P., 2009, ApJ, 699, L178

%\bibitem[Nelan et al.(2005)]{Nelan:05} Nelan, J.E., et al. 2005, ApJ,
%632, 137

%\bibitem[\protect\citeauthoryear{Oser et al.}{2010}]{oser10}
%Oser L., Ostriker J.~P., Naab T., Johansson P.~H., Burkert A., 2010,
%ApJ, 725, 2312

%\bibitem[\protect\citeauthoryear{Pasquali et al.}{2010}]{Pasquali:10}
%Pasquali, A., Gallazzi, A., Fontanot, F., van den Bosch, F.C., 
%De Lucia, G., Mo, H.J., Yang, X., 2010, MNRAS, 407, 937

\bibitem[\protect\citeauthoryear{Peletier et al.}{1990}]{PVJ90}
Peletier, R.F. et al. 1990, A\&A, 233, 62

%\bibitem[\protect\citeauthoryear{Pipino et al.}{2008}]{pip08} 
%Pipino, A., D\'Ercole, A., Matteucci, F., 2008, A\& A, 484, 679

%\bibitem[\protect\citeauthoryear{Pipino et al.}{2010}]{pip10} 
%Pipino, A., D\'Ercole, A., Chiappini, C., Matteucci, F.,
%2010, MNRAS, 407, 1347

%%\bibitem[Pipino et al.(2010)]{Pip:10} Pipino, A., D'Ercole, A.,
%%Chiappini, C., Matteucci, F., MNRAS, in press ($arXiv:1005.2154v1$)

%\bibitem[Poggianti et al.(2001a)]{PBM:01} Poggianti, B.M., Bridges,
%T.J., Mobasher, B., et al. 2001, ApJ, 562, 689

%\bibitem[Poggianti et al.(2001b)]{PBC:01} Poggianti, B.M., Bridges,
%T.J., Carter, D., et al. 2001, ApJ, 563, 118

\bibitem[\protect\citeauthoryear{Prochaska, Rose \& Schiavon}{2005}]{PRS:05}
Prochaska, L. C., Rose, J. A., Schiavon, R. P., 2005, AJ, 130, 2666

%\bibitem[\protect\citeauthoryear{Rayner et al.}{2009}]{jr09}
%Rayner et al.\ 2009, ApJS, 185, 289

\bibitem[\protect\citeauthoryear{Ricciardelli et al.}{2012}]{Ricciardelli:12} 
Ricciardelli, E., Vazdekis, A., Cenarro, A. J., Falc\'on-Barroso, J.,
2012, MNRAS, 424, 172 (MIUSCAT-II)

%\bibitem[\protect\citeauthoryear{Roche, Bernardi, Hyde}{2010}]{RBH:10}
%Roche, N., Bernardi, M., Hyde, J., 2010, MNRAS, 407, 1231

%\bibitem[\protect\citeauthoryear{Roediger et al.}{2011}]{Roediger:11}
%Roediger, J.C., Courteau, S., MacArthur, L.A., McDonald, M., 2011,
%MNRAS, 416, 1996

%\bibitem[\protect\citeauthoryear{Rogers et al.}{2007}]{br07} 
%Rogers B., Ferreras I., Lahav O., Bernardi M., Kaviraj S., Yi S.~K.,
%2007, MNRAS, 382, 750

\bibitem[\protect\citeauthoryear{Rogers et al.}{2010}]{br10} 
Rogers B., Ferreras I., Peletier, R., Silk, J., 2010, MNRAS, 402, 447

%\bibitem[Roche, Bernardi, and Hyde(2010)]{RBH:10} Roche, N.,
%Bernardi, M., Hyde, J. 2010, MNRAS, submitted (arXiv:0911.0044)

%\bibitem[\protect\citeauthoryear{Saglia et al.}{2000}]{SMG00}
%Saglia R.~P., Maraston C., Greggio L., Bender R., Ziegler B., 2000,
%A\&A, 360, 911

%\bibitem[Salim et al.(2005)]{sal05} Salim, S., et al.\ 2005, ApJL,
%619, L39

%\bibitem[Salim et al.(2007)]{sal07} Salim, S., et al.\ 2007, ApJS,
%173, 267

\bibitem[\protect\citeauthoryear{Salpeter}{1955}]{salp:55} 
Salpeter, E.~E. 1955, ApJ, 121, 161

\bibitem[\protect\citeauthoryear{S\'anchez-Bl\'azquez et al.}{2006}]{ps06} 
S\'anchez-Bl\'azquez, P. et al.\ 2006, MNRAS, 371, 703

\bibitem[\protect\citeauthoryear{Schiavon et al.}{1997a}]{SchiavonNaD:97} 
Schiavon, R., Barbuy, B., Rossi, S. C. F., Milone, A., 1997, ApJ, 479,
902

%\bibitem[Schiavon et al. (1997b)]{SchiavonFeH:97} Schiavon, R.,
%Barbuy, B. \& Singh, P.~D. 1997, ApJ, 484, 499

\bibitem[\protect\citeauthoryear{Schiavon et al.}{1997b}]{SchiavonFeH:97} 
Schiavon, R., Barbuy, B., Rossi, S. C. F., Milone, A., 1997, ApJ, 479,
902

\bibitem[Schiavon et al.(2002)]{Schiavon:02} Schiavon, R.P., Faber, S.M., Rose, J.~A., Castilho, B.~V., 2002, ApJ, 580, 873

\bibitem[\protect\citeauthoryear{Schiavon, Caldwell \& Rose}{2004}]{SCR:04}
Schiavon, R., Caldwell, N., Rose, J. A., 2004, AJ, 127, 1513

\bibitem[\protect\citeauthoryear{Serven et al.}{2005}]{Serven05} 
Serven, J., Worthey, G., Briley, M.~M., 2005, ApJ, 627, 754

\bibitem[\protect\citeauthoryear{Serven \& Worthey}{2007}]{SW:07} 
Serven, J., Worthey, G., 2010, AJ, 140, 152

\bibitem[\protect\citeauthoryear{Serra \& Trager}{2007}]{SerraTrager:07} 
Serra, P., Trager, S. C., 2007, MNRAS, 374, 769

\bibitem[\protect\citeauthoryear{Smith, Lucey \& Carter}{2012}]{Smith:12} 
Smith, R. J., Lucey, J. R., Carter, D., 2012, MNRAS, 426, 2994

%\bibitem[\protect\citeauthoryear{S\'anchez-Bl\'azquez et al.}{2006}]{PSB:06}
%S\'anchez-Bl\'azquez, P., Gorgas, J. \& Cardiel, N., 2006, A\& A, 457, 823


%\bibitem[\protect\citeauthoryear{Savoy et al.}{2009}]{SWF:09}
%Savoy, J., Welch, G.A., Fich, M., 2009, ApJ, 706, 21

%\bibitem[Schlegel, Finkbeiner and Davis(1998)]{SFD98} Schlegel, D.,
%Finkbeiner, D.P., \& Davis, M. 1998, ApJ, 500, 525 (SFD98)

%\bibitem[\protect\citeauthoryear{Silva \& Elston}{1994}]{SE94}
%Silva D.~R., Elston R., 1994, ApJ, 428, 511 

%\bibitem[\protect\citeauthoryear{Silva \& Wise}{1996}]{SW:96}
%Silva, D.R., \& Wise, M.W. 1996, ApJ, 457, 15

%\bibitem[\protect\citeauthoryear{Spolaor et al.}{2010}]{Spolaor:10}
%Spolaor, M., Kobayashi, C., Forbes, D.A., Couch, W.J., Hau, G.K.T.,
%2010, MNRAS, 408, 272

%\bibitem[Sorrentino et al.(2006)]{SAR06} Sorrentino, G.,
%Antonuccio-Delogu, \& V., Rifatto, A.  2006, A\&A, 460, 673

\bibitem[\protect\citeauthoryear{Spiniello et al.}{2012}]{Spiniello:12} 
Spiniello, C., Trager, S. C., Koopmans, L. V. E., Chen, Y. P., 2012, ApJ,
753, 32

\bibitem[\protect\citeauthoryear{Spinrad}{1962}]{Spinrad:62} 
Spinrad, H., 1962, ApJ, 135, 715 

%\bibitem[\protect\citeauthoryear{Steidel et al.}{2010}]{steidel10} 
%Steidel C.~C., Erb D.~K., Shapley A.~E., Pettini M., Reddy N.,
%Bogosavljevi{\'c} M., Rudie G.~C., Rakic O., 2010, ApJ, 717, 289

%\bibitem[\protect\citeauthoryear{Suh et al.}{2010}]{SUH:10} 
%Suh, H., et al. 2010, ApJS, 187, 374

\bibitem[\protect\citeauthoryear{Swindle et al.}{2011}]{Swindle} 
Swindle, R., Gal, R. R., La Barbera, F., de Carvalho, R. R., 2011, AJ,
142, 118S (Paper V)

%%\bibitem[Swindle et al.(2010)]{swi10} Swindle, R., Gal, R.~R., La
%%Barbera, F. \& de Carvalho, R.~R. 2010, {\em in prep}

%\bibitem[\protect\citeauthoryear{Tal \& van Dokkum}{2011}]{TalvanDokkum:11}
%Tal, T., van Dokkum, P.~G., 2011, ApJ, 731, 89

%\bibitem[Tamura et al.(2000)]{TKA00} Tamura, N., Kobayashi, C.,
%Arimoto, N., Kodama, T., Ohta, K. 2000, AJ, 119, 2134

%\bibitem[Tamura \& Ohta(2000)]{TaO00} Tamura, N., Ohta, K. 2000, AJ,
%120, 533

%\bibitem[Tamura \& Ohta(2003)]{TaO03} Tamura, N., Ohta, K. 2003, AJ,
%126, 596

%\bibitem[Tamura \& Ohta(2004)]{TaO05} Tamura, N., Ohta, K. 2004,
%MNRAS, 355, 617

%\bibitem[Taylor et al.(2010)]{TFG:10} Taylor, E.N., Franx, M., Glazebrook, K., Brinchmann, J.,% van der Wel, A., van Dokkum, P.G., 2010, ApJ, 720, 723 (TFG10)

%\bibitem[Terlevich \& Forbes(2002)]{TeF:02} Terlevich, A.I., Forbes,
%D.A. 2002, MNRAS, 330, 547

\bibitem[\protect\citeauthoryear{Thomas et al.}{2003a}]{TMB:03} Thomas D., Maraston C., Bender R., 2003, MNRAS, 339, 897

\bibitem[\protect\citeauthoryear{Thomas et al.}{2003b}]{TMB:03b} Thomas D., Maraston C., Bender R., 2003, MNRAS, 343, 279

\bibitem[\protect\citeauthoryear{Thomas et al.}{2004}]{Thomas:04} 
Thomas D., Maraston C., Korn A., 2004, MNRAS, 351, L19

\bibitem[\protect\citeauthoryear{Thomas et al.}{2005}]{Thomas:05}
Thomas, D., et al. 2005, ApJ, 621, 673

\bibitem[\protect\citeauthoryear{Thomas \& Davies}{2006}]{ThomasDavies:06} 
Thomas D., Davies R.L., 2006, MNRAS, 366, 510

\bibitem[\protect\citeauthoryear{Thomas, Johansson \& Maraston}{2011}]{TJM11} 
Thomas, D., Johansson, J., Maraston, C., 2011, MNRAS, 412, 2199

\bibitem[\protect\citeauthoryear{Thomas, Maraston \& Johansson}{2011}]{TMJ11} 
Thomas, D., Maraston, C., Johansson, J., 2011, MNRAS, 412, 2183

\bibitem[\protect\citeauthoryear{Thomas et al.}{2011}]{Thomas:11} 
Thomas, J., et al., 2011, MNRAS, 415, 545

\bibitem[\protect\citeauthoryear{Trager et al.}{1998}]{Trager98} 
Trager, S.~C., Worthey, G., Faber, S.~M., Burstein, D., 
Gonz\'alez, J.~J., 1998, ApJS, 116, 1

%\bibitem[Tinsley(1973)]{Tinsley73} Tinsley, B.M. 1973, ApJ 186, 35

%\bibitem[\protect\citeauthoryear{Tortora et al.}{2010}]{Tor:10}
%Tortora, C., Napolitano, N.R., Cardone, V.F., Capaccioli, M., Jetzer,
%P., Molinaro, R. 2010, MNRAS, in press ($arXiv1004.4896T$)

\bibitem[\protect\citeauthoryear{Tortora  et al.}{2012}]{Tortora:12}  
Tortora, C., La Barbera, F., Napolitano, N. R., de Carvalho, R.~R.,
Romanowsky, A. J., 2012, MNRAS, 425, 577 (Spider V)

\bibitem[\protect\citeauthoryear{Tortora et al.}{2013}]{Tortora:13}  
Tortora, C., Romanowsky, A.J., Napolitano, N., 2013, ApJ, in press
($arXiv1207.4475T$)

\bibitem[\protect\citeauthoryear{Trager et al.}{2000}]{Tra:00} 
Trager, S.~C., Faber, S.~M., Worthey, G., Gonz\'alez, J.~J. 2000, AJ,
120, 165

\bibitem[\protect\citeauthoryear{Treu et al.}{2010}]{Treu:10} 
Treu, T., Auger, M.W., Koopmans, L.V.E., Gavazzi, R., Marshall, P.J.,
Bolton, A.S., 2010, ApJ, 709, 1195


%\bibitem[Treu et al.(1999)]{Treu:99} Treu, T., Stiavelli, M.,
%Casertano, S., M\o ller, P., Bertin, G. 1999, MNRAS, 308, 1037

\bibitem[\protect\citeauthoryear{Trevisan et al.}{2012}]{Trevisan:12} 
Trevisan, M., Ferreras, I., de~la~Rosa, I.~G., La~Barbera, F., 
de Carvalho, R.~R. 2012, ApJ, 752, L27

%\bibitem[\protect\citeauthoryear{Trujillo et al.}{2006}]{truj06} 
%Trujillo I., et al., 2006, ApJ, 650, 18

%\bibitem[\protect\citeauthoryear{Trujillo et al.}{2011}]{truj11}
%Trujillo I., Ferreras I., de La Rosa I.~G., 2011, MNRAS, 415, 3903

%\bibitem[\protect\citeauthoryear{Valdes et al.}{2004}]{fv04}
%Valdes, F. et al.\ 2004, ApJS, 152, 251

%\bibitem[\protect\citeauthoryear{van Dokkum et al.}{2008}]{vdk08} 
%van Dokkum P.~G., et al., 2008, ApJ, 677, L5

%\bibitem[\protect\citeauthoryear{van Dokkum \& Conroy}{2011}]{vDC:11}
%van Dokkum, P.~G., Conroy, C., 2011, ApJ, 735, 13

%\bibitem[van Dokkum \& Stanford(2003)]{vDS:03} van Dokkum, P.G.,
%Stanford, S.A. 2003, ApJ, 585, 78

\bibitem[\protect\citeauthoryear{van Dokkum \& Conroy}{2010}]{vDC:10} 
van Dokkum, P.G., Conroy, C., 2010, Nature, 468, 940

\bibitem[\protect\citeauthoryear{van Dokkum \& Conroy}{2011}]{vDC:11} 
van Dokkum, P.G., Conroy, C., 2011, ApJ, 735, 13

\bibitem[\protect\citeauthoryear{Vazdekis et al.}{1996}]{Vazdekis:1996} 
Vazdekis, A., Casuso, E., Peletier, R.F., Beckman, J.E., 1996, ApJS, 106, 307 

\bibitem[\protect\citeauthoryear{Vazdekis et al.}{1997}]{Vazdekis:1997}  
Vazdekis, A., Peletier, R.F., Beckman, J.E., Casuso, E., 1997, ApJS,
111, 203

\bibitem[Vazdekis et al.(2001)]{Vazdekis:01} Vazdekis, A., Salaris, M., Arimoto, N., Rose, J.A., 2001, ApJ, 549, 274

\bibitem[\protect\citeauthoryear{Vazdekis et al.}{2003}]{Vazdekis:2003} 
Vazdekis, A., Cenarro, A.J., Gorgas, J., Cardiel, N., Peletier, R.F.,
2003, MNRAS, 340, 1317

\bibitem[\protect\citeauthoryear{Vazdekis et al.}{2010}]{Vazdekis10} 
Vazdekis, A., S\'anchez-Bl\'azquez, P., Falc\'on-Barroso, J., et
al. 2010, MNRAS, 404, 1639

\bibitem[\protect\citeauthoryear{Vazdekis et al.}{2012}]{Vazdekis:12} 
Vazdekis, A., Ricciardelli, E., Cenarro, A.J., Rivero-Gonz\'alez,
J.G., D\'iaz-Garc\'a, L.A., Falc\'on-Barroso, J., 2012, MNRAS, 424,
157 (MIUSCAT-I)

\bibitem[\protect\citeauthoryear{Yamada et al.}{2006}]{Yamada:06} 
Yamada, Y., Arimoto, N., Vazdekis, A., Peletier, R.F., 2006, ApJ, 637,
200

\bibitem[\protect\citeauthoryear{Warren et al.}{2007}]{War07} 
Warren, S.~J., Hambly, N.~C., Dye, S., et al. 2007, MNRAS, 375, 213

\bibitem[Wegner et al.(2012)]{WCT:12} Wegner, G.A., Corsini, E.M., Thomas, J., Saglia, R.P., Bender, R., Pu, S.B., 2012, MNRAS, 144, 78

\bibitem[\protect\citeauthoryear{Wing \& Ford}{1969}]{WingFord:69} 
Wing, R.F., Ford, Jr., W.K., 1969, PASP, 81, 527

\bibitem[\protect\citeauthoryear{Worthey, Ingermann \& Serven}{2011}]{worthey:11} 
Worthey, G., Ingermann, B.A., Serven, J., 2011, ApJ, 729, 148

\bibitem[\protect\citeauthoryear{Worthey}{1998}]{Worthey:98} 
Worthey, G., 1998, PASP, 110, 888

%\bibitem[\protect\citeauthoryear{Westera et al.}{2002}]{pw02}
%Westera, P. et al.\ 2002, A\&A, 381, 524

%\bibitem[White(1980)]{White:80} White, S.D.M. 1980, MNRAS, 191, 1

%\bibitem[\protect\citeauthoryear{Wise \& Silva}{1996}]{Wise:96}
%Wise, M.~W., \& Silva, D.R. 1996, ApJ, 461, 155

%\bibitem[Worthey, Trager, \& Faber(1995)]{WTF:95} Worthey, G.,
%Trager, S.C., Faber, S.M. 1995, ASPC, 86, 203

%\bibitem[\protect\citeauthoryear{Worthey}{1994}]{wo94} 
%Worthey G., 1994, ApJS, 95, 107 

%\bibitem[\protect\citeauthoryear{Wu et al.}{2005}]{wu05} 
%Wu H., Shao Z., Mo H.~J., Xia X., Deng Z., 2005, ApJ, 622, 244

%\bibitem[\protect\citeauthoryear{Zhu et al.}{2010}]{Zhu:10}
%Zhu, G., Blanton, M.R., Moustakas, J., 2010, ApJ, 722, 491

%\bibitem[\protect\citeauthoryear{Zibetti et al.}{2012}]{zib12}Zibetti, S., Gallazzi, A., Charlot, S., Pierini, D., Pasquali, A., 2012, MNRAS, submitted ($arXiv:1205.4717$)


\end{thebibliography}
\end{document}